\documentclass[CBM,manyauthors]{cbm_notes}

\usepackage{rotating}
\usepackage{wrapfig}
\usepackage[english]{babel}
\usepackage{blindtext}
\usepackage{amssymb}
\usepackage{amsmath}
\usepackage{upgreek}
\usepackage{colortbl}
\usepackage[per-mode=symbol]{siunitx}
\usepackage{chemformula}
\usepackage[table]{xcolor}    
\usepackage{listings}
\usepackage{hyperref}
\hypersetup{
    unicode=false,          
    pdftoolbar=true,        
    pdfmenubar=true,        
    pdffitwindow=false,     
    pdfstartview={FitH},    
    pdfnewwindow=true,      
    colorlinks=true,       
    linkcolor=red,          
    linkbordercolor=red,
    citecolor=magenta,        
    filecolor=magenta,      
    urlcolor=cyan           
}

\DeclareSIUnit\clight{\text{\ensuremath{c}}}
\DeclareSIUnit\eVperc{\eV\per\clight}

\definecolor{jocol}{rgb}{0.0,0.0,0.5}

\newcommand\NOTENUMBER{25001}

\begin{document}%

\begin{titlepage}
%
%
\PHnumber{CBM-TN-\NOTENUMBER} 
\PHdate{\today}
%
\title{
Performance and quality testing of CBM RICH front-end electronics}
\ShortTitle{RICH FEE testing}   

\author{%
P.~Subramani$^{1}$,
C.~Pauly$^{1}$,
J.~Förtsch$^{1}$,
D.~Pfeifer$^{1}$,
K.-H.~Kampert$^{1}$
}
\author{
1. Bergische Universit\"at Wuppertal, Germany
}
\author{Email: subramani@uni-wuppertal.de}
%
\ShortAuthor{CBM-TN-\NOTENUMBER}      
\begin{abstract}
This note describes the laboratory setup at the University of Wuppertal and its use to measure charge sharing crosstalk, perform high rate and high occupancy tests of MAPMT-DIRICH readout electronics using a realistic detector setup, and also includes the analysis of induced electronic noise characteristics of a new iteration of the DIRICH power module to be used in the CBM RICH readout chain.

\end{abstract}
\end{titlepage}
\tableofcontents
\clearpage
\section{Ring Imaging Cherenkov detector front-end electronics}
\label{rich-fee}
The CBM Ring Imaging Cherenkov Detector (RICH) is designed for the efficient separation of electrons from pions up to a momentum of $\sim\SI{6}{\giga\eVperc}$ at SIS100 energies.
Charged particles moving faster than the speed of light in the medium produce Cherenkov photons~\cite{Cherenkov_effect}.
In the CBM RICH detector, \ch{CO2} gas is used as the radiator medium.
The Cherenkov threshold for electrons in CBM RICH is $\SI{16}{\mega\eVperc}$, whereas for pions it is $\SI{4.7}{\giga\eVperc}$~\cite{co2_threshold}.
The Cherenkov photons are emitted in the shape of a cone around the charged particle trajectory inside the radiator volume.
Spherical aluminum mirrors of high reflectivity reflect- and focus the emitted photons onto two photon detection cameras.
The angle of the cone depends on the momentum of the particle, which then translates to the radius of the ring in the photon detection plane.

The CBM RICH uses H12700 multi-anode photo multiplier tubes(MAPMT) manufactured by Hamamatsu as spatially resolved photon sensors.
Each H12700 MAPMT has 64 pixels ($8 \times 8$) of size $\SI{6}{\milli\meter} \times \\ \SI{6}{\milli\meter}$, a peak quantum efficiency of $33\%$ and a large effective area of $87\%$.
The gain per photoelectron is in the order of $\sim 10^6$, with the dark count rate of $<\SI{100}{\hertz}$ per pixel. 
Each set of six photon sensors is connected to a backplane, which also hosts all necessary front-end readout components. 
Furthermore, the backplane also provides a light- and gas-tight seal of the radiator volume and incorporates all connections to power and signal lines.

Signals from the MAPMTs are read out via the DIRICH front-end board (FEB). 
Each channel of the DIRICH FEB incorporates a single-stage transistor-based inverter amplifier with a gain of $\sim\SIrange{10}{20}{}$ in order to provide sufficient amplification to the input signal.
The native PMT single photon signal is a short, negative pulse of $\SIrange{3}{5}{\nano\second}$ width and an amplitude in the range of $\SI{10}{\milli\volt}$ per photon (depending on the individual PMT gain).
After analog pre-amplification, the signal is directly fed to one of the two inputs of a differential line receiver port of the onboard ECP5 FPGA.
Here it is compared to a channel-individual threshold voltage applied to the second input of the differential receiver.
This threshold voltage is generated via Pulse-Width Modulation (PWM), using two Mach XO3 FPGAs on the same DIRICH FEB. 
In this approach, the differential line receivers onboard the ECP5 FPGA are used as signal discriminators, minimizing the need for additional components.

The times of both threshold crossings (leading- and trailing edges of the signal) are subsequently measured in a high-precision FPGA-TDC implemented onboard the ECP5 FPGA.
The data flow in the TDC is explained in section~\ref{section-dataflow}.
Each DIRICH FEB has 32 analog input channels.
A single backplane can host up to six MAPMTs and twelve DIRICH FEBs (two DIRICH FEBs per PMT).

The data from all individual DIRICH FEBs on a given backplane are transferred via individual $\SI{2}{Gbps}$ digital links to the DIRICH combiner/concentrator module hosted on the same backplane.
The DIRICH Combiner acts as a simple data hub, combining data stream of the 12 input links from one backplane to a single $\SI{2}{Gbps}$ optical output link.
The data protocol on all these data links is defined by the TRBnet specifications~\cite{Michel_2017}.
The combiner module can also receive external trigger signals, which are distributed simultaneously via the backplane to each DIRICH module. An additional, special TDC channel on each DIRICH assigns timestamps to the incoming trigger signal.
This trigger signal is needed for later hit assignment and event building across multiple DIRICH backplanes.

\begin{figure}[h!tb]
    \centering
    \includegraphics[width=10cm]{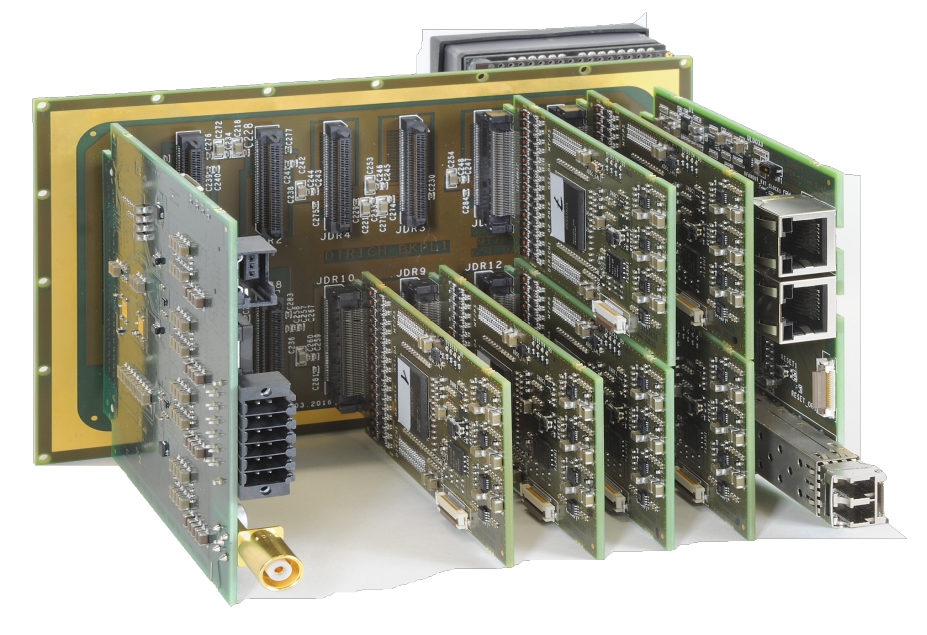}
    \caption{Backplane with MAPMT and readout components. Picture Courtesy: G. Otto (GSI)}
    \label{fig:full-backplane}
\end{figure}

The DIRICH-Power module (one per backplane) provides all low voltage power supply lines ($\SI{1.1}{\volt}$, $\SI{1.2}{\volt}$, $\SI{2.5}{\volt}$, and $\SI{3.3}{\volt}$) generated from a single input ($\SIrange{18}{36}{\volt}$) using DC/DC converters. Special attention is put to minimizing the amount of electronic noise (EMC) induced on the analog components on the backplane.
In addition, the DIRICH-Power module also transfers a single HV input supply line ($\sim\SI{-1}{\kilo\volt}$, Lemo-01 input connector) via the backplane to all 6 MAPMTs.
A picture of the backplane with readout components is shown in figure~\ref{fig:full-backplane}.

\section{Motivation}
\label{motivation}
Both the upgraded HADES RICH and the mini-CBM (mCBM) setups at GSI SIS18 use the same DIRICH front-end electronics, which are also projected for the CBM RICH readout, along with the already acquired H12700 MAPMTs.
Initial testing and detailed characterization of all MAPMTs has already been completed, the obtained series data are used for gain matching sets of six MAPMTs on one backplane sharing the same HV supply channel~\cite{mapmt_series_testing}. 
The DIRICH FEBs have already been successfully tested for average input hit rates up to $\SI{100}{\kilo\hertz}$ per channel~\cite{Michel_2017}. 
In the CBM RICH, when running at maximum heavy ion interaction rates of up to $\SI{10}{\mega\hertz}$ and at highest operational energies, one expects maximum hit rates in the order of a few $\SI{100}{\kilo\hertz}$ per pixel ($\SI{6}{\milli\meter} \times \SI{6}{\milli\meter}$) in the \lq\lq{}hottest\rq\rq{} regions of the photon detector.

Emission of Cherenkov photons is a quasi-instantaneous process, emitted photons travel basically with the same speed as the emitting electron (i.e., speed of light in the radiator medium), and the track length of all photons emitted from the same lepton is quasi-identical. As a consequence, all Cherenkov photons belonging to the same Cherenkov ring do simultaneously arrive at the MAPMT.
They result in a larger hit multiplicity (\lq\lq{}occupancy\rq\rq{} if normalized to the total number of MAPMT pixels) during a very short time duration (typically  $< \SI{100}{\pico\second}$)~\cite{lebedev_collab_meet_2020}.
The dimension of the active area of a H12700 MAPMT is $\SI{48.5}{\milli\meter} \times \SI{48.5}{\milli\meter}$.
Based on the average number of photons per ring ($\sim 30$) and the expected ring radius ($R \sim \SI{50}{\milli\meter}$) the number of simultaneous photon hits per MAPMT (i.e., PMT occupancy) is estimated to be 6 -- 9 simultaneously incident photons per MAPMT.
Incident light on one MAPMT channel can induce signals in other channels, a phenomenon known as crosstalk.
Crosstalk can be of two types, neighboring channel charge sharing crosstalk and capacitive crosstalk.
The neighboring channel crosstalk is caused by parts of the electron avalanche leaking into neighboring pixels, producing a signal above threshold.
While capacitive crosstalk is the result of the parasitic conduction between the channels due to large signals at the common photocathode \footnote{The H12700 MAPMT is designed with common dynode grids in combination with a segmented anode}.
These crosstalk signals are typically short and can be removed by imposing a Time over Threshold cut (discussed in sec~\ref{section-high-occupancy-result}).
Larger occupancies may increase the probability of crosstalk, hence the characterization of the MAPMT as a function of photon occupancy is paramount.

The large maximum expected hit rate also implies a large data rate at the readout- and combiner boards.
All 12 DIRICH FEBs on a given backplane transmit data to a single combiner, which forwards it over a single data link to subsequent DAQ readout stages.
Therefore, this combiner upstream link is one of the current bottlenecks for the data flow.
In order to maximize the data rate performance of the existing hardware chain, individual channel- and event buffer sizes on DIRICH- and Combiner modules can be optimized (in software).
In a second step, also the uplink speed of the existing combiner modules might be further increased (currently \SI{2}{Gbps}, FPGA limit is \SI{10}{Gbps}) using the existing hardware.
Ultimately, a new version of the combiner module, possibly with more than one output link, might follow.
However, the analog part of the DIRICH front-end modules (2500 modules in total) is fixed already and could possibly pose a hard limit to the achievable maximum hit rates for many years to follow.
This is why a detailed study of the hardware limitations in terms of hit rate and data quality is of utmost importance and needed to be done before the start of series production.
Since HADES and mCBM, in general, both do not provide the environment for such tests, a laboratory setup was prepared to study the high-rate behavior of the DIRICH readout chain,  and also to define the optimal readout parameters.
The results of these studies are summarized in the subsequent sections.

\section{Laboratory setup}
\begin{figure}[htb]
    \centering
    \includegraphics[width=0.55\textwidth]{setup-schematics.pdf}
    \includegraphics[width=0.35\textwidth]{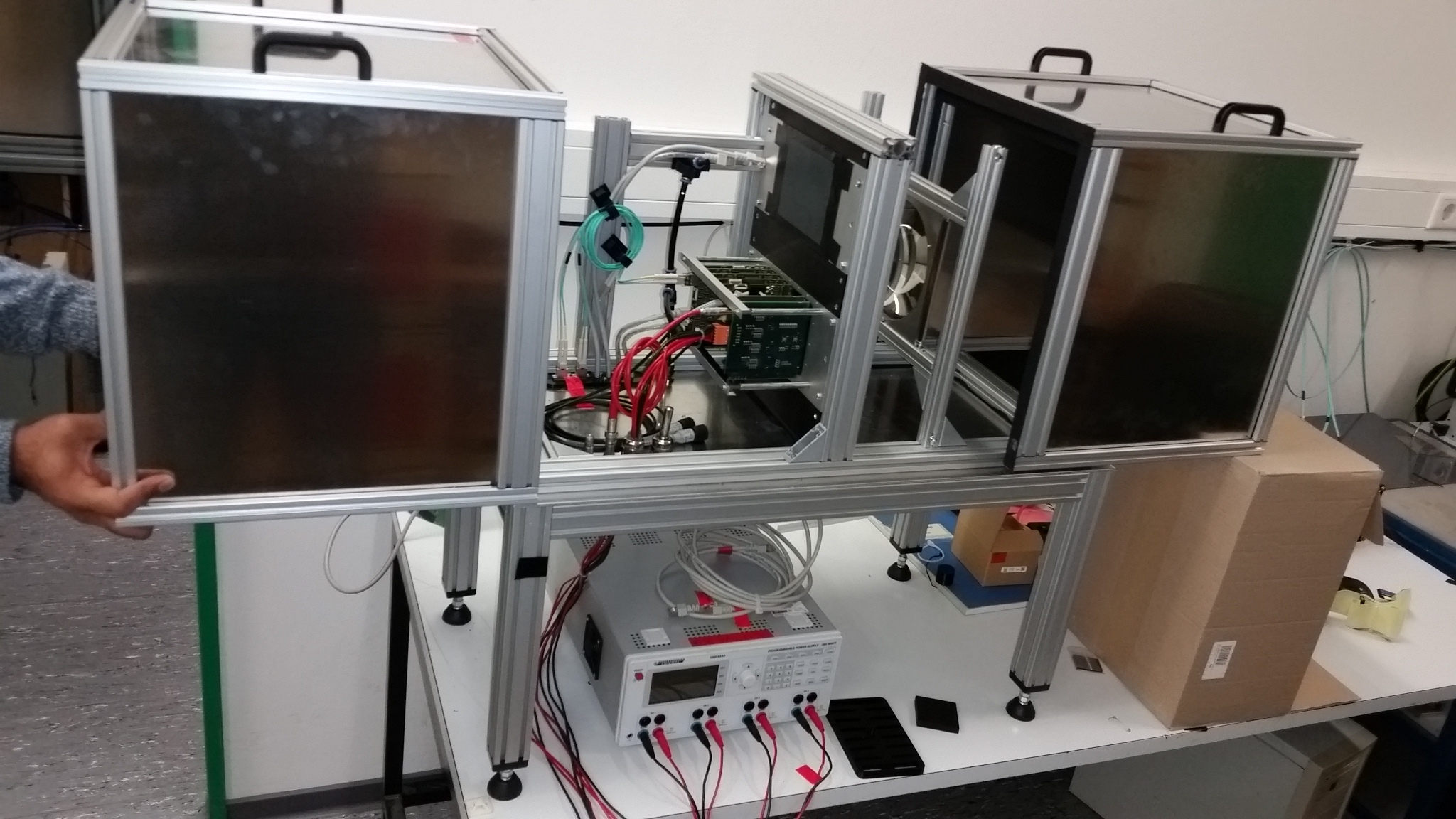}
    \caption{Left panel: Schematic showing components of the laboratory setup. The MAPMT is illuminated by a LED and a picosecond laser source. The signal from the MAPMT is processed by the DIRICH-TRB3 readout chain. Right panel: Photograph of the laboratory setup. 
    }
    \label{labsetup}
\end{figure}

The setup basically consists of a single readout module, equipped with a single MAPMT and two DIRICH modules for readout, and two different light sources.
One is a pulsed laser, and the second is a CW LED driven by a DC current source operated at very low current in the order of \SI{}{\micro\ampere} in order to achieve statistical single-photon detection on the sensor.
This setup is enclosed in a light-tight box (originally built for a beam test at the COSY accelerator at FZ Jülich cf.~\cite{pauly_2017}). The main goal is to operate MAPMT and DIRICH readout under realistic high-rate conditions, as similar to the final CBM RICH high-rate operation as possible.
The schematic and photograph of the setup are shown in figure~\ref{labsetup}.

Readout of the data is done using a TRB3 (Trigger Readout Board: V3), connected to the upstream optical link of the DIRICH combiner module on the backplane.
In this setup, the TRB3 board, in addition to the pure data transport, also incorporates a Central Trigger System (CTS) that generates regular readout triggers fed to each individual DIRICH in the setup.
Later in CBM, this part of the readout will be replaced by a CRI-based readout of the CBM DAQ concept which will not be studied here.
Data is sent via a single GBit Ethernet connection from the TRB3 to a DAQ PC for disk storage. 
A pulsed laser light source produces ultra-short picosecond pulses (width $<$\SI{50}{\pico\second} in Violet, $\lambda \sim$ \SI{400}{\nano\meter}) which are passed through an optical cable into a collimator with a conically opening slit structure.
At the input of the collimator, the incoming light is diffused by using a white glass diffuser and homogeneously illuminates the collimator slit entry.
Light from the collimator is then projected onto the MAPMT, generating a ring-like image of numerous simultaneous photons, mimicking a Cherenkov ring in the real detector.
A pulse-synchronous reference signal, generated by the laser pulser, is measured additionally using a dedicated TDC on the TRB3, which allows isolating PMT signals generated by the pulse laser (so-called “signal” hits) from hits by other sources (“background”).
The photon yield per laser pulse of the laser can be adjusted (0 – 30 photons/pulse) in order to vary the MAPMT occupancy.
In general, the large interaction rate at the future CBM experiment, and possibly also fluorescence effects, will produce a large photon background that is not time-correlated to signal photons in a given event.
This background must be incorporated into the setup.
This is achieved by implementing a second light source into the setup: A LED driven by a constant current source, causing an uncorrelated, statistically distributed photon background. 
Combining both light sources, the setup provides a controlled signal photon rate together with a controlled stochastic background noise, which is not possible in a test beam environment seemingly closer to the final application.

\section{Charge sharing crosstalk measurement}
In the first commissioning measurement, this setup was used to re-evaluate the neighboring-channel charge-sharing crosstalk of the H12700 MAPMTs in comparison to its predecessor H8500.
A similar study has been done previously \cite{Reinecke:2016grs} using CERN test beam data but using different (nXYter-based~\cite{BROGNA2006301}) readout electronics and slow signal shaping.
The new results presented here are based on the DIRICH readout to be used in the later CBM RICH detector.
The measurement aims to quantify the amount of crosstalk (to be included in final Monte Carlo simulations of the entire RICH detector in the experiment) and to study the influence of operating threshold settings on the noise suppression and crosstalk. 

\subsection{Measurement principle}
The measurement presented here is based on the illumination of the MAPMT with low-intensity short laser pulses ($\sim 5$\, ps).
A reference signal from the pulse generator is fed into the data and used to select “signal” hits based on a time cut around the prompt peak.
The setup depicted in Figure~\ref{labsetup} is used without a collimator and also the LED is switched off.
The charge-sharing crosstalk into a neighboring channel is then estimated as follows:
\begin{itemize}
\item Only events with two simultaneous “signal” hits on the MAPMT are selected (characterized by a relative time difference between reference pulse and PMT signal of less than 10\, ns, and a ToT (Time over Threshold) $\ge$ 3\, ns).
\item The distance between the positions of the two hits (counted in the number of pixels) is histogrammed.
\item The obtained 1D distance distribution is compared to a simple Monte Carlo simulation where two hits are randomly thrown on the MAPMT ($8 \times 8$ pixels).
\item Non-uniform illumination of MAPMTs and efficiency variations of the pixels might distort the statistical distribution of the measured distances.
In order to account for such effects in simulation, an efficiency matrix was derived for MAPMT based on the normalized absolute number of hits in each pixel,
\begin{equation}
\textrm{efficiency} = \frac{\textrm{Hit multiplicity in a pixel}}{\textrm{Total number of events}}.
\end{equation}
This is then applied to each pair of hits as a correction factor.
\item The simulated distance distribution is scaled to the measurement such that optimal agreement for all distances $d \ge 2$ is achieved. Any excess yield of measured distances $d=1$ can now be attributed to the charge-sharing crosstalk.
\item Finally, the relative crosstalk contribution is quantified using the following formula:
   \begin{equation}
       F_{cross\,talk} = \frac { N_{data,d=1} - N_{sim,d=1} } {N_{sim,all}} \times 100 \%
   \end{equation}
Where $N_{data,d=1}$ and $N_{sim,d=1}$ are the number of entries in data and simulation histograms corresponding to distance $d=1$. $N_{sim,\ all}$ is the total number of simulated pair events.
\item The procedure is repeated for four different PMTs and five different threshold values.
\end{itemize}

\subsection{Results and Discussion}
\begin{figure}[htb]
    \centering
    \includegraphics[width=0.45\textwidth]{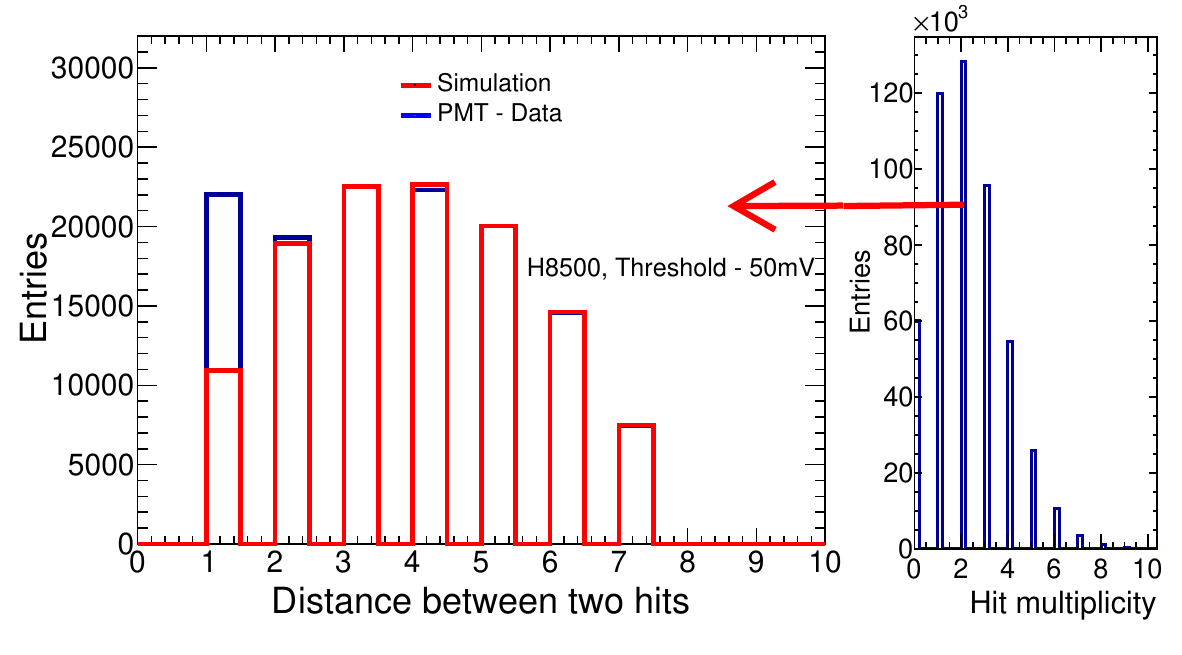}
    \includegraphics[width=0.45\textwidth]{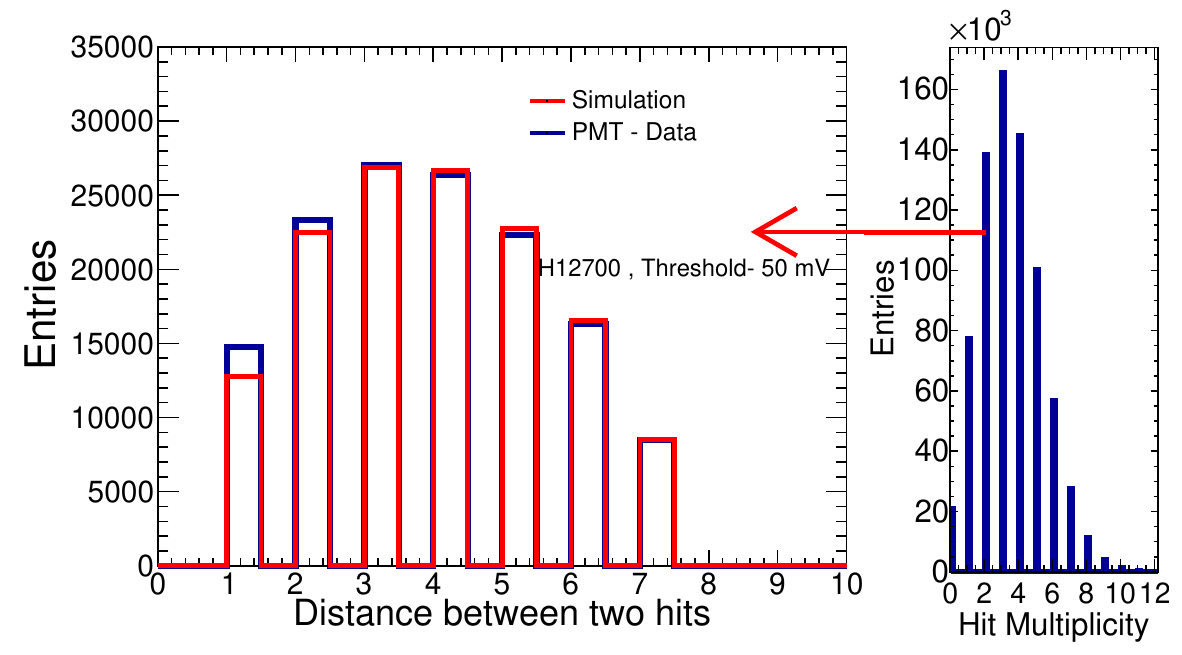}

    \caption{Measured distance between two hits for data (blue) and simulation (red) and the hit Multiplicity of detected photons per event. Left: H8500 MAPMT, Right: H12700 MAPMT. Threshold: \SI{50}{\milli\volt} (after pre-amplification).
    }
    \label{fig:crosstalk_distance_distribution}
\end{figure}

The measurement results are shown in figure~\ref{fig:crosstalk_distance_distribution}.
The measured distance distribution matches very well with the simulation results, with the only sizable difference being observed in the first bin, corresponding to the crosstalk into neighboring channels. 
The H8500 MAPMT (figure~\ref{fig:crosstalk_distance_distribution} (left)) shows a sizable amount of crosstalk which is much reduced in the case of the H12700 MAPMT (figure~\ref{fig:crosstalk_distance_distribution} (right)).

\begin{figure}[!h]
\centering
  \includegraphics[width=0.45\textwidth]{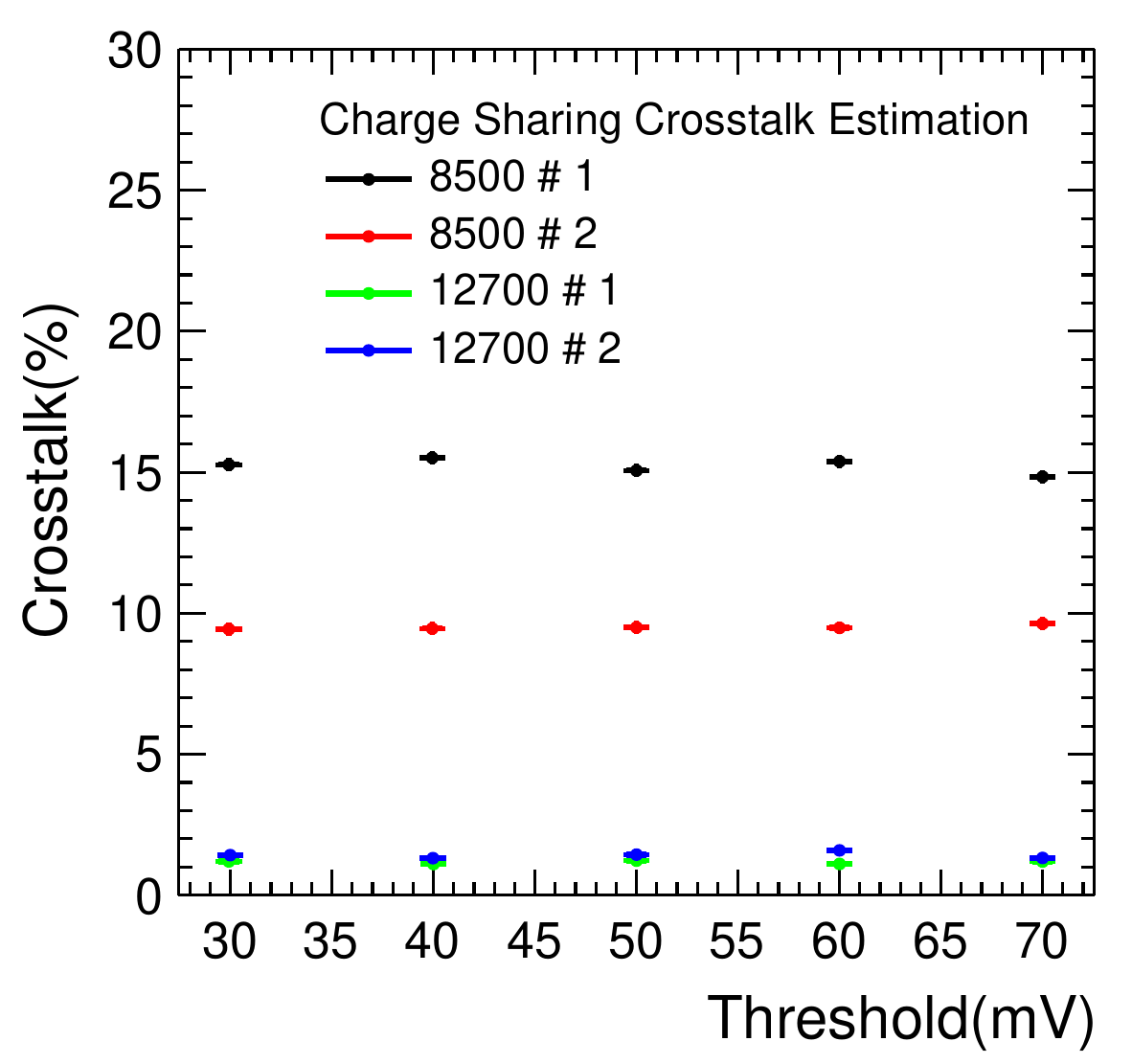}
  \caption{Neighboring channel charge sharing crosstalk is plotted against noise rejection threshold voltage. Different colors indicate different MAPMTs used for measurement.
  H12700 MAPMTs have significantly lower neighboring channel crosstalk compared to H8500 variant.}
  \label{crosstalk-2}
\end{figure}

Results for measurements on four different MAPMTs (2x H8500 and 2x H12700) are summarized in figure~\ref{crosstalk-2}, with relative crosstalk being plotted as a function of applied noise reduction threshold value (threshold after pre-amplification).

The observed crosstalk for H8500 MAPMTs is in the order of 10-15 $\%$, which is in good agreement with the earlier measurements.
In contrast, the crosstalk for both H12700 MAPMTs is only in the order of 1-2$\%$. 
This is expected based on the MAPMT series testing results mentioned earlier but could now be confirmed with the full DIRICH readout chain. 

As a second interesting result, the crosstalk contribution does not depend on the applied noise threshold. 
This comes as a bit of a surprise, since the total amount of charge generated by a photon is fixed, and sharing this charge over two channels should result in reduced signal amplitude and thus reduced charge-sharing contribution at higher threshold values. 
However, the range of threshold values covered in this study resembles realistic threshold values as would be used in the real experiment. 
All studied thresholds are still well below the average single photon signal, in the order of 10\,\% to 20\,\% of the typical single photon charge. 
The expected reduction of charge sharing would probably be observed at much higher threshold values only.

\section{High occupancy analysis}
\label{section-high-occupancy-result}
A good understanding of possible capacitive crosstalk and time over threshold (ToT) spectra at high occupancy (as described in sec.~\ref{motivation}) is required for implementing it properly in the detector simulation and optimal selection of ToT threshold for the operation of MAPMT-DIRICH readout.
It is stipulated that the charge sharing crosstalk between neighboring channels, as discussed in the preceding section, is independent of occupancy in the MAPMT, as it persists whenever there is at least a single hit in a pixel.
In this section, the measurement principles and results of the high occupancy test of MAPMTs are presented.
The capabilities of the readout chain to handle possible crosstalk resulting from these higher occupancies are discussed.
\subsection{Measurement principle}
\begin{figure}[h!tb]
\centering
  \includegraphics[width=0.65\textwidth]{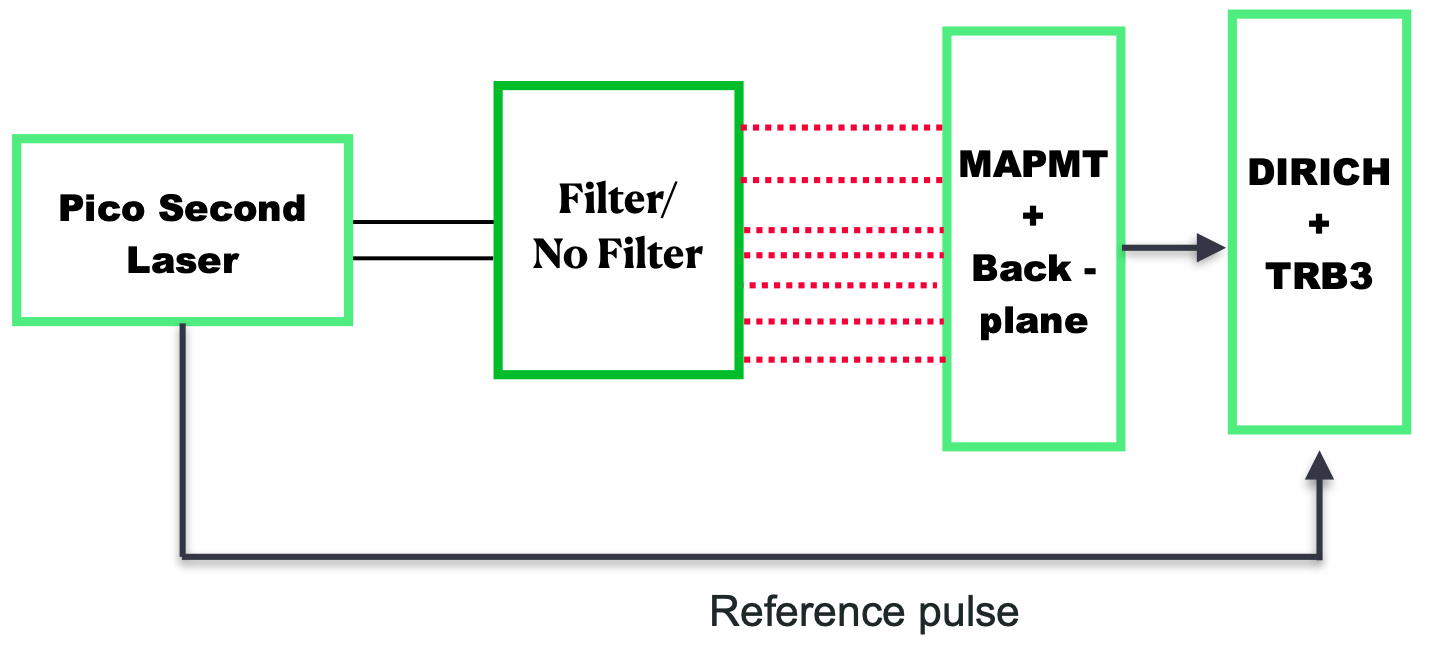}
  \caption{Schematic of the setup for the high occupancy measurement. No LED is used for the measurement. A set of neutral density filters is used for attenuating the photon intensity from the laser.}
  \label{highoccupancy setup}
\end{figure}

A picosecond laser is used to produce high photon occupancy. To separate the signal from the background, a reference pulse is used as described in the previous sections.
The hits that are correlated in time ($ \leq$ \SI{10}{\nano\second}) are selected.
The schematic of the setup used for the study is depicted in figure~\ref{highoccupancy setup}.
Photon emission from laser is a stochastic process.
It is quite challenging to accurately measure the photon flux from the laser (on the level of few photons)  with a basic laboratory setup. 
Hence, in order to have a measurable control over the photon yields, a set of neutral density (ND) filters (with well-known attenuation factors) are used as a calibration device.
Using this setup, instead of quantifying the number of photons per single pulse the measurement is reduced to counting the number of detected single photons in a large given number of pulses, with well less than one photon per pulse. Difficult absolute quantification is reduced to simple binary counting.

The measurement principle for this analysis is as follows:
\begin{enumerate}
    \item The neutral density (ND) filters to be used are calibrated, i.e., the transmittance $T$ of the ND filter is calculated .
    \item An ND filter is placed in between laser and MAPMT, thereby attenuating the laser yield. 
    \item With the measured transmittance of the filter, expected hit multiplicity without filter is calculated. 
    \item The measurement is repeated by removing the filter while maintaining the same intensity of laser.
    \item The observed hit multiplicity is compared with the expected result, any excess measured is attributed to crosstalk resulting from the high occupancy setting.
    \item The intensity dependent crosstalk contribution is estimated by repeating steps from 2 to 5 for different intensities of the laser.
\end{enumerate}
An explanation of the measurement principle is provided in the subsequent sections.
\subsection{Calibration of filters}
\label{section-filter-calibration}
\begin{figure}[h!tb]
\centering
  \includegraphics[width=0.75\textwidth]{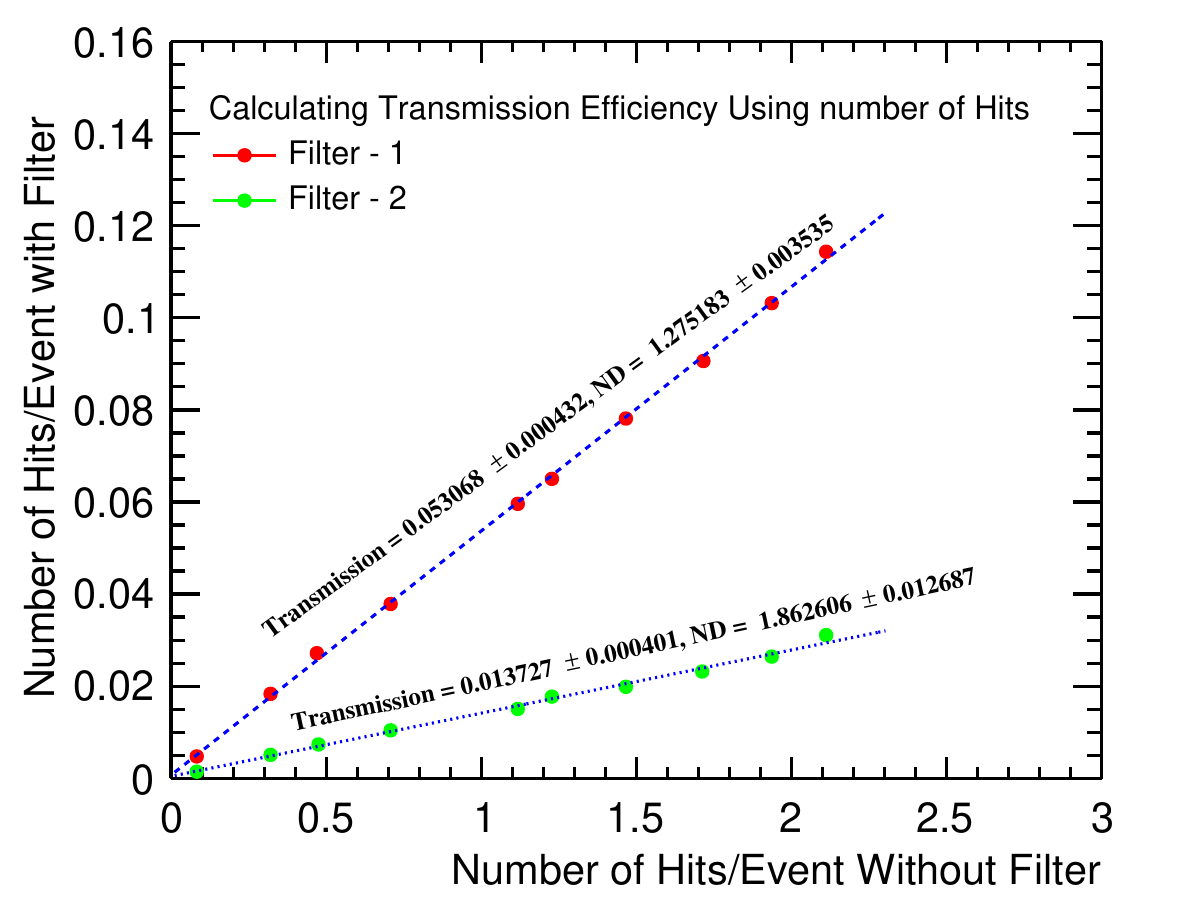}
  \caption{Hit multiplicity with filter is plotted against hit multiplicity without filter. The resultant graph is fitted with line and the extracted slope is considered as the transmittance of the filter.}
  \label{highoccupancy-calibration-without-tot}
\end{figure}

In order to calibrate the ND filters, the intensity of the laser pulses is reduced to produce low hit multiplicity in an MAPMT (hits $< 3$).
In this setting the capacitive crosstalk contribution is assumed to be minimal.
The hit multiplicity per event without filter ($N_{without-filter}$) is measured and plotted on the $X$ axis.
Further the neutral density filter is placed between the laser and MAPMT.
The resulting reduced hit multiplicity is measured and plotted on the $Y$ axis ($N_{with-filter}$).
The number of dark hits ($N_{dark-hits}$), which is the number of hits measured when the laser is off (because of stray light and/or other sources), is subtracted from both values.
The procedure is repeated for 10 different laser intensities.
The resultant graph (shown in figure~\ref{highoccupancy-calibration-without-tot}) is fitted with a straight line,
\begin{equation}
    N_{without-filter} \,=\, N_{with-filter} \times T\, +\, N(0),
\end{equation} where $N(0)$ is the $Y$ intercept and $T$ is the slope.
The slope derived by line fit is considered as the transmission probability ($T$) of the ND filter.
To reduce bias, the measurement is repeated using two different filters with different transmittance
 and the results are tabulated in~\ref{table:filter}.
 A similar procedure is repeated by imposing ToT-cut (ToT $>$ \SI{3}{\nano\second}) on the hit selection (appendix~\ref{app-calibration-with-tot}).
\begin{table}[h!tb]
\caption{Transmission probability and neutral density are calculated for the two different filters and compared with their nominal value (taken from ~\cite{thor-labs-nominal}).
The measured neutral density is in agreement with the nominal value .
}
\label{table:filter}
\begin{tabular}{|p{2.5cm}|p{3cm}|p{3cm}|p{2.8cm}|}
\hline
\textbf{Filter Number} & \textbf{Transmission probability (T)} & \textbf{Neutral density} (ND = $\log_{10}\frac{1}{T}$) & \textbf{Nominal ND} \\ \hline
1        & 0.0531 ± 0.0004             & 1.2751 ± 0.0035 & 1.3 $\pm$ 0.02\\
2      & 0.0137 ± 0.0004             & 1.8626 ± 0.0126  & 2 $\pm$ 0.2\\
\hline
\end{tabular}
\end{table}

\subsection{Estimation of crosstalk at high occupancy conditions}
The calibrated filters are placed in between the laser and MAPMT.
The hit multiplicity with filter ($N_{with-filter}$) is measured and the intensity of the laser is adjusted such that multiplicity does not exceed 2 hits (crosstalk is assumed to be minimal).

\begin{figure}[h!tb]
    \centering
    \includegraphics[width=0.49\textwidth]{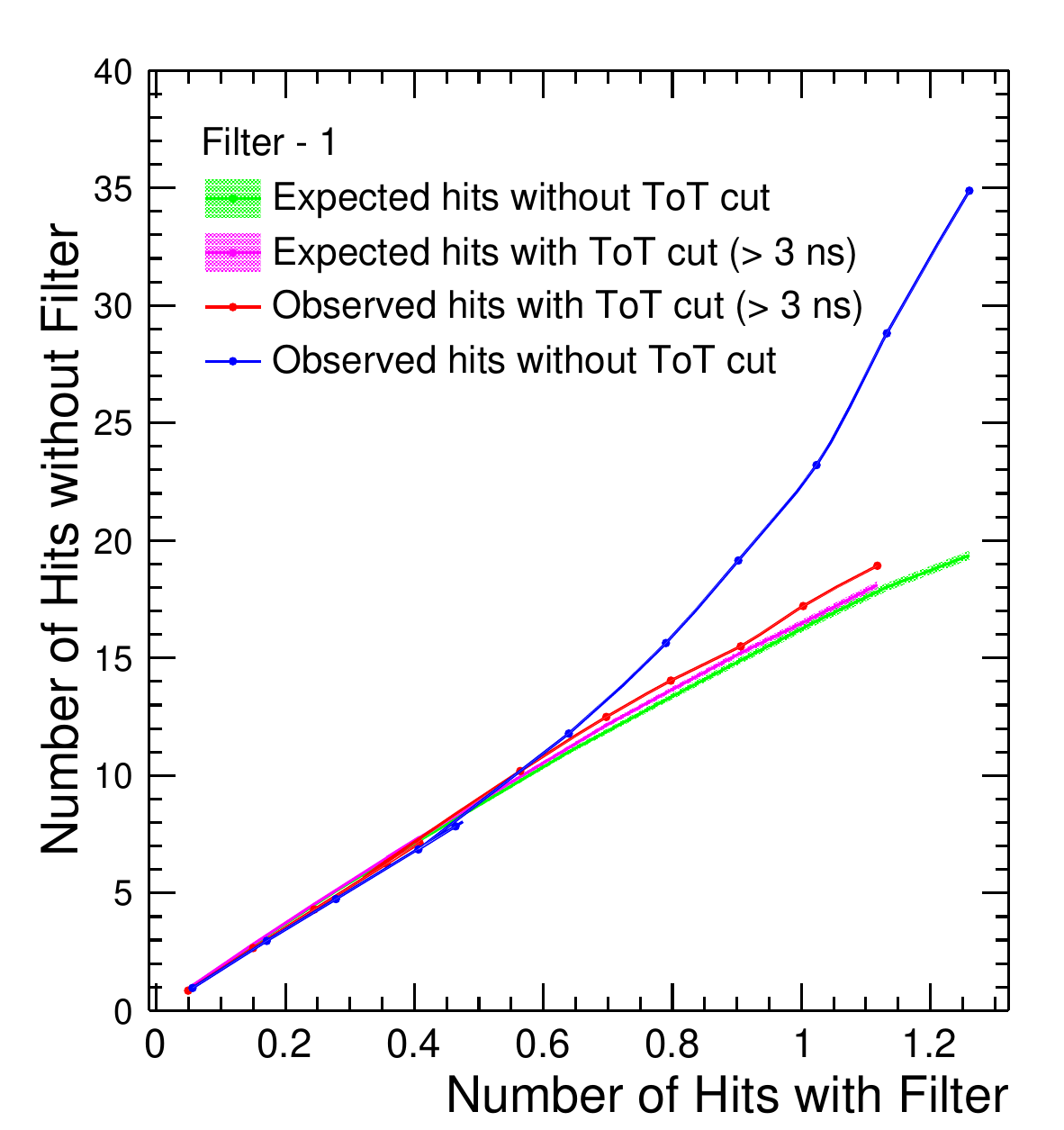}
    \includegraphics[width=0.49\textwidth]{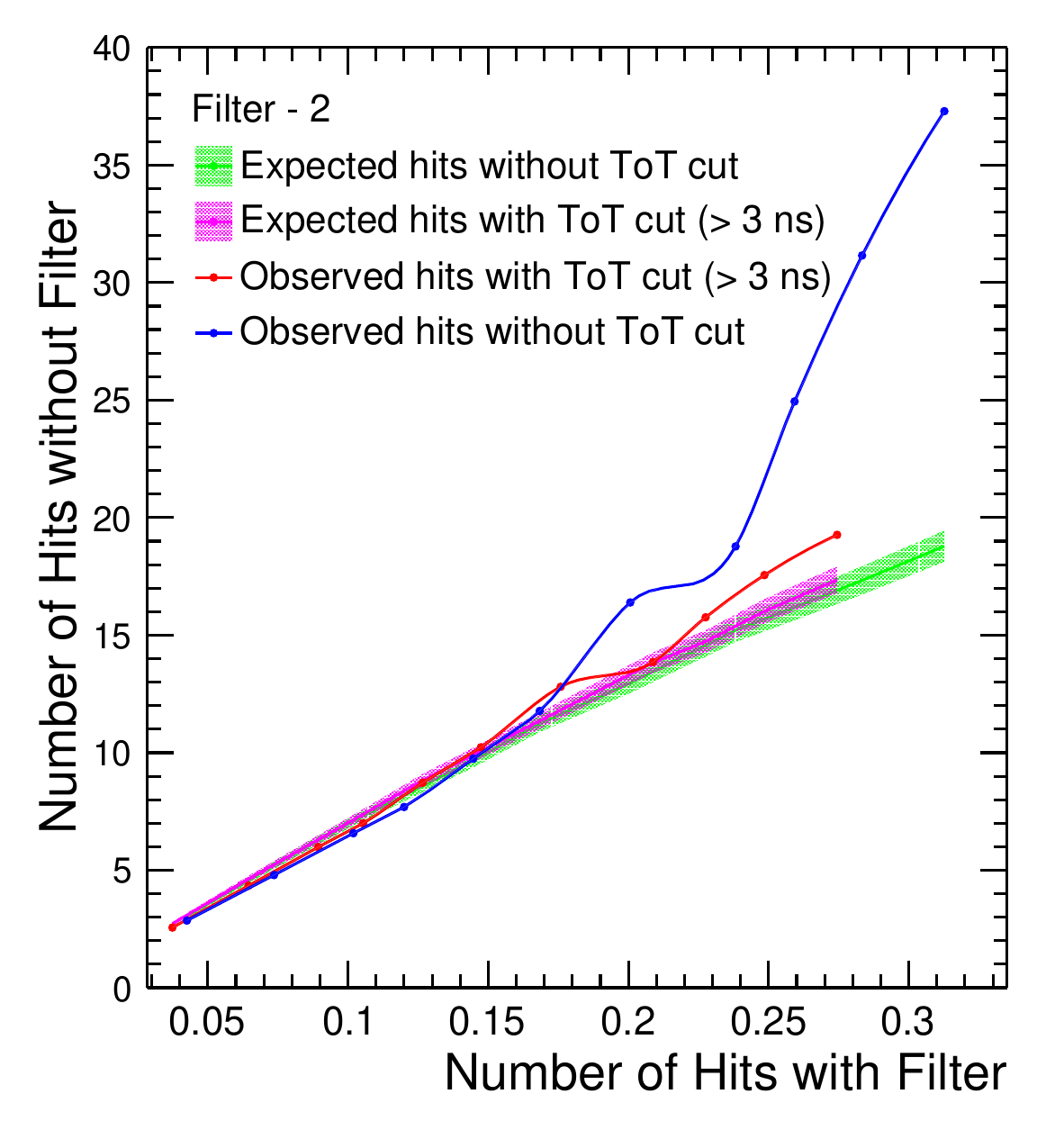}
    \caption{Expected number of hits vs. observed number of hits for filter 1 (Left) and filter 2 (Right).  Without any ToT-cut the observed number of hits increases with increasing hit multiplicities. However, the observed number of hits with ToT-cut is in agreement with the expected number of hits. Since filter 2 has transmission close to 1\%, the error in estimation of hits in higher multiplicities is higher than when using filter 1 which has about 5\% transmittance. 
    }
    \label{high-occ-result-1}
\end{figure}

The expected hit multiplicity without filter (adjusted for the dark hits) is calculated as,
\begin{equation}
    N_{Expected}\, =\,N_{with-filter}\,\times \frac{1}{T}-\,N_{dark-hits}\,-\,N_{double-photons}.
\end{equation}
An additional correction is applied to correct for the increased probability of double photon hits (two real single photons reaching the same channel) within the same pixel at increasing occupancy:
Such contribution from double photon ($N_{double-photon}$) is simulated (appendix~\ref{app-sim-double-photon}) and subtracted accordingly.
Calculated expected number of hits are plotted in figure~\ref{high-occ-result-1}.
The expected hit multiplicity without ToT-cut is slightly higher than the same with filter in figure~\ref{high-occ-result-1}, this is caused by the estimation of transmittance $T$ with and without ToT-cut (appendix~\ref{app-calibration-with-tot}).
The difference in the estimation of the T might be due to the negligible crosstalk associated with low hit multiplicities, which is assumed to be zero in these studies.

The filter is removed and hit multiplicity $N_{without-filter}$ is measured.
Measured hit multiplicities are adjusted for dark hits as,
\begin{equation}
    N_{Observed}\, =\,N_{without-filter}\,-\,N_{dark-hits}.
    \label{eqn:nobserved}
\end{equation}
The measurement procedure is repeated for 12 different laser intensities, the results are plotted in figure~\ref{high-occ-result-1}.
Each data point in the analysis is averaged over $10^6$ regular readout trigger events (Readout trigger — \SI{10}{\kilo\hertz}, laser frequency — \SI{10}{\kilo\hertz}). 
In the expected range of hit multiplicities of 6–9 hits / event / MAPMT (8–14\% occupancy), the additional number of hits due to crosstalk is minimal (less than 1) as shown in figure~\ref{high-occ-result-1}.
Significant crosstalk hits are observed for the expected hit multiplicities greater than 10 hits.
Similar results were observed for both the filters under test.
As anticipated, the contribution of capacitive crosstalk hits increases with an increase in hit multiplicity.
However, once the ToT-cut of \SI{3}{\nano\second} is employed on the hits, the observed hit multiplicities follow the trend of expected hit multiplicities.
This indicates that the ToT-cut plays a vital role in eliminating capacitive crosstalk signals.

\begin{figure}[h!tb]
    \centering
    \includegraphics[width=0.45\textwidth]{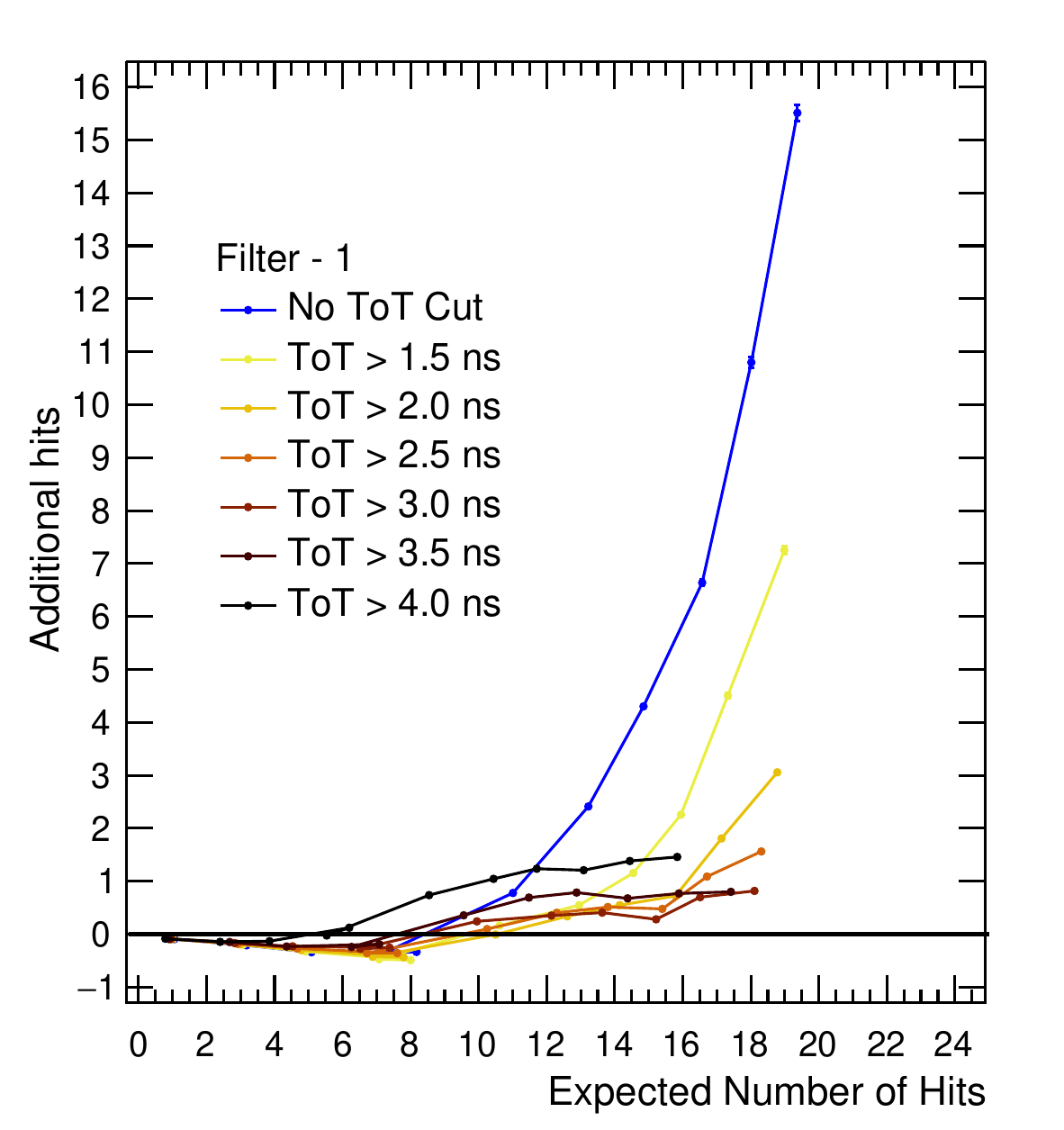}
    \includegraphics[width=0.45\textwidth]{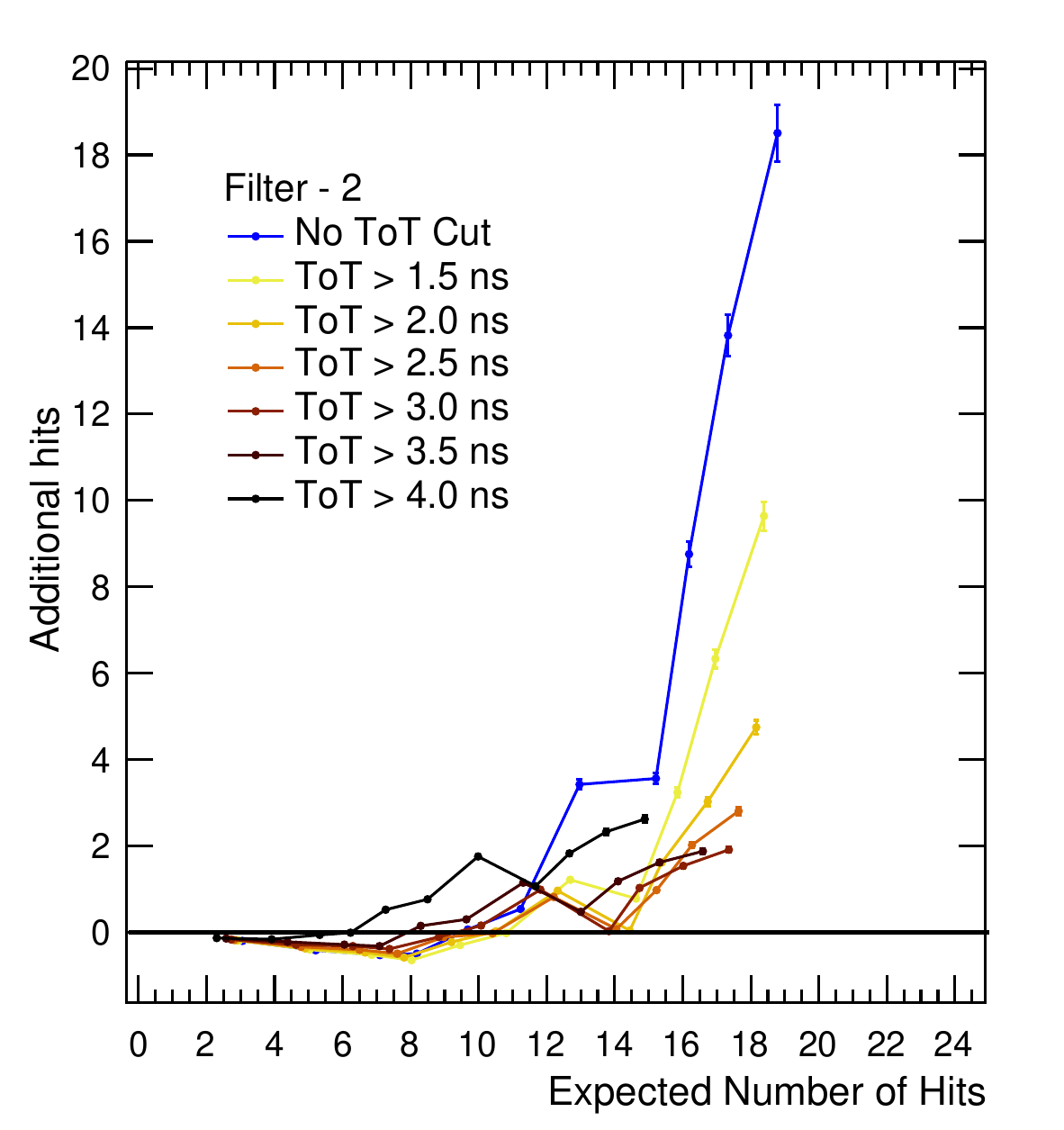}

    \caption{ Excess hits (observed-expected) as a function of the expected hit multiplicity for different ToT-cuts for the filter 1 (left panel) and filter 2 (right panel).
    }
    \label{high-occ-result-2}
\end{figure}

Choosing the optimal ToT-cut depends on its capacity to eliminate crosstalk hits as well as preserving true photon hits.
Quantifying the impact of the ToT-cut on selecting the true photon hits is quite intricate.
As a general rule, a lower ToT is better for efficiency of selecting true photon hits.
Hence, a minimal ToT-cut for maximum rejection of capacitive crosstalk is employed as a compromise strategy.
In order to test the impact of different ToT-cuts on the effect of the capacitive crosstalk, an extended version of the analysis presented in figure~\ref{high-occ-result-1} was performed.
Here, the expected and observed hit multiplicities are calculated for different ToT-cuts.
Figure~\ref{high-occ-result-2} displays the excess hits which are calculated by subtracting observed hits and expected hits, as a function of expected hits for different ToT-cuts.
The performance of lower ToT-cuts, less than \SI{3}{\nano\second}, is worse in higher occupancies.
More than 30\% additional hits are produced as compared to expected hits for the ToT-cut of \SI{1.5}{\nano\second}.
For CBM RICH operation at twice the maximum expected occupancy of 18 hits (9 (maximum) $\times 2)$ which translates to $\sim28\%$ pixels, a ToT-cut of \SI{3}{\nano\second} is optimal.

\begin{figure}[ht]
    \centering
    \includegraphics[width=0.32\textwidth]{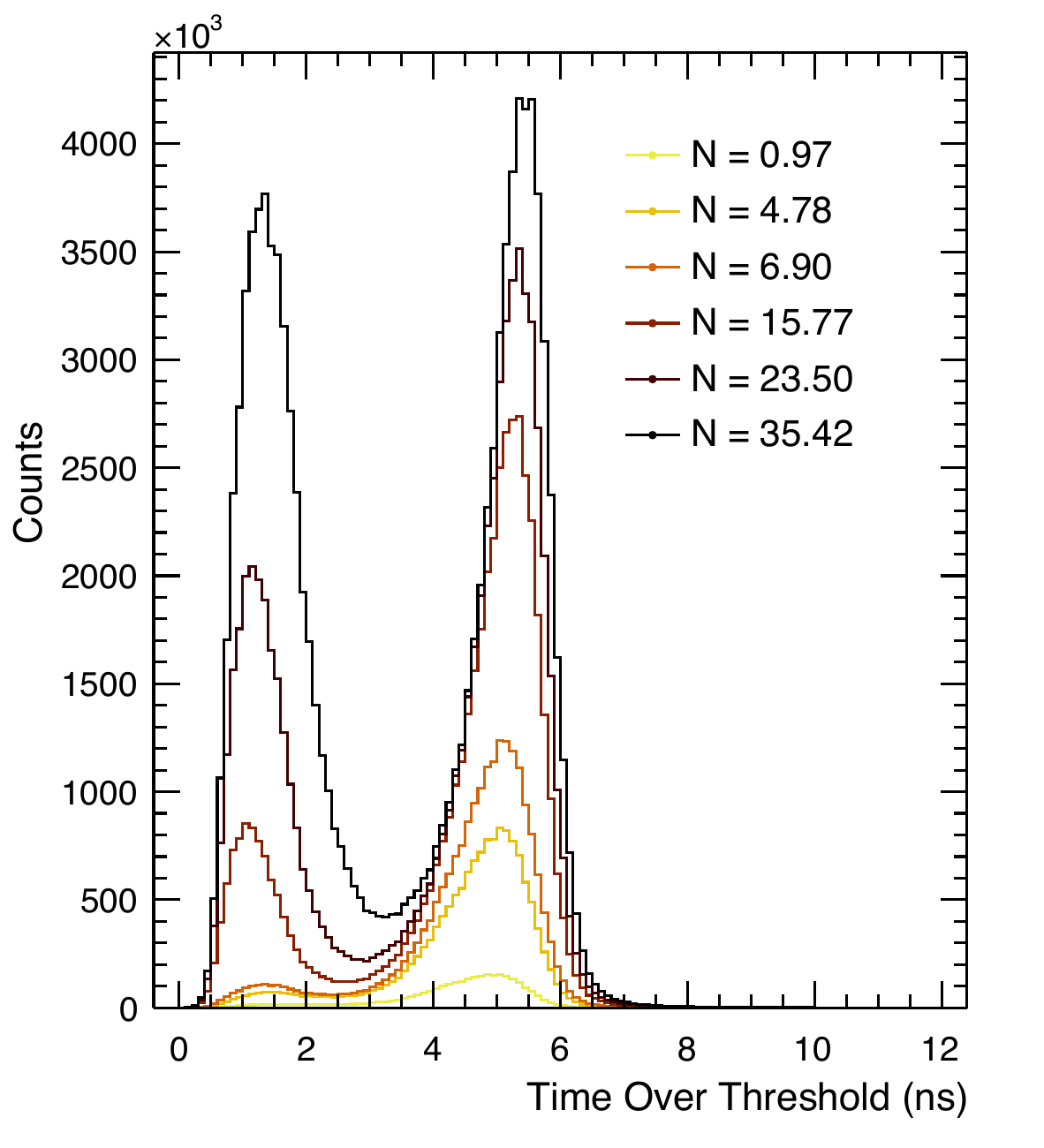}
    \includegraphics[width=0.32\textwidth]{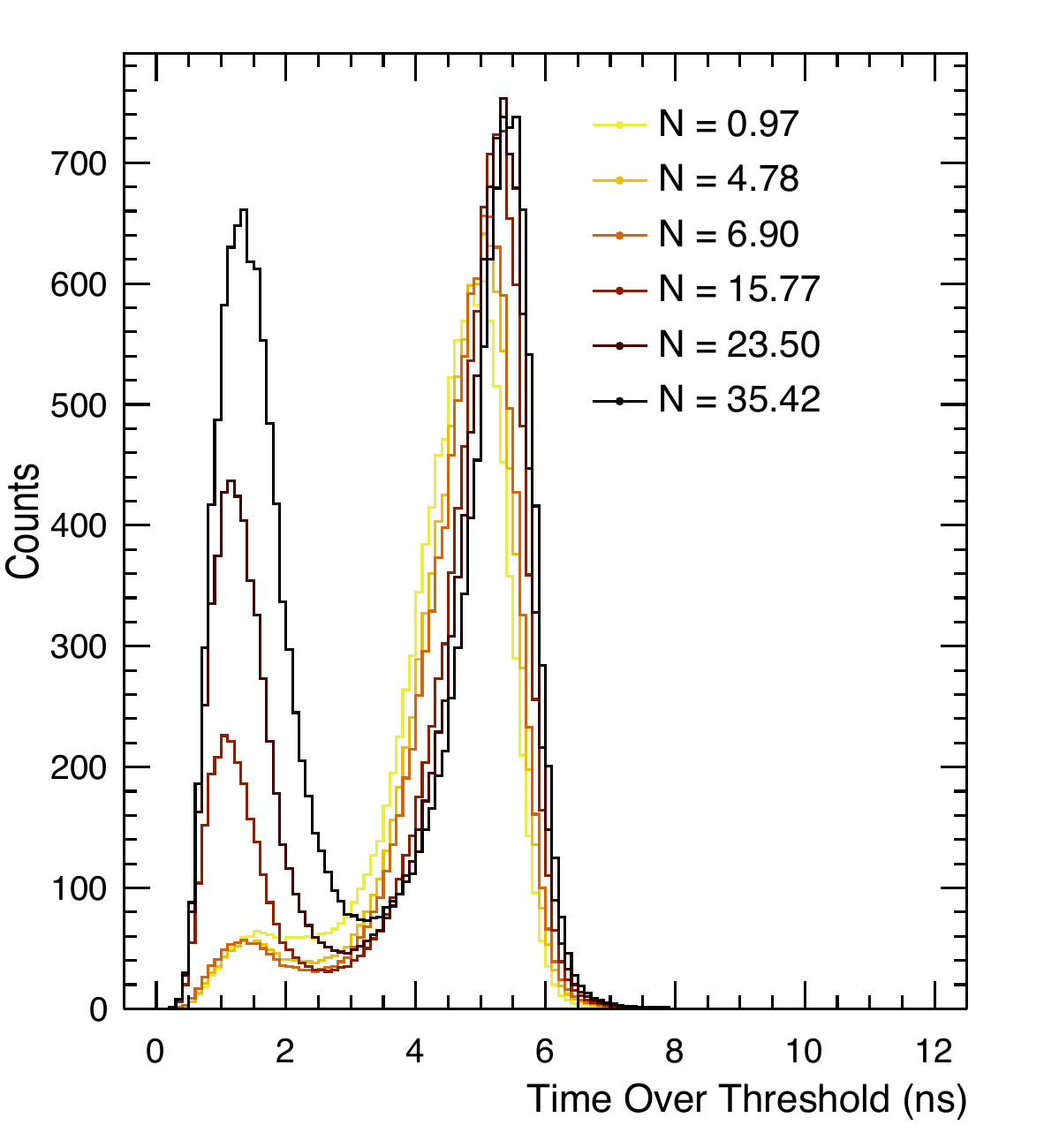}
    \includegraphics[width=0.32\textwidth]{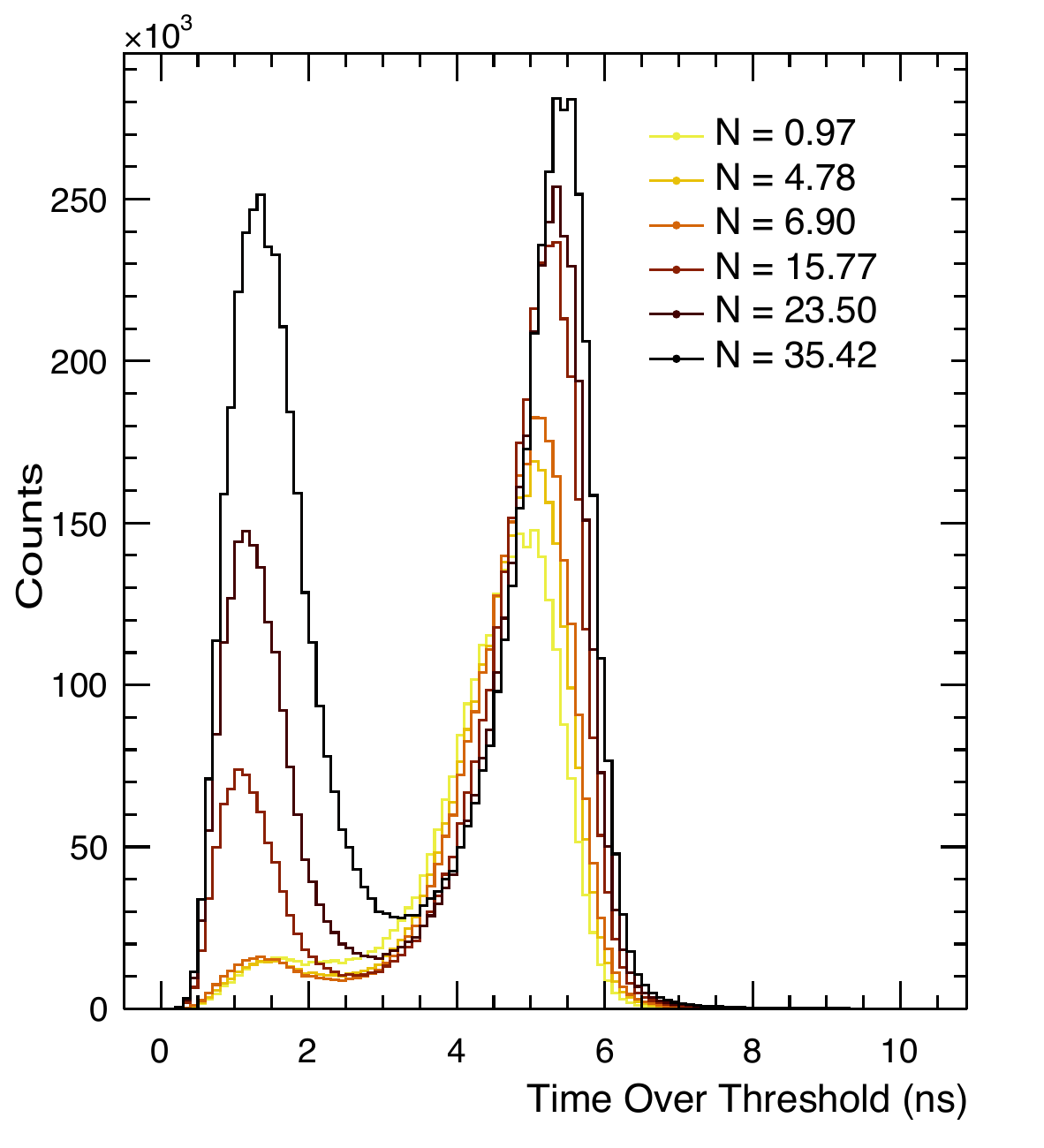}

    \caption{ ToT spectra for different hit multiplicities (N),
    left panel - No scaling, \,
    middle panel - Scaled to Integral of spectra for ToT $>$ 3ns, \,
    right panel - Scaled to single photon contribution.
    }
    \label{high-occ-result-3}
\end{figure}

The Time over threshold (ToT) spectra for different multiplicities are plotted in figure~\ref{high-occ-result-3}.
One can observe a rise in the low ToT peak as the hit multiplicity increases, indicating additional crosstalk. 
Also, the valley between the lower ToT and higher ToT peaks shifts gradually towards the higher ToT region and tends to get sharper with the increase in multiplicity, indicating that the ToT usage for crosstalk signal suppression has certain limitation 

The histogrammed ToT spectra in figure~\ref{high-occ-result-3} are shown in three different variants, using different choices of scaling factors of the individual curves of increasing illumination.
The first plot uses no additional scaling, i.e., the number of entries in each curve increases simply with increasing hit multiplicity.
The second plot is scaled such, that all individual ToT distributions have the same area of the signal peak, at ToT $>$ \SI{3}{\nano\second}.
This is done to visualize the proportional increase of the peak on the left (Crosstalk).
The third iteration of this ToT plot is scaled in order to better express the single photon contribution.
A scaling factor, defined as,
\begin{equation}
    \textrm{Scaling factor} = N_{expected} - N_{double-photons} = N_{with-filter}\,\times \frac{1}{T}-\,N_{dark-hits}\,-\,2\,\times\,N_{double-photons}
\end{equation}
is used to scale the ToT spectra in the region above ToT $>$ \SI{3}{\nano\second}.
The effective charge deposited on the photocathode due to the incidence of two photons is twice as large as for single photons.
Thus, it is expected that the double photon hits will have a larger ToT.
At high occupancies, which are facilitated by the high intensity of laser pulses, the likelihood of encountering double photons in a pixel is higher.
Hence, at higher hit multiplicities, one would expect distinct peaks of ToT spectra for single photon and double photon hits.
On contrary, figure~\ref{high-occ-result-3} (right) indicates that the separation is weak for discrimination of single and double photon hits.
On comparing the lowest hit multiplicity (N $\sim$ 0.97) and highest hit multiplicity (N $\sim$ 35), the most probable value of the ToT spectra ($>$ \SI{3}{\nano\second}) is about \SI{1}{\nano\second}.
Thus, the DIRICH front-end board is not very well suited to separate single- and double photon hits within the same channel only based on ToT information.

\section{Leading edge- and Time-over-Threshold timing characteristics of the DIRICH FEB}
In the CBM experiment, due to the high interaction rate and large track multiplicity, multiple electron/positron tracks (mostly stemming from photon conversion in the detector materials) might enter the RICH detector at close vicinity in space and time.
Since the emission of Cherenkov photons is a rather instantaneous physical process, this can result in multiple simultaneous single-photon hits into the same readout pixel.
The probability of multi-photon hits might be further enhanced due to additional scintillation light caused by the large charged track multiplicity for a given event inside the RICH gas radiator and surroundings.

In this section, we discuss the capability of the DIRICH readout chain to differentiate simultaneous multi-hits (which result in larger PMT pulse charge compared to single photon hits) only based on the measured leading-edge time and ToT information.
The evaluation is based on a systematic study of the DIRICH FEB response on input pulses generated by a pulse generator, systematically varying both amplitude and width of the input pulses.
Since the DIRICH FEB does not provide a direct measurement of the input pulse charge, its time response is the only criterion available to distinguish photon multiplicity.

\subsection{ToT response from DIRICH FEB}
Already from figure~\ref{high-occ-result-3}, it is evident that the separation of double and single photon hits using ToT information is rather weak, as the valley between the one- and the two- photon pulse is not very deep. 
A detailed characterization of the leading-edge and ToT response to well-controlled input pulse shapes can help to better understand this behavior.

A pulse generator was used to generate distinct negative pulses with \SI{1}{\nano\second} rise time (RT) and \SI{1}{\nano\second} fall time (FT), similar to PMT pulses.
The signal from the pulse generator is fanned out into 16 DIRICH input channels in parallel using a resistor divider.
The DIRICH threshold (after pre-amplification stage) is set to \SI{100}{\milli\volt}.
Two measurements were conducted, one involving the variation of the amplitude of the signal at the input of the DIRICH FEB at a constant pulse width, and the other involving the variation of the width of the signal while fixing the amplitude of the signal.
Each data point is analyzed for $10^6$ regular readout triggers.

\begin{figure}[h!tb]
\centering
  \includegraphics[width=0.50\textwidth]{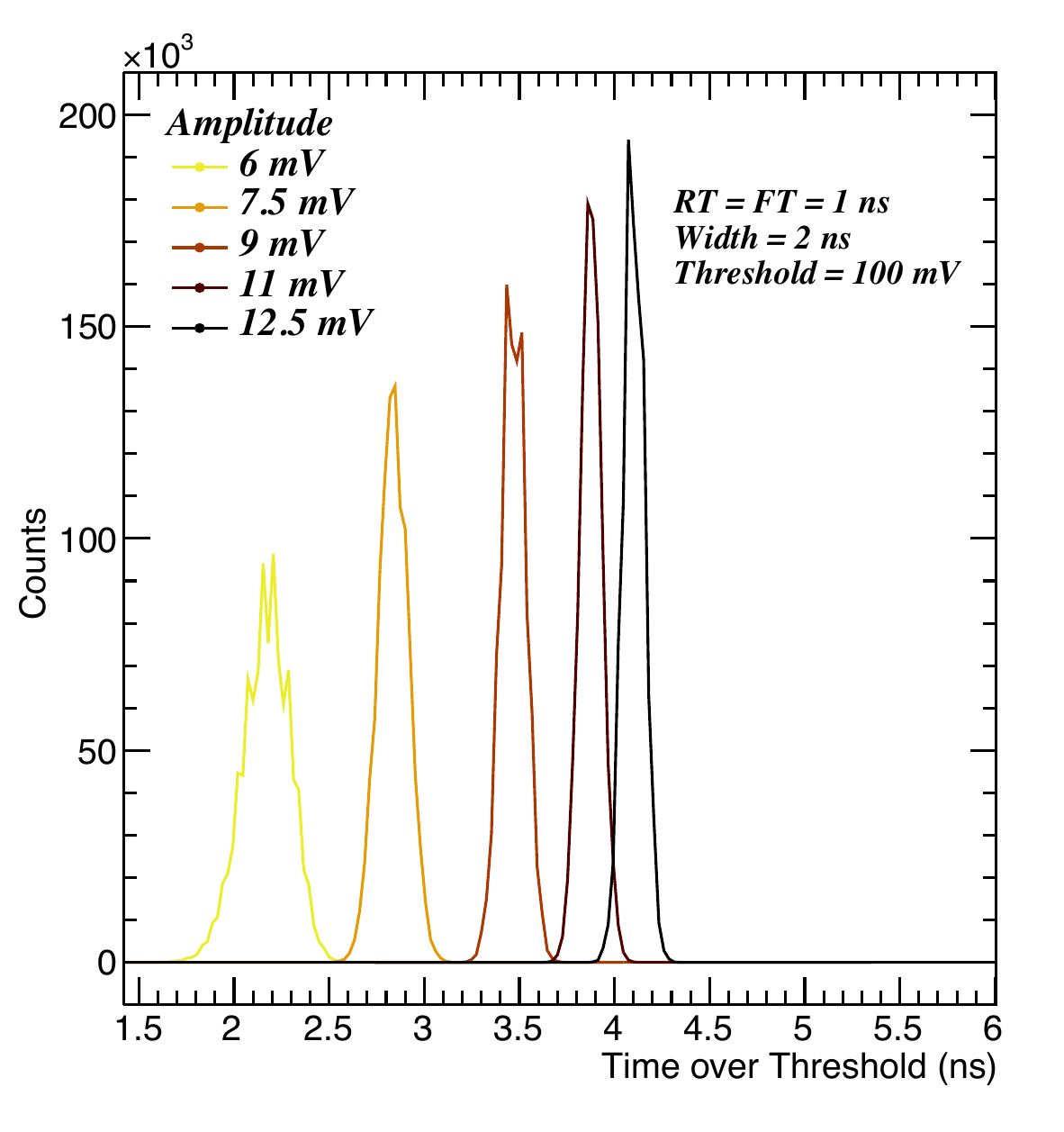}
  \includegraphics[width=0.46\textwidth]{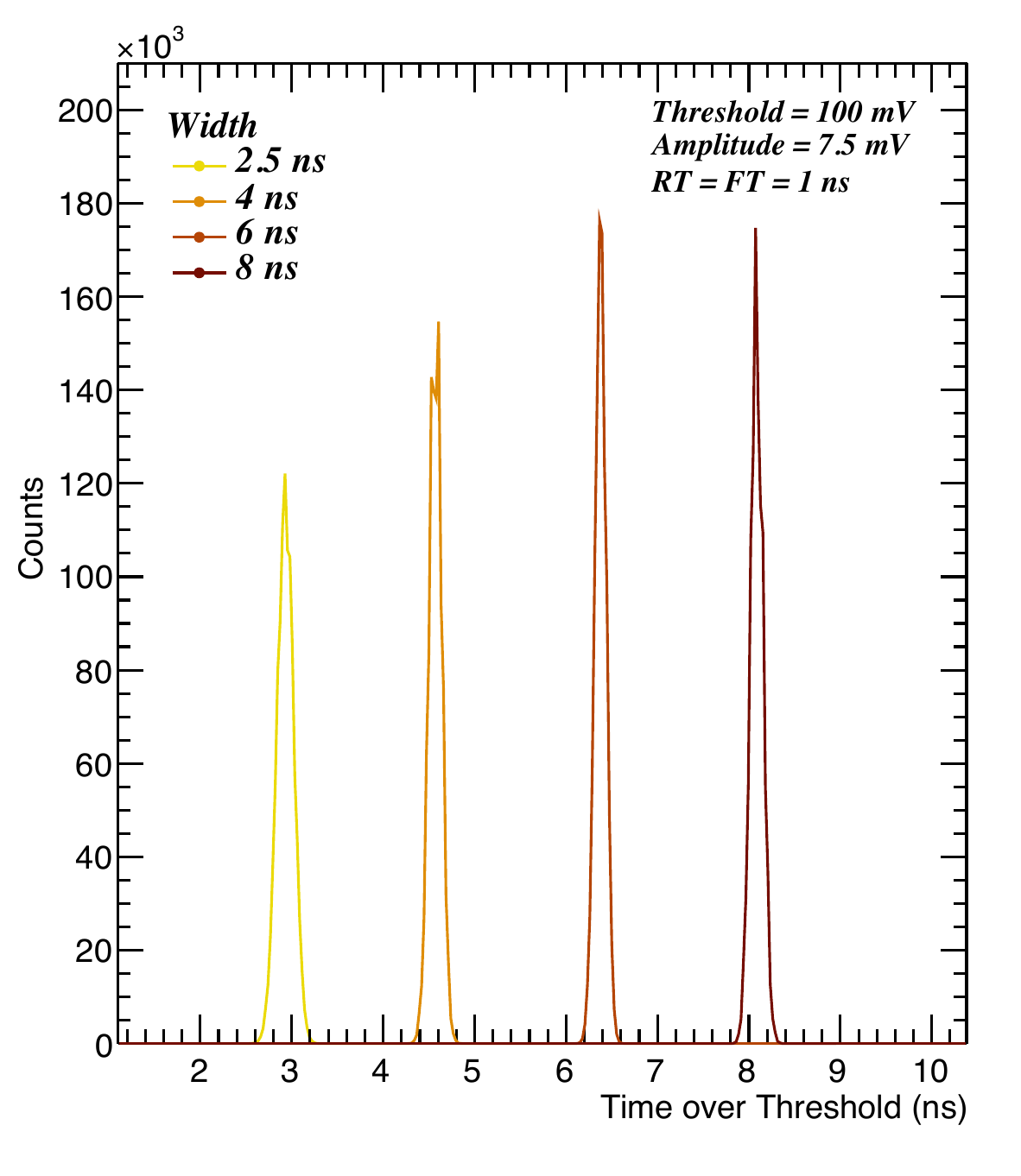}
  \caption{Left panel:  Time over threshold (ToT) distribution measured for different amplitude of input signal by maintaining a constant width. Right panel:  Same for different width of input signal at constant amplitude.}
  \label{tot-diff-width-amplitude}
\end{figure}

In the first measurement, the input pulse amplitude is systematically incremented in steps of \SI{6}{\milli\volt}, keeping the width of the input signal constant at \SI{2}{\nano\second}.
The result of this measurement is plotted in figure~\ref{tot-diff-width-amplitude} (left):
Increasing amplitude of the input signal translates to higher ToT values.
However, a clear saturation effect is observed for signal amplitudes beyond \SI{10}{\milli\volt}. 
In addition, larger amplitude input signals exhibit a more narrow peak in the ToT distribution in comparison to low amplitude signals. 
These observations are to be expected given the fact, that the DIRICH time measurement is based on a simple leading / trailing edge discrimination at a given threshold.
For low threshold (relative to the pulse height), discrimination happens early in the signal rise-, and late in the signal tail, with low timing jitter due to steep signal rise / fall.
For relatively large threshold / low pulse amplitude, discrimination happens in the less steep top part of the signal waveform, resulting in larger timing jitter and broader ToT peaks.

For the second measurement, the amplitude of the input signals was fixed at \SI{7.5}{\milli\volt} while their width was varied, the result is plotted in figure~\ref{tot-diff-width-amplitude} (Right).
Here, a nice linear dependence of measured ToT on the input pulse width can be observed, proving the good pulse-width measurement capability of the DIRICH. 
The analog DIRICH input stage comprises an inductive signal transformer for galvanic isolation of the input, which is known to cause a bipolar signal shaping prior to discrimination. However, for the covered range of pulse widths, this clearly does not hinder the pulse width measurement.

These results suggest that the PMT pulse itself already exhibits only a weak correlation between pulse charge (photon multiplicity) and signal width. Combined with the saturation effects in the amplitude-to-ToT correlation as observed in the first measurement, this can explain the limited capability to distinguish photon multiplicity based on measured ToT alone.

\subsection{Leading edge timing precision from DIRICH FEB}
Another measurement is performed to understand the DIRICH FEB response to distinguish quasi simultaneous hits in time.
To understand the leading edge (LE) precision of the DIRICH FEB, a measurement was made using the same setup as described in the previous subsection.
For a comparative study, two active channels in the same and different DIRICH FEBs are used.
The LE time difference between two hits is calculated to eliminate any path length effects of the fan-out module to the signal.
\begin{figure}[!ht]
    \centering
    \begin{subfigure}{0.52\textwidth}
    \centering
        \includegraphics[width=\linewidth]{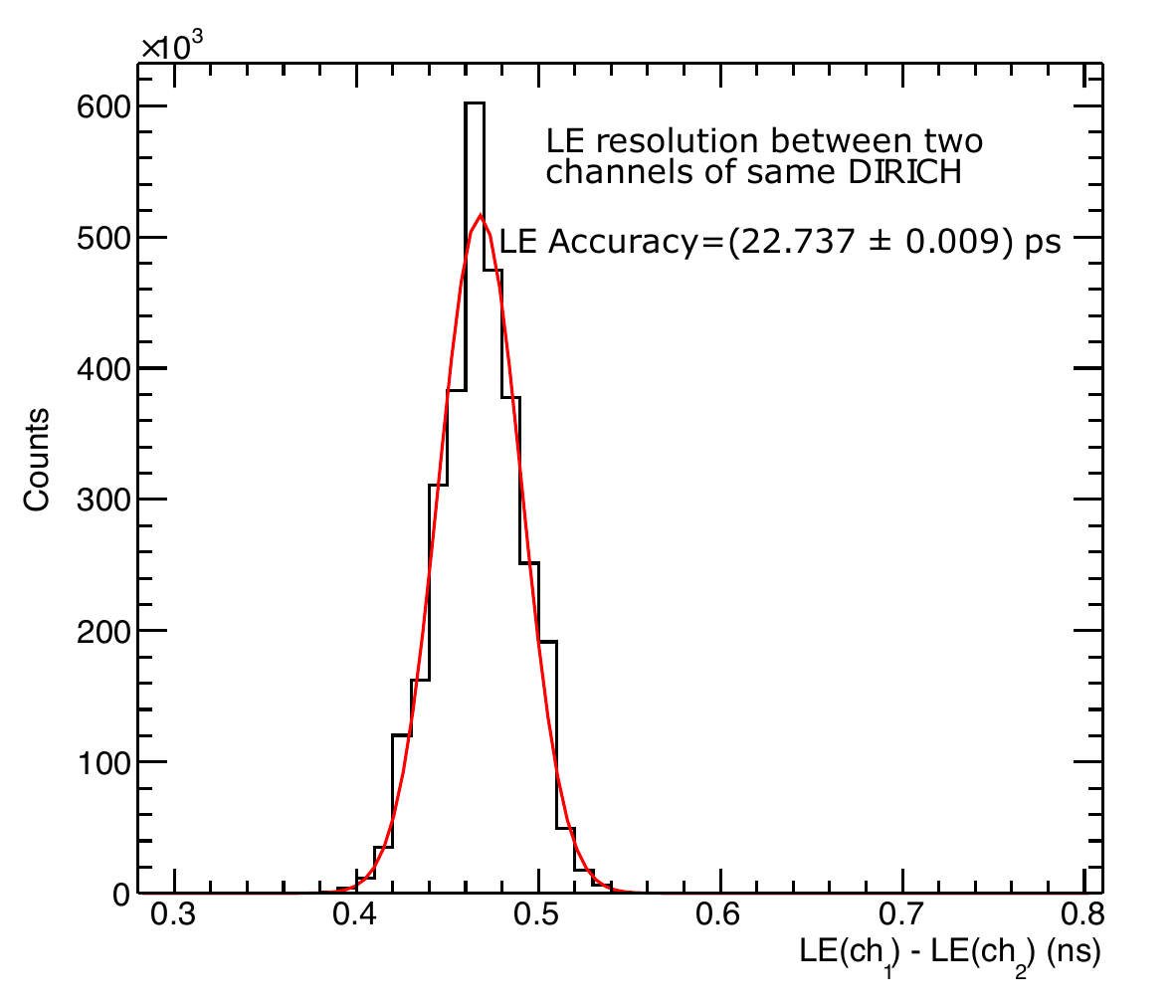}
    \end{subfigure}%
    \begin{subfigure}{0.44\textwidth}
    \centering
    \includegraphics[width=\linewidth]{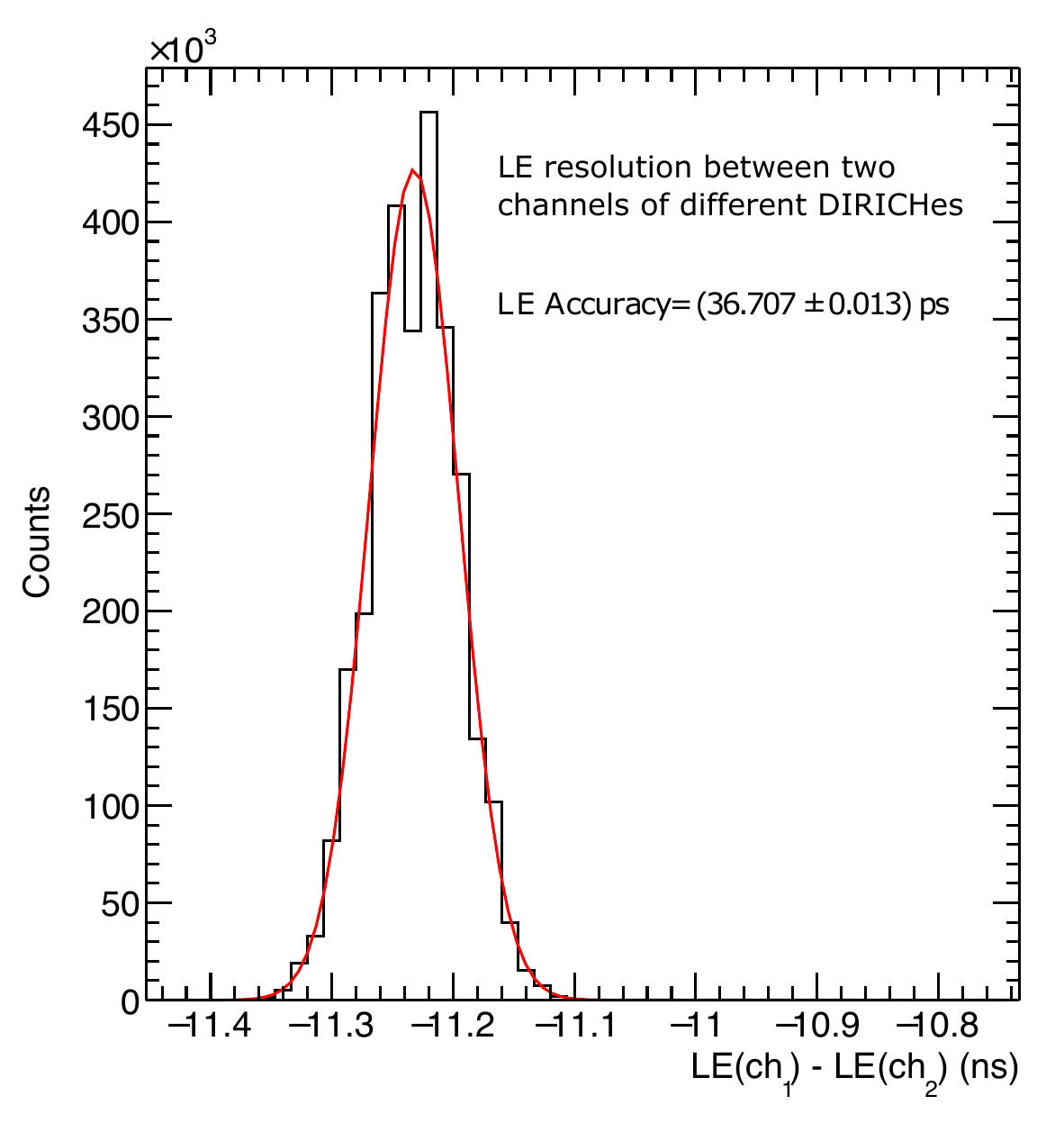}
    \end{subfigure}
    \caption{Leading edge accuracy measured with two active channels of the same DIRICH (left) and two active channels of two different DIRICHes (right).}
    \label{le-same-diff-DIRICH FEB}
\end{figure}%
The difference in the LE time between the hits of the two active channels in the same DIRICH FEB is plotted in figure ~\ref{le-same-diff-DIRICH FEB} (left). 
The resulting distribution is fitted with a Gaussian profile, and the width of the profile is the LE spread, and a measure of time measurement precision of the DIRICH.
Similarly, the LE time difference  between hits in two active channels in two different DIRICH FEBs is histogrammed and fitted with a Gaussian profile, and the LE spread is calculated.

For two channels in the same DIRICH FEB,
\begin{equation}
\label{eqn:result_le_same}
    \sigma_{\textrm{LE}} = (22.7377 \pm 0.009) \; \textrm{ps.}
\end{equation}
For two channels in different DIRICH FEB,
\begin{equation}
\label{eqn:result_le_different}
    \sigma_{\textrm{LE}} = (36.7071 \pm 0.013)\; \textrm{ps.}
\end{equation}
If the registered time of two hits of the same DIRICH (different channels) are measured and compared, then both time measurements are derived from the same clock signal, and can be directly compared.
The time difference $\delta t $ and the uncertainty ($\sigma$) in measuring time for two hits ($t_1$, $t_2$) in different channels of the same DIRICH can be expressed as,
\begin{alignat}{2}
\label{eqn:le-same-dirich}
    \delta  t &=t_1 - t_2,\\
    \sigma_{same-DIRICH} &=\sqrt{
    \sigma_{t_1}^2 + \sigma_{t_2}^2}.
\end{alignat}

However, if two hits in two different DIRICH modules are compared to each other, then the time measurement in both DIRICH modules is based on two individual clocks, running asynchronously.
In order to derive an absolute time difference, first, both measurements have to be synchronized. This is achieved by measuring the arrival time of the readout trigger signal (distributed to both DIRICH modules) using a dedicated extra TDC channel (channel 0) on each of the DIRICH modules.
Since, each DIRICH measures the time of the hit relative to the measured trigger arrival time, the absolute trigger arrival time cancels out.
Let $t_0$ be the absolute time in channel 0, the time difference $\delta t $ and the uncertainty ($\sigma$) in measuring time for two hits ($t_1$, $t_2$) in two channels of the different DIRICH can be expressed as,
\begin{alignat}{2}
\label{eqn:le-different-dirich}
    \delta  t &=(t_1- t_0) - (t_2 - t_0) = t_1 - t_2,\\
    \sigma_{different-DIRICH} &=\sqrt{
    \sigma_{t_1}^2 + \sigma_{t_2}^2 + \sigma_{t_0}^2+\sigma_{t_0}^2}.
\end{alignat}

As is shown, the absolute value of the $t_0$ measurement cancels out, however, the measurement uncertainty of the reference channel 0 on each DIRICH adds to the overall uncertainty.
Opposite, if both hits were measured on the same DIRICH, there is no additional penalty for the $t_0$ measurement.
This explains the lower precision in time difference measurement, if the two hits are spread over different modules (see figure~\ref{le-same-diff-DIRICH FEB}).
The results (equations~\ref{eqn:result_le_same},~\ref{eqn:result_le_different}), indicate that the LE timing precision of the DIRICH FEB is an order of magnitude better than the transit time spread (TTS) from the MAPMTs (The TTS for the H12700A Hamamatsu MAPMTs is about $\sim$\SI{350}{\pico\second} FWHM), indicating that the DIRICH FEB is not the limiting factor in measuring / comparing photon arrival time of different hits.

\section{Test of DIRICH maximum hit rate capability}
The maximum hit rate capability of the DIRICH module, and understanding its performance limiting elements, is another important aspect of the DIRICH readout chain qualification.
This section describes the measurement principle and results of a high rate test of the complete MAPMT-DIRICH readout chain.
Inside the DIRICH FEB, incoming analog input signals are first amplified, and then fed into differential line receivers of the FPGA for signal discrimination (sec.~\ref{rich-fee}).
Inside the FPGA, individual scaler entities are implemented and connected to the output of each comparator, counting directly the number of detected edges in each channel. 
These scaler values can be read out via a slow control data channel, and allow counting hits and measure input rates completely independent of all further data digitization and transfer.
The slow control path is logically separated from the main data stream, and as such not affected by loss of data or data quality due to data overload, buffer overflows, or high rate conditions in general.
This feature allows for a sensitive test of high rate stability of the DIRICH, by comparing the number of detected hits per channel as obtained from the slow-control scalers to the number of hits found in the output data stream written to the file.

\subsection{Data flow in the TRB-based lab setup compared to later (m)CBM operation}
\label{section-dataflow}

Data measured by the DIRICH TDC is organized in hit messages containing several data words of 32 bit each. 
Usually, a single hit consists of a maximum of 3 data words: the leading edge time, the trailing edge time, and an additional occasional epoch message containing the absolute time information.
For a given readout trigger, each channel can produce several such hit messages, depending on the number of hits registered since the last readout, and depending on a possible readout window which can be defined relative to the trigger time.
Different buffers in the data flow provide the capability of a quasi free-streaming data taking operation despite a fixed-period readout initiated by the regular readout trigger.
The first stage of the data storage happens at individual DIRICH FEB channel buffers which (in the present DIRICH firmware) can store a maximum of up to 123 words.
Upon receiving a readout trigger, the data from all 32+1 channel buffers are sequentially shifted into the DIRICH FEB main readout buffer, starting with channel 1.
This common readout buffer can store up to 499 words, limiting the maximum number of individual hits which can be read out by a single readout trigger.
The data from each DIRICH on a single backplane module (containing up to 12 FEBs) are streamed to the DIRICH combiner module, which is acts as a data hub, combining the individual sub events from each FEB to a single output event.
The combiner output link is a standard TRBnet optical link of \SI{2}{Gbps} (\SI{2.4}{Gbps} in case of the CRI readout used in mCBM and later CBM operation).

The further data transfer from the combiner onwards is different for the lab setup and the future CBM experiment.
\begin{figure}[h!tb]
\centering
  \includegraphics[width=0.49\textwidth]{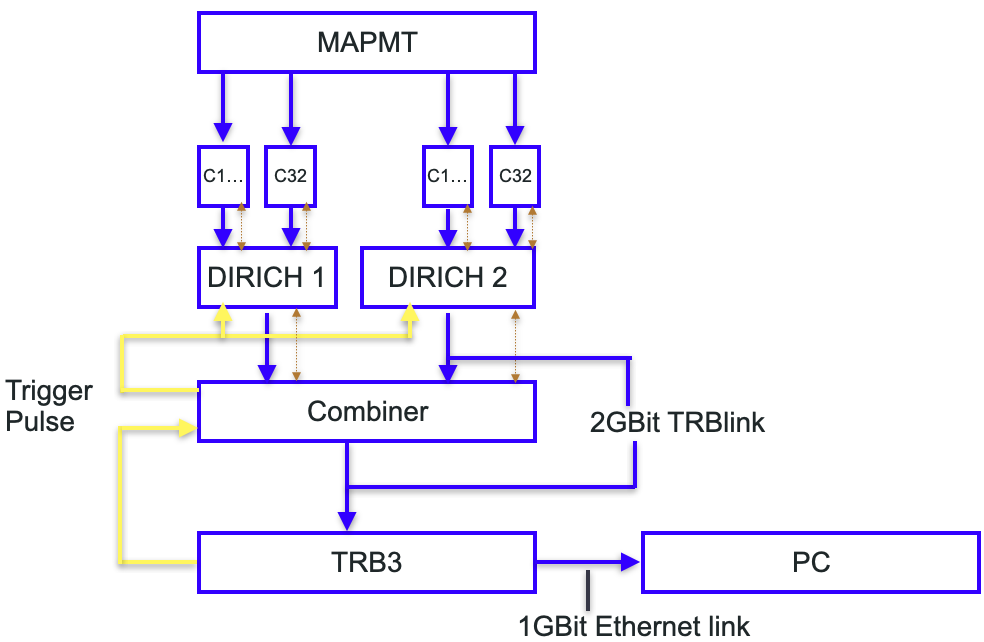}
  \includegraphics[width=0.49\textwidth]{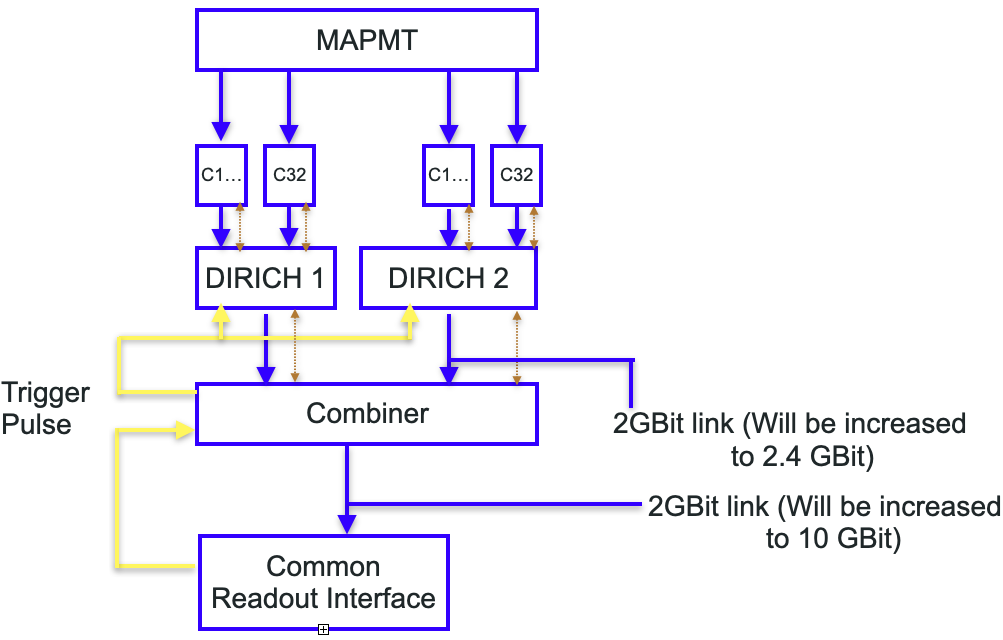}
  \caption{Schematic showing data flow in the lab (left) and future CBM (right). The blue arrow indicates the direction of the data stream and the dotted brown indicates the slow control link.
          Yellow lines indicate the flow of the trigger signal.}
  \label{highrate-data-flow}
\end{figure}
In the lab setup, a TRB3 is used as the DAQ master. 
Its task is to collect data from up to 4x7=28 combiner modules, and to send these data via its single GBit Ethernet interface to a DAQ PC for data storage and analysis. 
The TRB3 in addition facilitates the Central Trigger System “CTS”, generating and sending out the regular readout trigger which is then distributed via the Combiner modules to all DIRICH frontend FEBs. 
It is obvious, that the Ethernet link is the slowest bottleneck in this setup, followed by the optical link between individual Combiners and the TRB3.
These bottlenecks in the lab setup make it challenging to operate the system at large data rates for evaluating data rate capabilities of the DIRICH frontend.
In the future CBM DAQ setup, as well as in mCBM already today, a modified data transport scheme is implemented:
Here, the single TRB3 DAQ master is replaced by several “Common Readout Interface” (CRI) boards.
Each CRI board provides up to 47 serial data links of up to \SI{10}{Gbps} each, and sends collected data via its multi-lane PCI-express interface to its corresponding high-performance DAQ PC, and further - via InfiniBand network - to the Greencube compute cluster. 
This CRI-based readout chain is currently further developed and tested in the mCBM detector setup at GSI. Details can be found in~\cite{adrian_weber}.

Despite these data rate constraints, all test measurements described in this chapter were conducted using the TRB3 based laboratory readout.
This is because the main objective is to study the analog components of the DIRICH FEB, which is unaffected by the data rate constraints.
Overall schematics of both data transport variants are sketched in figure~\ref{highrate-data-flow}.

\subsection{Analysis method}
\begin{figure}[h!tb]
\centering
  \includegraphics[width=0.35\textwidth]{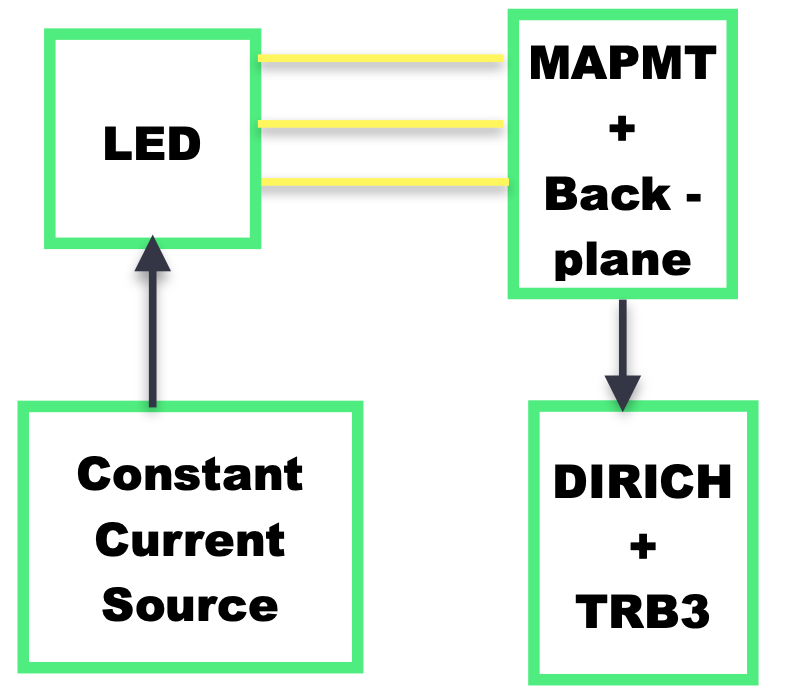}
  \includegraphics[width=0.30\textwidth]{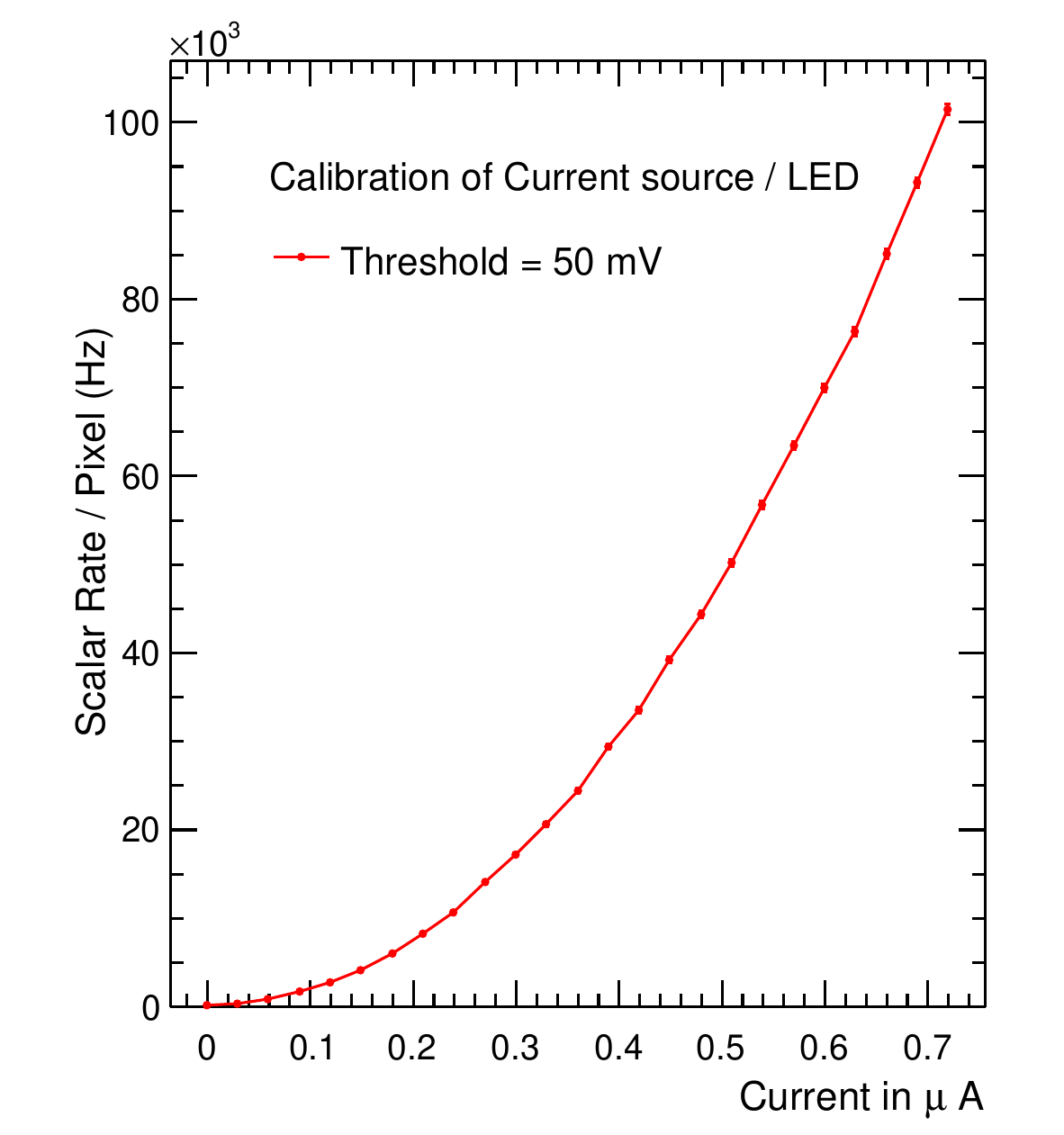}
  \includegraphics[width=0.30\textwidth]{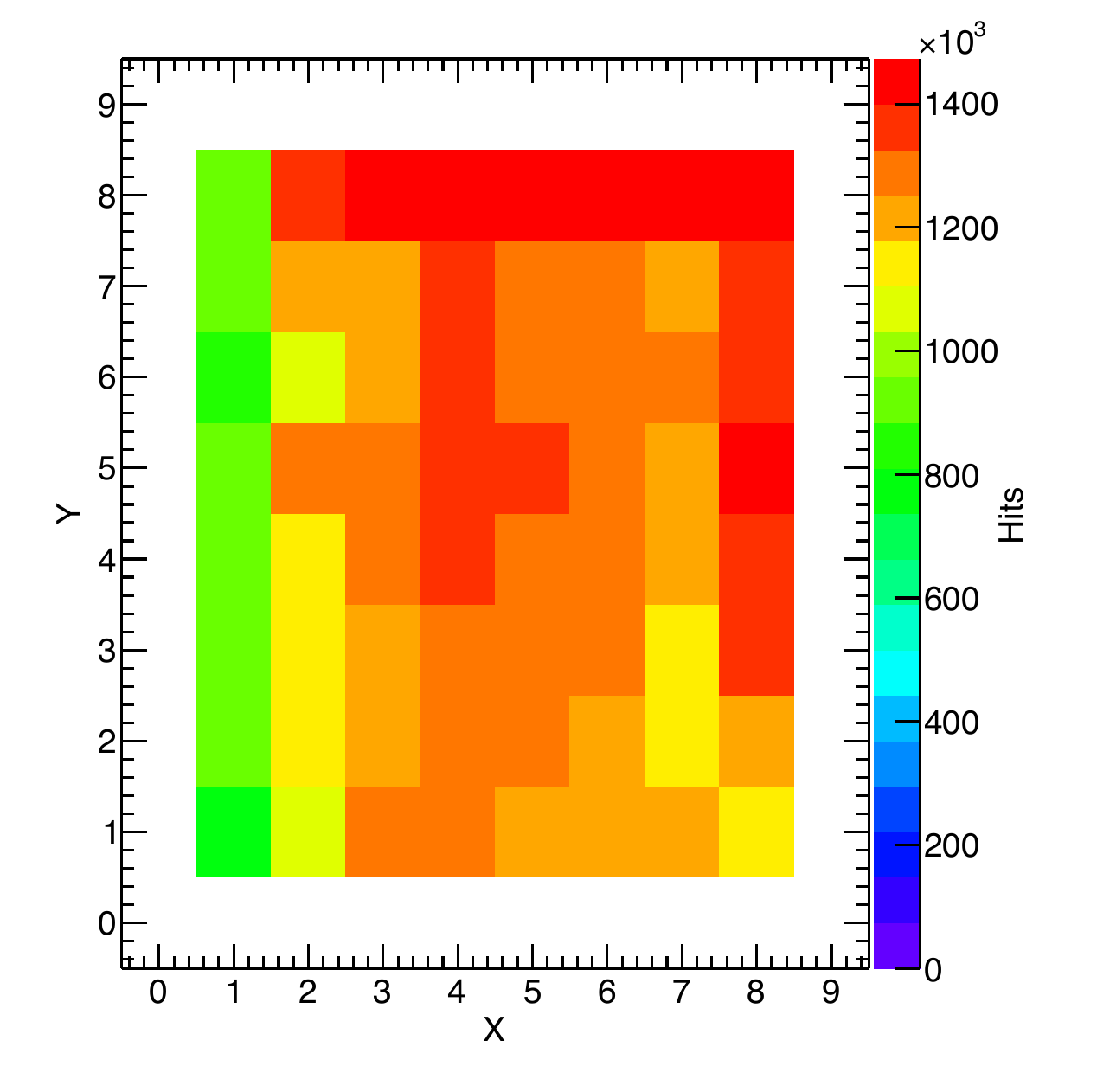}
  \caption{Left panel: Schematic depicting laboratory setup for performing the high rate capability test on DIRICH readout electronics. Middle panel: Relationship between scaler rate and operating current for LED. Right panel: Distribution of hits on MAPMT (each bin of this 2D histogram corresponds to a pixel of MAPMT).}
  \label{highrate-labsetup}
\end{figure}

The measurement described below uses a single H12700 MAPMT connected to two DIRICH FEBs.
In order to produce a large  hit rate at the DIRICH FEB, the MAPMT is illuminated by an LED (enclosed in a light-tight box) that is powered by a constant current source (figure~\ref{highrate-labsetup}).
The DIRICH threshold was set at standard \SI{50}{\milli\volt} (unless mentioned otherwise), providing good single-photon detection efficiency.
The variation of hit multiplicity in the MAPMT is achieved by changing the intensity of the LED controlled by the current source.
In figure~\ref{highrate-labsetup} (middle), the LED driving current is plotted versus the average photon hit rate as measured by the scaler counters on each DIRICH channel. 
No data acquisition is required to obtain this plot, only slow control communication is used to periodically read out the scaler values. 
A quadratic relation between output rate and current is observed and the position of the LED in the box is adjusted to produce a nearly homogeneous illumination of the MAPMT (see figure~\ref{highrate-labsetup} (right)).
For testing the rate capability of the DIRICH digitization chain, this scaler rate is now compared to the number of fully digitized hits per time in the DAQ data file after activating DIRICH digitization.
Each data point in the scaler rate is averaged over 10 seconds, and each data point in DAQ data is averaged over $10^6$ regular readout trigger events.

\subsection{DIRICH performance at high rate}
\label{section-highrate}
To perform the high rate test, the DIRICH FEB channel buffer size is fixed at 120 words and the main ring buffer size at 499 words.
\begin{figure}[h!tb]
\centering
  \includegraphics[width=0.48\textwidth]{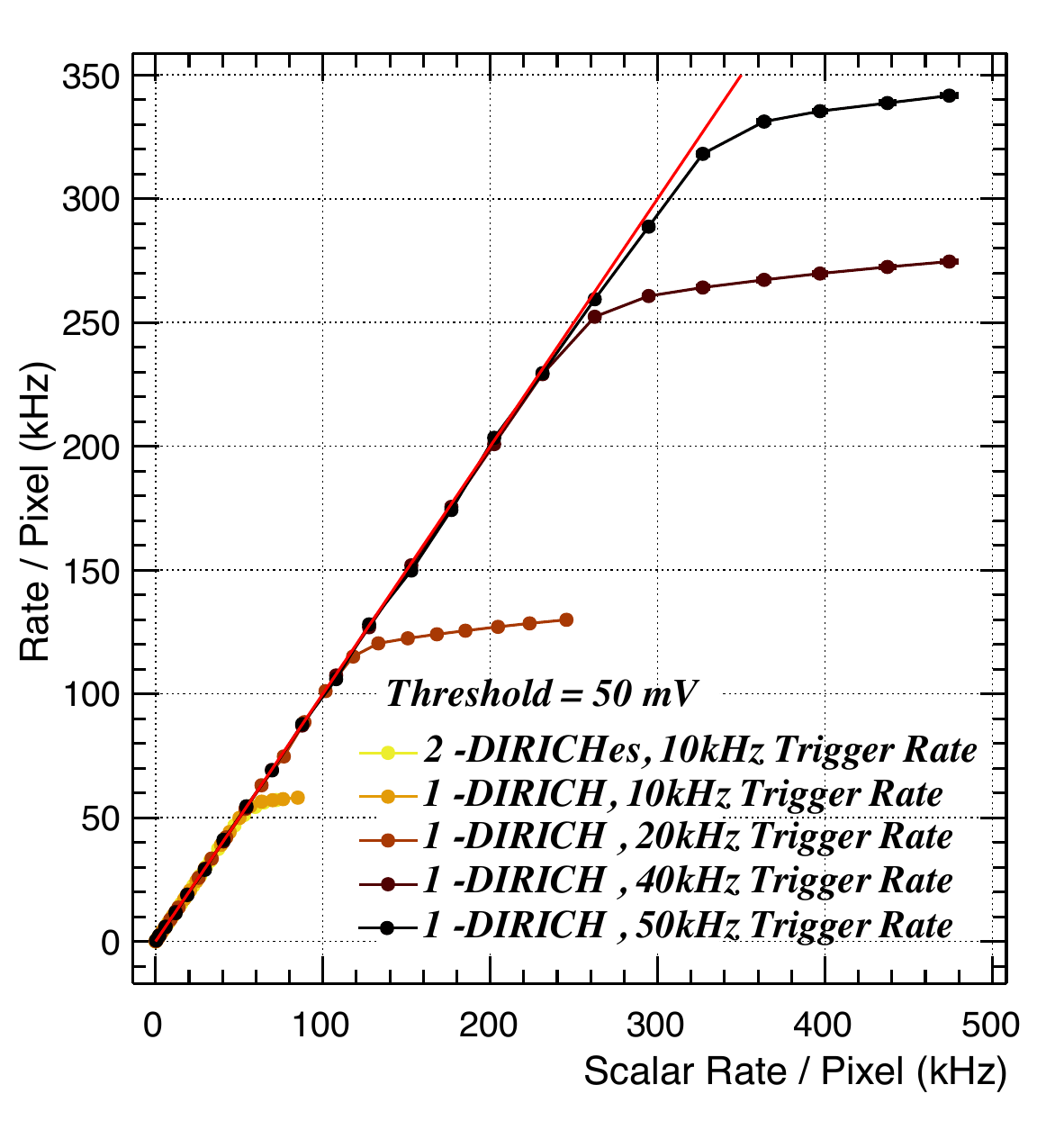}
  \includegraphics[width=0.48\textwidth]{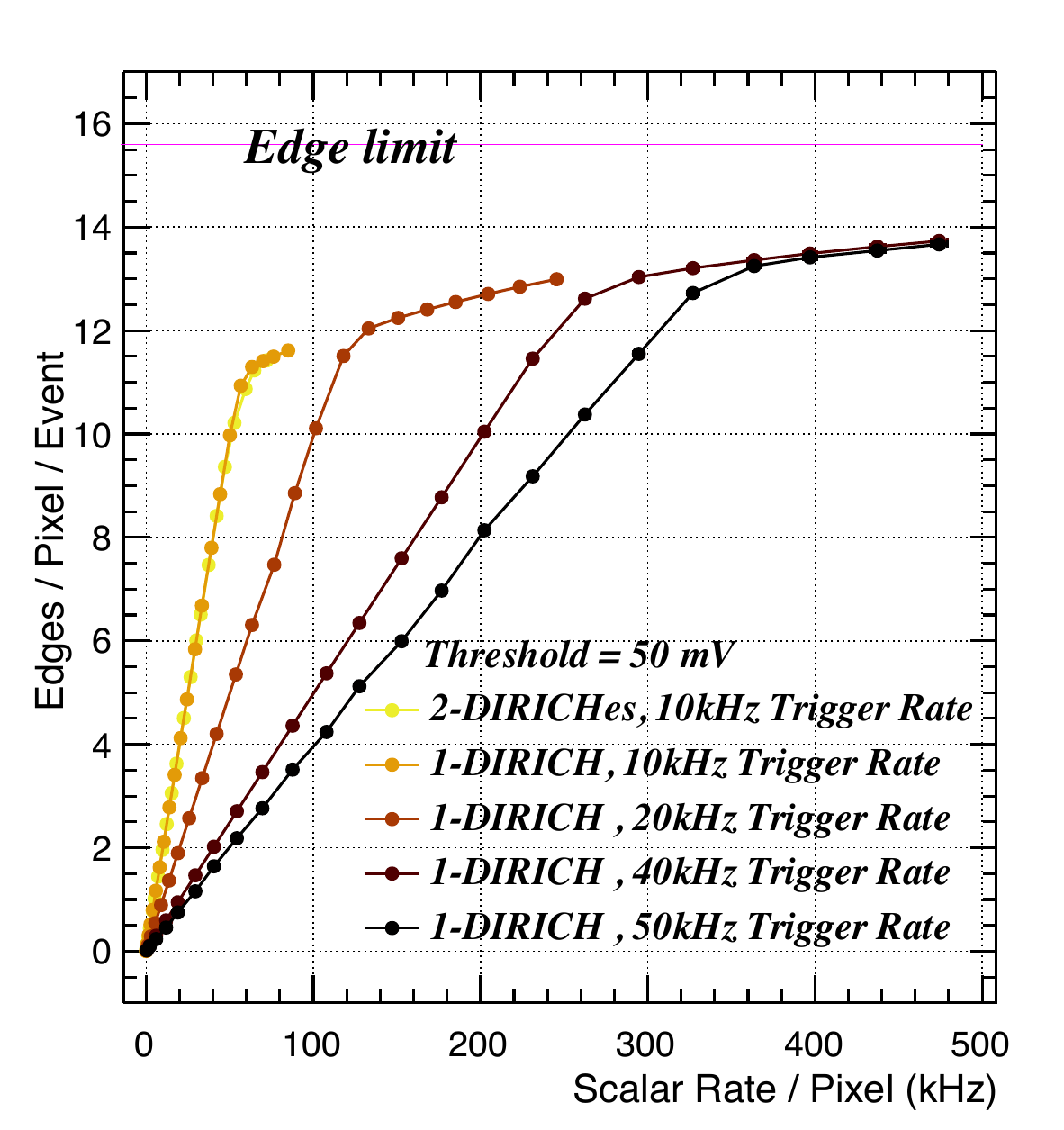}
  \caption{Left panel: Hit rate recorded at DAQ is plotted against the scaler rate per pixel, for measurements carried out with different number of active DIRICH FEBs and different readout frequency.
  Right panel: For the same measurement setup, the number of recorded edges (only LE and TE) in DAQ is plotted against scaler rate (Edge limit - Maximum number of edges that can be recorded per pixel when there is overflow at the DIRICH FEB main buffer).}
  \label{highrate-scalar-vs-rate-all}
\end{figure}
In the first measurement, this test was carried out using two DIRICH FEBs at \SI{10}{\kilo\hertz} regular readout trigger frequency (shown in figure~\ref{highrate-scalar-vs-rate-all} (Yellow)).
A linear relationship between the scaler rate and the rate of DAQ per pixel is seen, as it is expected because they are identical.
However, after a break-even point in the recorded scaler rate, the rate measured at DAQ saturates indicating data loss (see figure~\ref{highrate-scalar-vs-rate-all}). 

In order to understand the reason for saturation, the measurement is repeated using only one active DIRICH FEB.
Both measured curves overlap (Yellow and orange curves in figure~\ref{highrate-scalar-vs-rate-all}), indicating that the combiner output link or the TRB3 Ethernet link is not the limiting factor.
If the combiner or Ethernet link were overloaded by the data from 2 DIRICH modules / 64 pixels one would expect the saturation to shift to higher rate when the number of channels is reduced by half.
The identical data values for both cases show that the combiner link is not the limiting factor.
Hence, the DIRICH FEB channel buffer or the DIRICH FEB main buffer must be the limiting factors in these measurements.

To identify which buffer on the DIRICH (channel- or ring buffer) is limiting the data rate, the number of recorded edges (only leading edge and trailing edge) per trigger is plotted in figure~\ref{highrate-scalar-vs-rate-all} (right).
If the DIRICH channel buffer is the limitation, then the saturation of the edge multiplicity must be near the maximum buffer size set for the channel, which is 120 words.
This is not observed here, the edge multiplicity saturates close to 12 edges per channel per trigger (at \SI{10}{\kilo\hertz} readout frequency, yellow and orange curves in figure~\ref{highrate-scalar-vs-rate-all}). 
This indicates the rate saturation is due to the size of DIRICH main ring buffer.

In order to increase the dynamic range of the measurement, the readout trigger frequency is increased up to \SI{50}{\kilo\hertz}.
The results are plotted in figure~\ref{highrate-scalar-vs-rate-all} (red, brown, black).
This increase in readout frequency reduces the hit multiplicity in the queue at the buffer, thereby increases the bandwidth.
As the readout frequency increases, the saturation rate at the DAQ increases, this is reflected in the number of edges measured.
A single-channel hit rate of up to \SI{320}{\kilo\hertz} per pixel is observed in the DAQ data stream if only a single DIRICH FEB with all 32 channels active is read out.
Further increase of the readout frequency (beyond 50 kHz) and measurement of even higher rates caused an overload of the Ethernet link from the TRB3 to the DAQ PC.

Maximum size of the data package that could be transferred from a DIRICH per second = maximum hit rate per channel (\SI{320}{\kilo\hertz}) $\times$ number of channels in DIRICH (32) $\times$  maximum size of single hit (2 edges + 1 epoch =  3 words (4 bytes per word)) = \SI{1.144}{Gbps} (assuming 8b/10b encoding).
This data size is below the saturation limit of the DIRICH-combiner link (TRB link) which is at \SI{2}{Gbps}.
Therefore, the DIRICH-combiner link is not a limitation for the data transmission at such high rates.

\begin{figure}[h!tb]
\centering
  \includegraphics[width=0.48\textwidth]{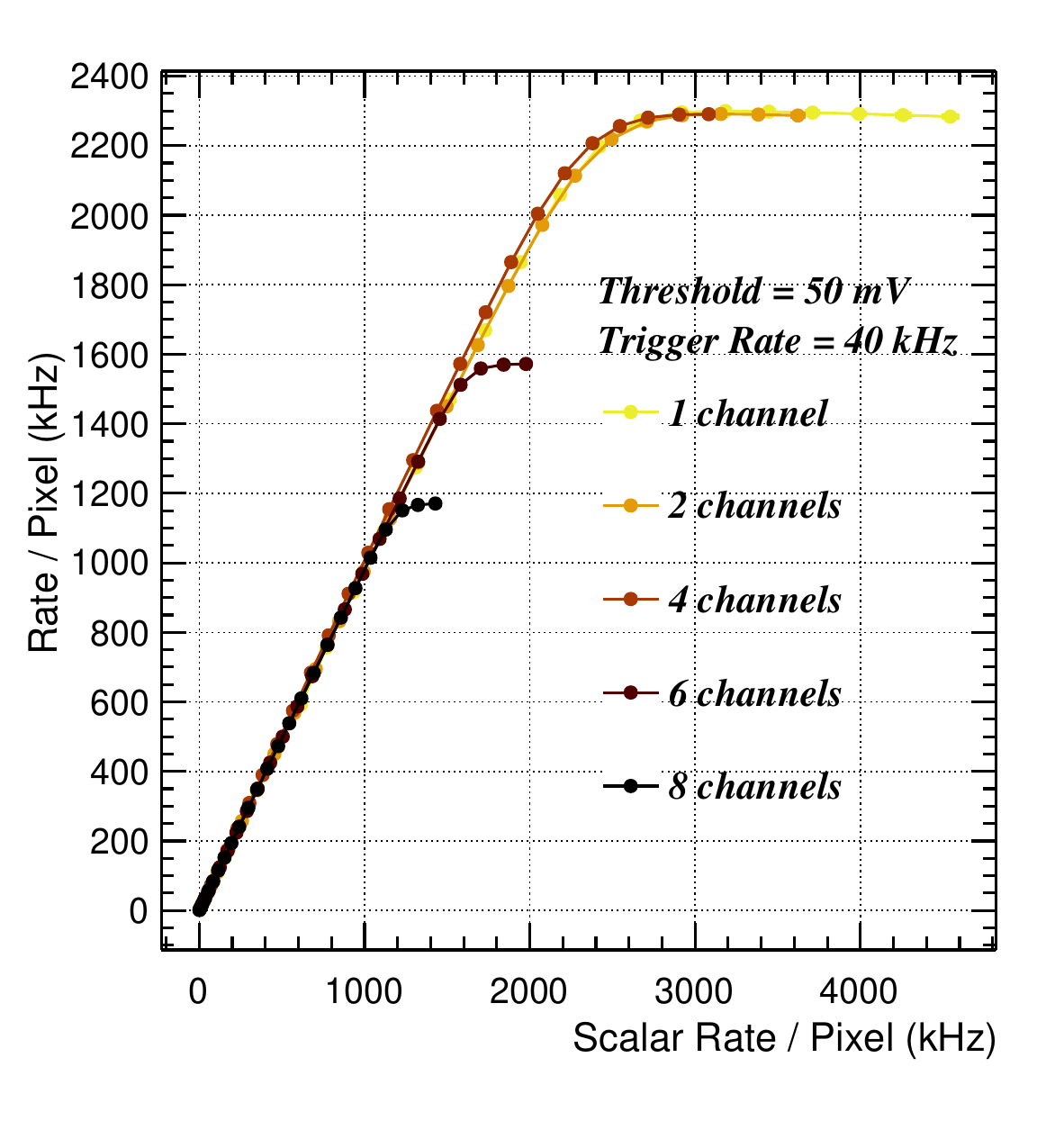}
  \includegraphics[width=0.48\textwidth]{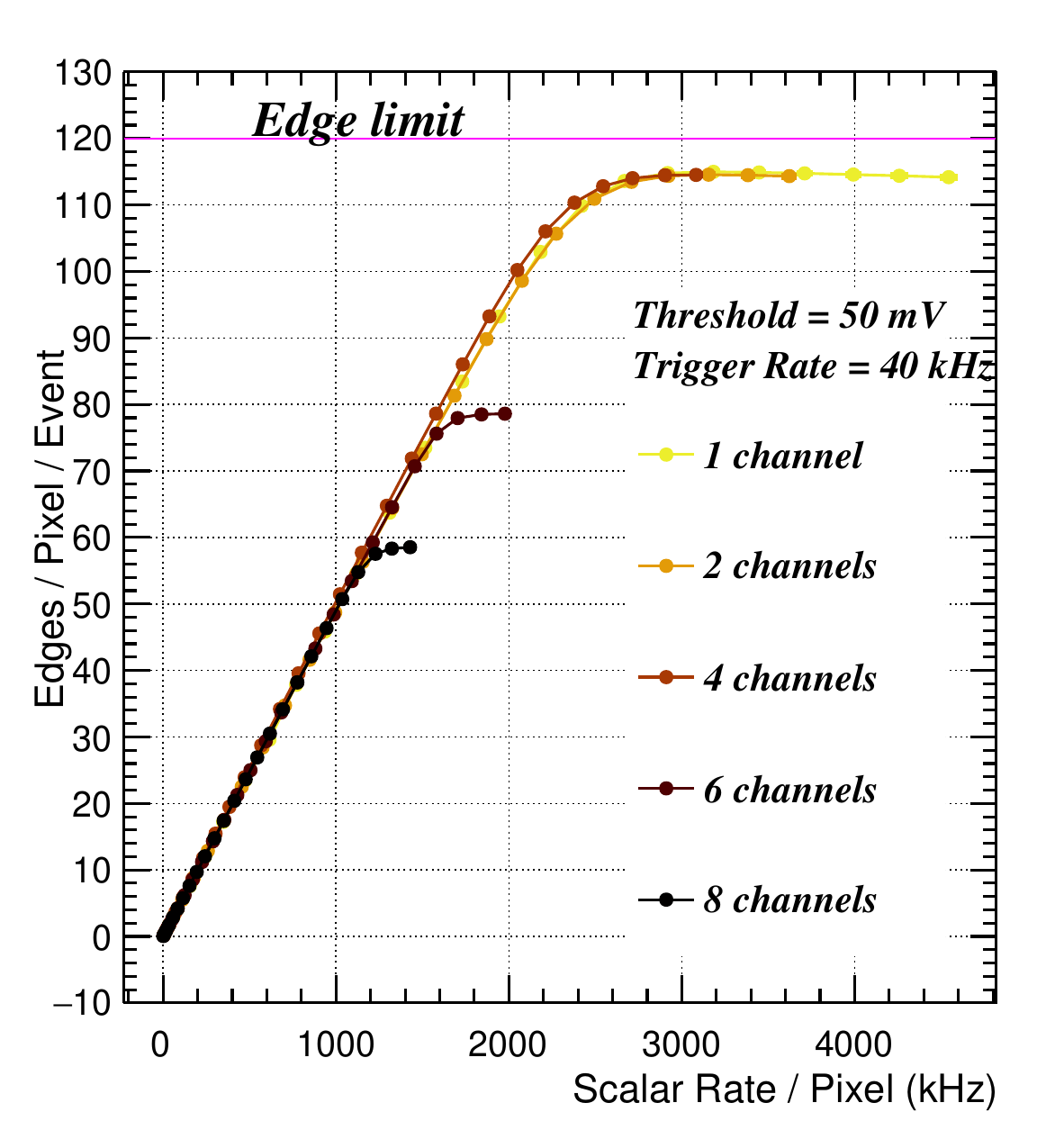}
  \caption
  {Left panel: Hit rate recorded at DAQ is plotted against the scaler rate per pixel, for measurements carried out with reduced number of active channels of a DIRICH FEB at \SI{40}{\kilo\hertz} readout frequency.
  Right panel: For the same measurement setup, the number of recorded edges (only LE and TE) in DAQ is plotted against the corresponding scaler rate (Edge limit - Maximum number of edges that can be recorded per pixel when there is overflow at the DIRICH channel buffer).}
  \label{highrate-scalar-vs-rate-edges}
\end{figure}

Now that the maximum attainable per-channel rate from a fully active DIRICH FEB (32 channels) using the lab setup is established (for a maximum readout rate of 50 kHz), further tests are done to test the rate capability of an individual channel.
This is achieved by reducing the number of active channels, also fixing the readout frequency at \SI{40}{\kilo\hertz} in order to prevent the Ethernet bottleneck.
The measurement mentioned above is repeated for the reduced number of active channels, and the results are shown in figure~\ref{highrate-scalar-vs-rate-edges}.
 Here, a saturation rate of up to \SI{2.2}{\mega\hertz} per pixel is achieved for the 1, 2, and 4 active channels.
 This higher saturation rate up to 4 channels is due to individual channel buffer size.
 Since individual channel buffer size is set at 120 words, for up to 4 channels the DIRICH FEB buffer will not be maxed out (4 channels $\times$ 120 words per channel = 480 $<$ 499  words per DIRICH FEB).
 When the number of active channels is increased to 6 and 8, a proportional drop in the overall saturation rate is observed, here the limitation is in the DIRICH FEB main buffer (6 channels $\times$ 120 words per channel  $>$ 499  words per DIRICH FEB).
 This is reflected in the number of leading and trailing edges derived, as shown in figure~\ref{highrate-scalar-vs-rate-edges} (right), where the edge multiplicity saturates close to the limit of the channel buffer size.

 Sharp saturation effects are not observed in either 64, 32, 8, or even only 6 active channels, in contrast to 1, 2, and 4 active channels because of the in-homogeneity of the hit distribution in the different channels (individual channel rate is defined by in-homogeneity of illumination and variation in PMT efficiency).
 For 64, 32 channels the edge multiplicity does not saturate at the maximum DIRICH ring buffer size (edge limit), which is $\frac{499\ (\textrm{DIRICH ring buffer size)}}{32\ \textrm{(number of channels)}}\ =\ $15.6 edges / trigger/ channel (cf., figure~\ref{highrate-scalar-vs-rate-all} (right)).
 Similarly, for 1,2, and 4 channels the saturation is not at the maximum channel DIRICH buffer size of 120 words (cf., edge limit in figure~\ref{highrate-scalar-vs-rate-edges}). 
This is due to the addition of epoch messages along with the LE and TE messages.
All DIRICH and combiner on one backplane share a common \SI{100}{\mega\hertz} clock onboard.
For approximately \SI{100}{\micro\second}, each channel sends out its epoch signal which contains the absolute time of the onboard clock.
The edges recorded in between two epochs are assigned time relative to the epoch, thereby reducing the data overhead of sending absolute time for each edge.
 This is also why the saturation limit of edges per readout is higher at a higher readout rate in figure~\ref{highrate-scalar-vs-rate-all} (right).
 Because the probability of receiving multiple edges in between the epoch markers is higher at high hit rates, the edges/epoch ratio per readout is higher compared to lower hit rates.  

 Apart from the buffer limitations in the DIRICH FEB, there are no severe limitations in the analog signal processing and subsequent digital processing of the signals.
 With the current setup, a single channel of the DIRICH can handle hit rates of at least \SI{2.2}{\mega\hertz}. This is a much higher rate than the expected photon rate at hot regions of the CBM RICH, and is limited by the maximum tested readout trigger rate of 50 kHz. For even larger readout rates (which could not be implemented in the TRB lab setup) even higher rates/channel seem feasible.
 
\subsection{Data quality test}
After evaluation of the readout chain performance in terms of data loss under high rate conditions, a natural next step is to check the quality of data transmitted at these rates.
The setup shown in figure~\ref{labsetup} is used for this measurement.
The measurement procedure is structured as follows,
\begin{itemize}
    \item Using the circular collimator, a ring like contour of hits is produced on the MAPMT plane. Since the opening width of the collimator slit is about \SI{1}{\milli\meter}, in order to avoid the divergence of light, the distance to MAPMT is adjusted such that the ring width is one pixel.
    \item A \SI{20}{\nano\second} selection window relative to the laser sync signal is applied to separate laser hits from uncorrelated background.
    Even within the correlation window, there are hits due to stray light, although they are minimal. 
    Therefore, in order to trace the ring contour, a minimum hit multiplicity per pixel condition is used.
    \item  The hit multiplicity in the pixels of the ring contour without any background illumination as well as the dark hit rate $N(0)$ is measured.
    The dark hits are detected in the absence of laser and LED, primarily due to stray light and these are subtracted for all measurements.
    \item The MAPMT is now homogeneously illuminated additionally to the laser reference pulse using an LED (in DC mode) as background source.
    Hence, the LED acts as the uncorrelated background source and the laser as the correlated signal source as elaborated above.
    \item The hit multiplicity in the ring for different background illumination is measured, $N(r)$ where $r$ is the overall rate in Hz.
    \item The ratio of hit multiplicities with and without LED background is measured, $\frac{N(r)}{N(0)}$.
    In the absence of loss of any data due to high rate, the ratio should be equal to one. 
\end{itemize}

\begin{figure}[h!tb]
\centering
  \includegraphics[width=0.64\textwidth]{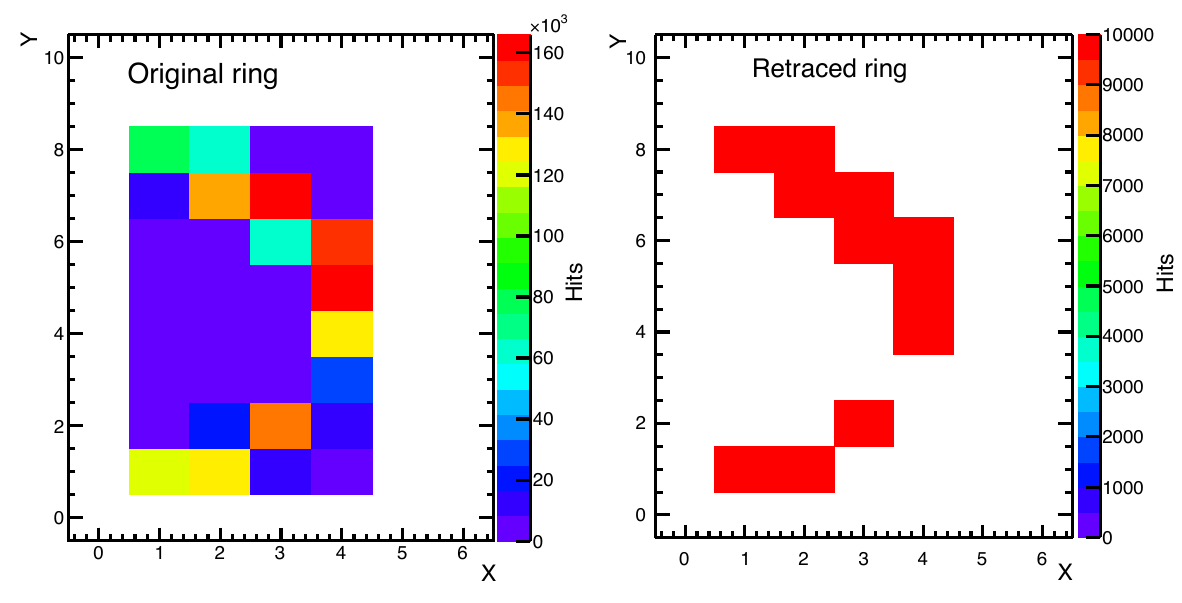}
  \includegraphics[width=0.33\textwidth]{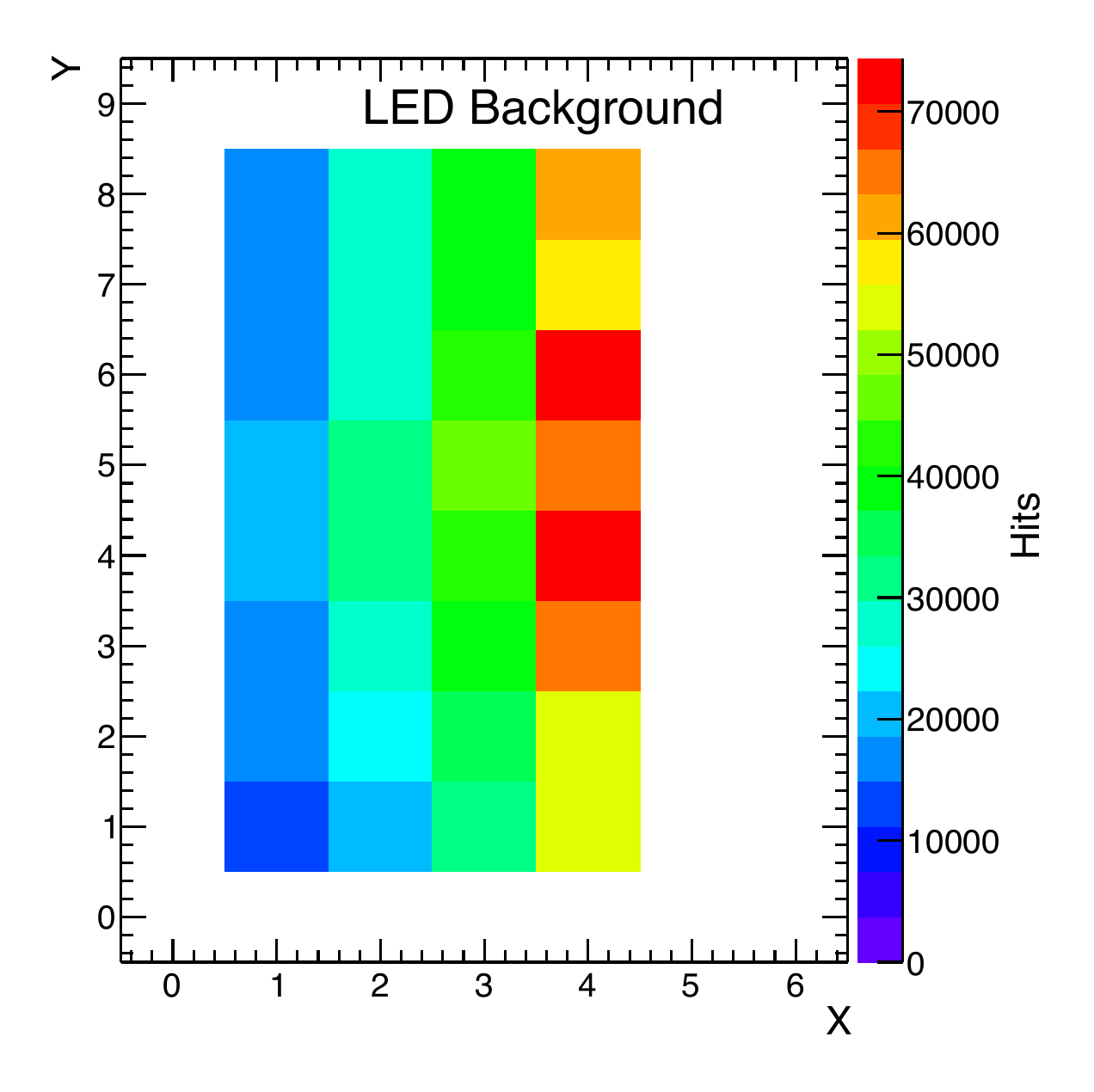}
  \caption{Left panel: The hits correlated to the laser signal are shown, where the arc of the ring on MAPMT is visible. 
  Middle panel: The ring contour is retraced by specifying a minimum hit per pixel criteria on the histogram from left panel.
  Right panel: The uncorrelated hits showing the background illumination by the LED (applying an exclusion window for the laser reference signals).}
  \label{highrate-ring-test-contour}
\end{figure}

One DIRICH FEB is used for this analysis, the measurement set up is depicted in figure~\ref{highrate-ring-test-contour}. 

\subsection{Results and discussions}
The average signal hit multiplicity per laser event within the correlation window is varied up to 6 hits.
During the test, the total hit rate (coming from the LED background hits) was gradually increased from ~3 Hz to ~300 kHz.
A regular readout trigger rate of \SI{40}{\kilo\hertz}, together with a laser pulse frequency of \SI{10}{\kilo\hertz} was used during these tests.
A total of $3 \times 10^6$ readout trigger events are analyzed for each data point.
\begin{figure}[h!tb]
\centering
  \includegraphics[width=0.85\textwidth]{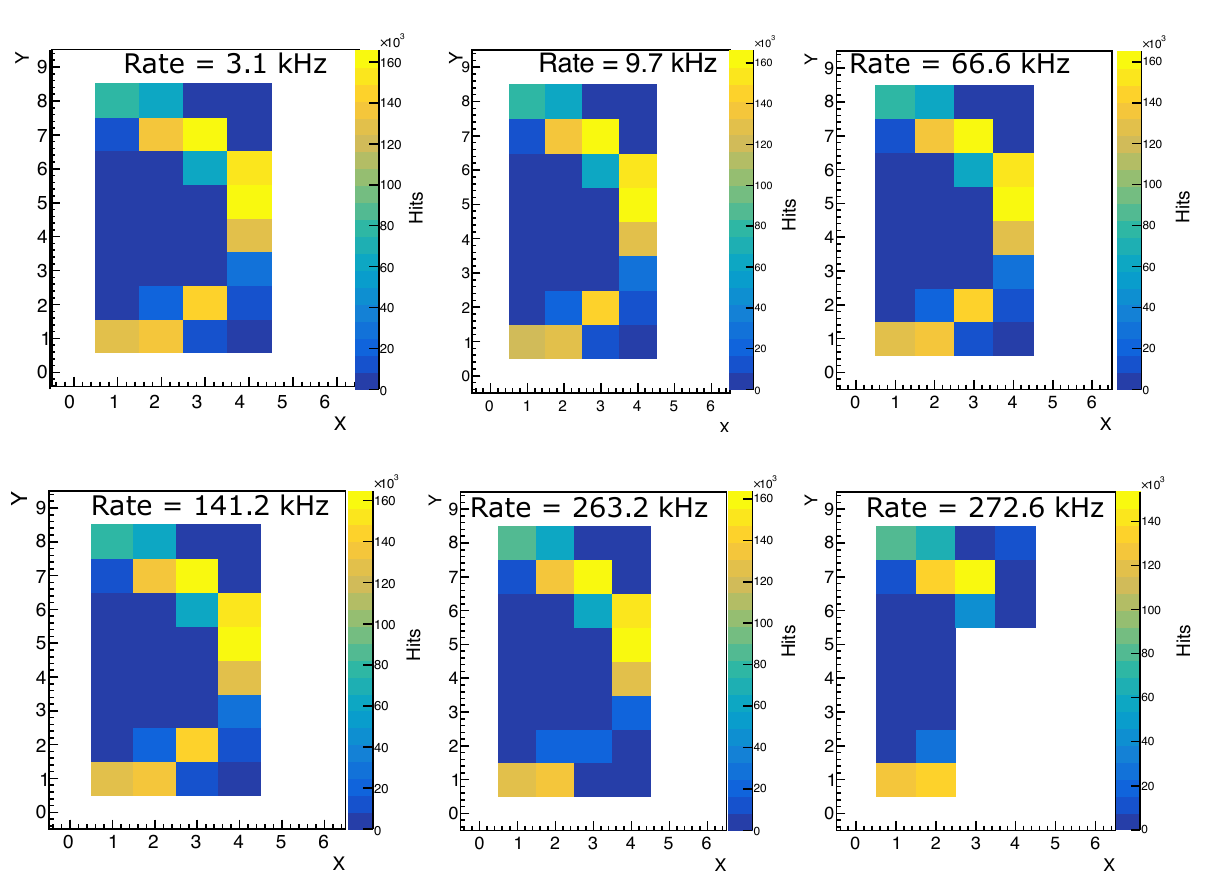}
  \caption{Snapshot of the reconstructed ring image for different overall hit rates as observed in the DAQ data stream. Laser correlated hit multiplicity per event = 5.94 hits.}
  \label{highrate-ring-test-snapshot}
\end{figure}
A snapshot of the ring image as reconstructed at different overall hit rates in the DAQ stream is shown in figure~\ref{highrate-ring-test-snapshot}.
It shows the image of laser-correlated hits after applying the \SI{20}{\nano\second} coincidence time cut.
No data loss is observed up to an average photon hit rate of $\sim$\SI{240}{\kilo\hertz}/channel.
Above these rates, data loss is expected (section ~\ref{section-highrate}) which is attributed to a buffer overflow in the DIRICH FEB.
The DIRICH FEB main buffer overflow due to background LED hits results in a complete data loss for a few channels including the channels where the ring is projected.
The channels sequentially transmit the data to the main ring buffer. Upon reaching its full capacity, the buffer discards data from a few channels first in the queue.
This explains the complete loss of data for some channels.
Even when some data from a few channels is lost, the data in the remainder of the channels remains identical for all hit rates.
This indicates there is no deterioration in  the quality of the data transmitted. 

\begin{figure}[h!tb]
\centering
  \includegraphics[width=0.49\textwidth]{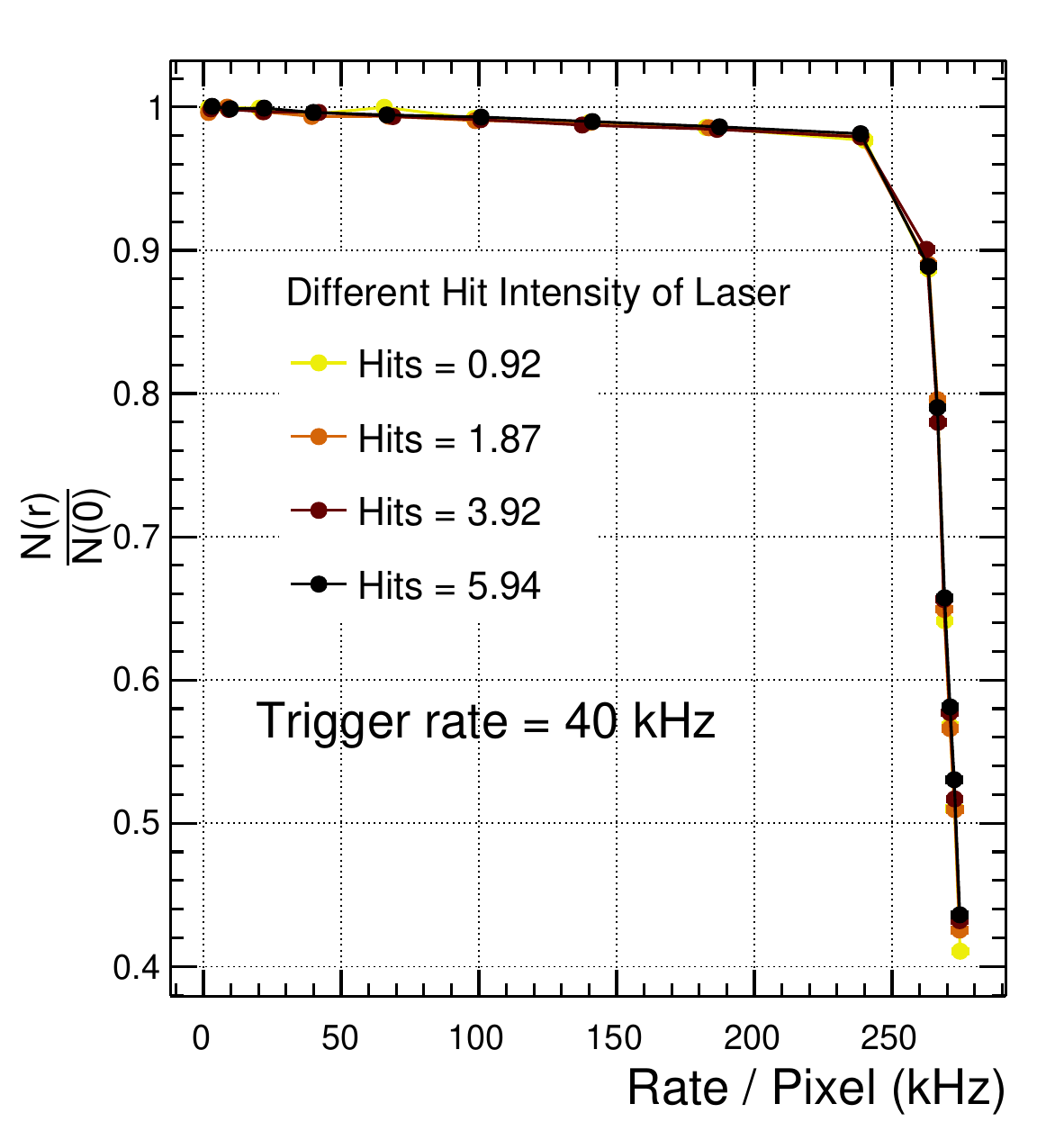}
  \includegraphics[width=0.49\textwidth]{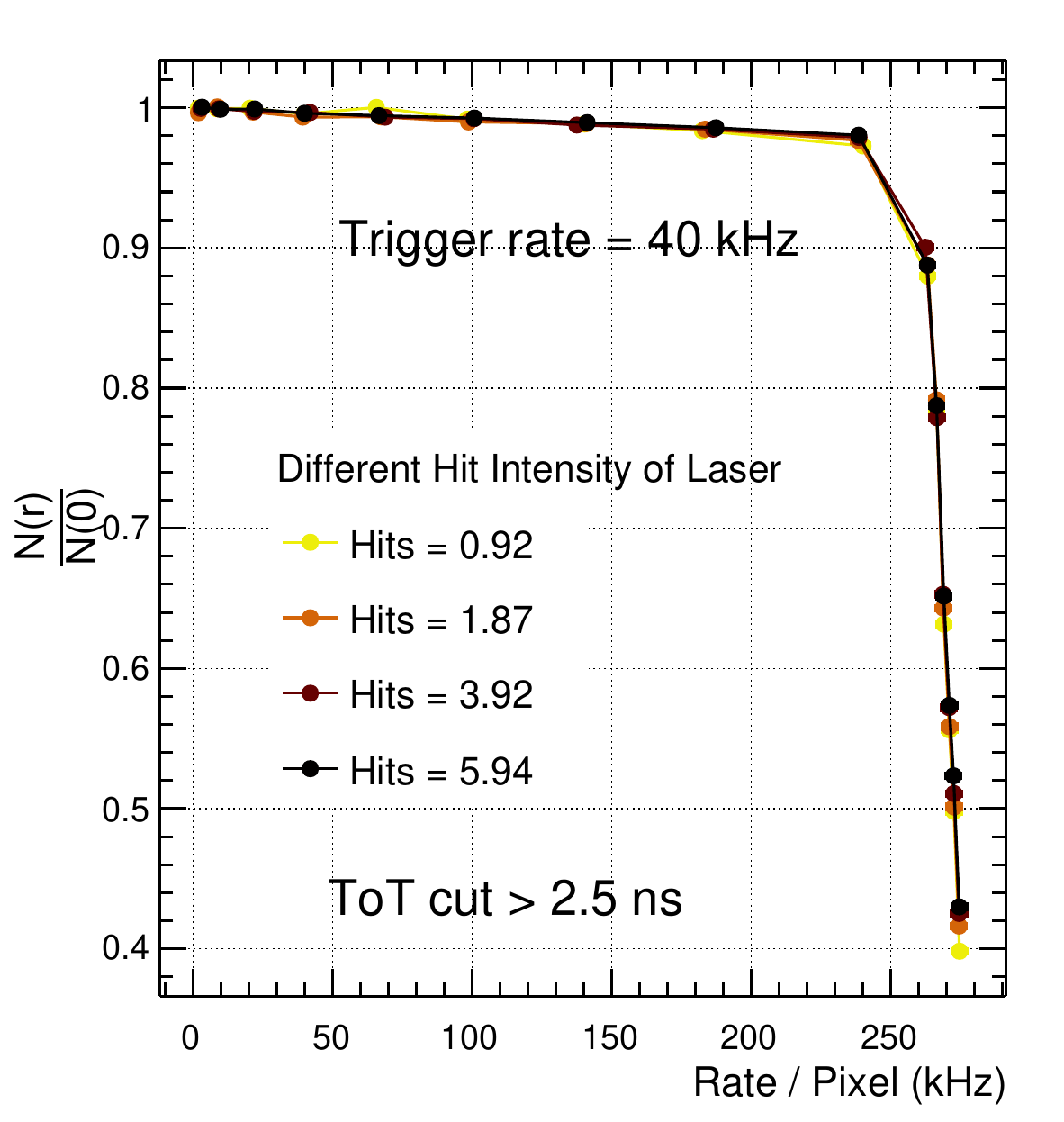}
  \caption{$\frac{N(r)}{N(0)}$ as a function of rate/pixel for different hit multiplicity in the ring. Left panel:  Without ToT-cut. Right panel:   With ToT-cut (ToT $>$ 2.5 ns). It can observed that the data quality remains relatively unaffected for input rates up to \SI{240}{\kilo\hertz} per pixel. The minor loss in data  (<1\%) at higher rates is not recoverable even after using a ToT cut (right). The detailed description is given in the text.}
  \label{highrate-ring-test-nr-n0}
\end{figure}

As a quantitative measure of the quality of the ring, the normalized number of hits per ring is calculated as $N(r)/N(0)$, with
\[
    N(r) \,=\,\sum_{Ring}{(N_{hits} \,- \,N_{hits-uncorrelated}- \,N_{dark-hits})}
\]
being the number of laser-induced signal hits per ring for different values of the background rate $r$, and
\[
    N(0) \,=\,\sum_{Ring}{(N_{hits-LED-off} \,- \,N_{dark-hits})}\;
\]
being the number of signal hits without background rate, with the LED being switched off.
Since LED hits are spread uniformly in the event trigger window, an uncorrelated contribution from the LED is estimated assuming a \SI{20}{\nano\second} wide readout window ($N_{hits-uncorrelated}$) and subtracted from $N(r)$ to correct for the contribution from the coincidence of LED hits onto the ring. 
The results are plotted in figure~\ref{highrate-ring-test-nr-n0}.

Nearly no loss of hits is observed in the ring up to the afore determined buffer overflow limit, where the ratio drops drastically.
Using a readout rate of \SI{40}{\kilo\hertz} this indicates good data quality for background rates up to at least \SI{240}{\kilo\hertz}.
The response is the same for different hit multiplicities in the ring.
Due to the presence of LED hits with laser hits, one expects a few additional hits due to crosstalk. 
In another iteration of the analysis, in order to reduce this crosstalk, a ToT-cut ($>$ \SI{2.5}{\nano\second}) is imposed on all hits registered.
Even with the ToT-cut, there is still a sub-percent shift from ideal linear behavior prior to the saturation region.
This shift increases with an increase in the overall hit rate, suggesting a direct correlation between them.
The cause of the shift can be understood as additional neighboring channel crosstalk, which has the ToT characteristics similar to that of a real photon hit.

\begin{figure}[htb]
\centering
  \includegraphics[width=0.43\textwidth]{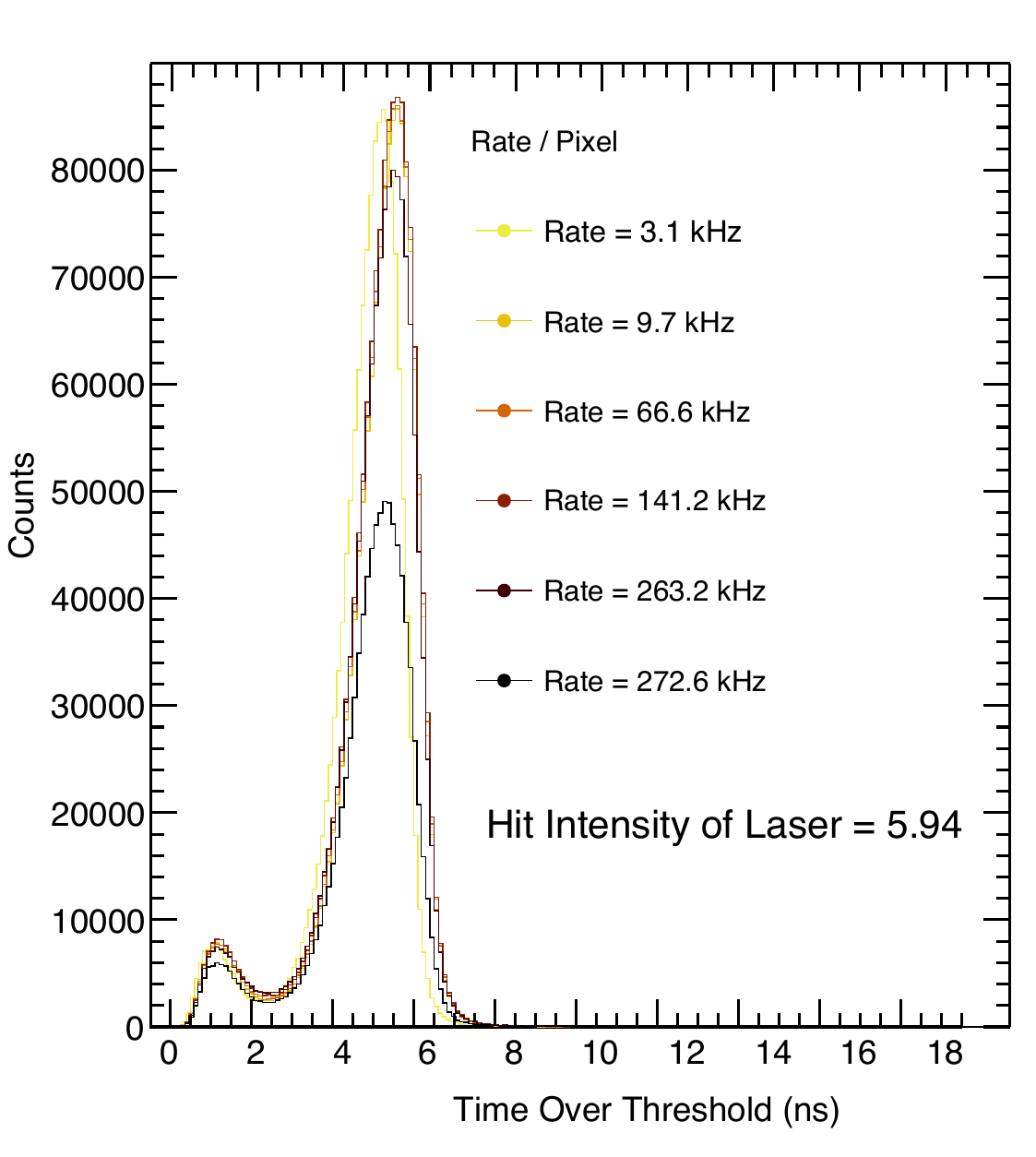}
  \includegraphics[width=0.49\textwidth]{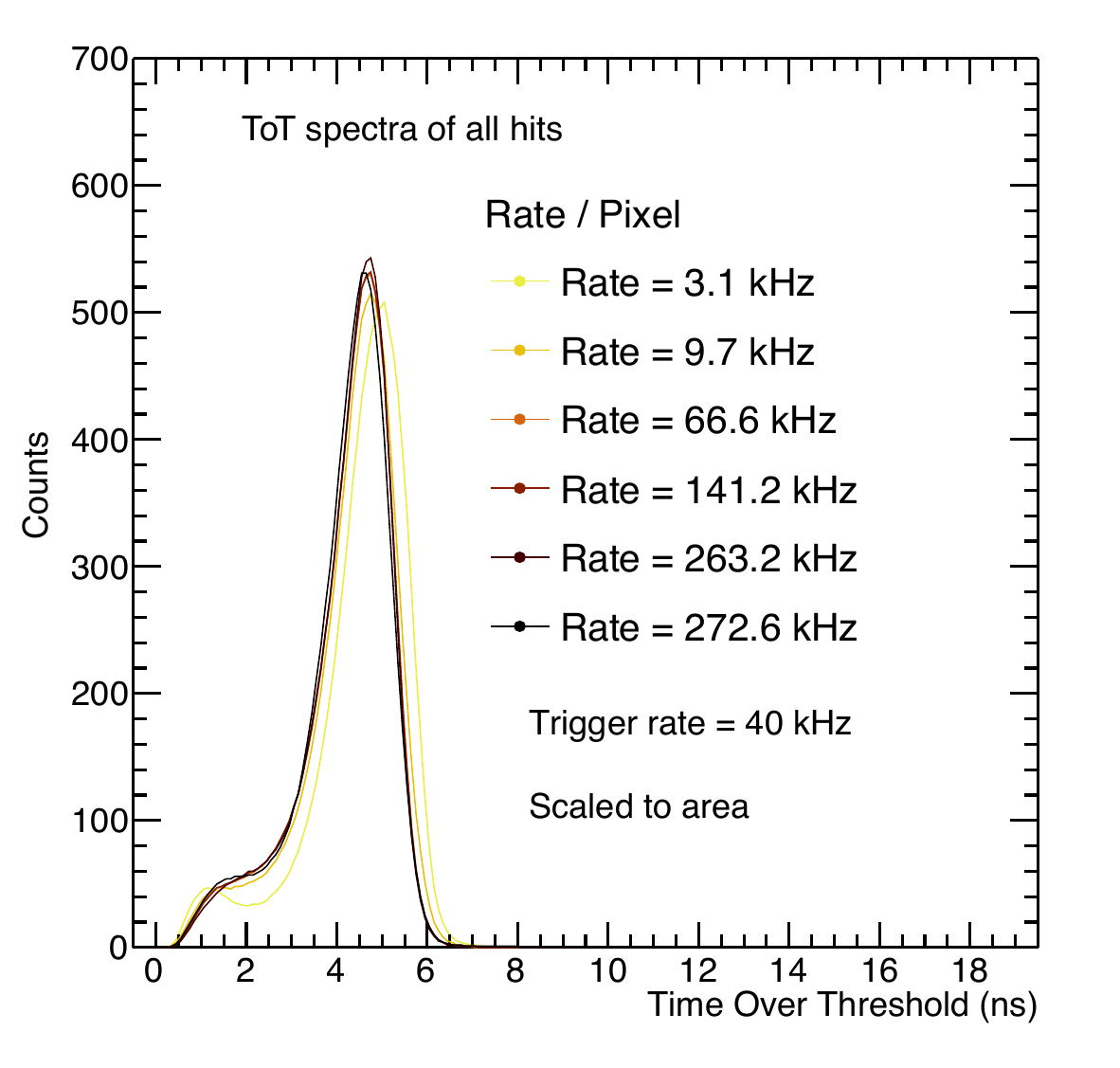}
  \caption{ Left panel: ToT spectra of correlated laser hits in the ring.
  Right panel: ToT spectra of integrated hits from all pixels, scaled to area.}
  \label{highrate-ring-test-tot-spectra}
\end{figure}

Another dedicated analysis was carried out in order to assess the possible effect of the high hit rate on the timing performance of the DIRICH by analyzing the ToT spectra of laser correlated hits in the ring and integrated hits in all pixels of the MAPMT. 
The results are plotted in figure~\ref{highrate-ring-test-tot-spectra}.
The ToT spectra of the laser correlated hits resemble a typical ToT spectra of six simultaneous hits in the MAPMT which was derived earlier in the previous section~\ref{section-high-occupancy-result}.
The ToT spectra of the laser correlated hits (figure~\ref{highrate-ring-test-tot-spectra} (left)) exhibit no significant qualitative difference for different hit rates.
However, the ToT spectra of integrated hits from all pixels resemble single hit ToT spectra for higher hit rates ($>$ \SI{60}{\kilo\hertz}).
This is to be expected because the hits are mostly dominated by uncorrelated LED hits, which are isolated hits spread over time.
For the lower hit rates ($<$\SI{10}{\kilo\hertz}), the hits are dominated by the laser correlated hits (since the pulse rate of the laser is \SI{10}{\kilo\hertz}).
This is reflected in the ToT spectra (figure~\ref{highrate-ring-test-tot-spectra} (right, yellow legend)), which bear resembles the ToT spectra of the correlated laser hits (figure~\ref{highrate-ring-test-tot-spectra} (left)).\\

In conclusion, the shape of ToT spectra does not change with the increase in hit rate, indicating that large hit rates have no effect on the DIRICH timing performance.

\section{Investigation of the noise induced by DC/DC converters of a new DIRICH Power module}
The MAPMT signal amplitude for single photons is rather small, in the order few mV only.
In the DIRICH FEB, these signals are first amplified in the front-end amplifier (roughly by a factor of 20) and subsequently discriminated (threshold in the range of \SIrange{30}{100}{\milli\volt} after amplification) to measure individual photons.
Preserving a good signal-to-noise ratio throughout the analog part of the digitization chain is important in order not to trigger on noise, and to obtain clean digitized signals despite small amplitudes.
A major source of noise and S/B degradation had been observed earlier to be caused by the usage of DC/DC converters on the DIRICH Power module. 
DC/DC converters are notoriously known to be a source of EMC noise, both electrically radiated and magnetically radiated, due to the large switching currents with fast rise / fall times. 

\subsection{The DIRICH LV power supply scheme}
Up to 12 DIRICH FEBs and the DIRICH Combiner module on a given backplane need to receive electrical power on several different voltage levels:
\SI{1.1}{\volt} and \SI{2.5}{\volt} are needed for the FPGA Core and I/O banks of the DIRICH FEB, and \SI{1.1}{\volt} as well for powering the FEB analog stage. 
Another \SI{1.2}{\volt} is needed in addition for powering the FPGA core on the DIRICH Combiner module, and \SI{3.5}{\volt} are necessary for powering the SFP fiber interface on the Combiner.
It is the task of the DIRICH Power module (one per backplane) to provide and regulate these 4 different voltage levels, which are further distributed via the DIRICH backplane to all modules.
In addition, the power module also receives the HV for the PMTs and delivers this also via the common backplane.
Two alternative powering concepts were initially implemented on the first iteration of the Power module: 
\begin{itemize}
\item All 4 individual LV supply levels can be either provided via a multi-pin connector, using external LV power supplies to provide all different voltage levels.
Here, the DIRICH Power module only provides additional filtering and stabilization using linear Low-Dropout Voltage regulators (LDOs). 
\item All 4 individual LV supply voltages are generated on the Power module itself using several individual DC/DC switching converters, fed by a single, common external supply line of up to \SI{36}{\volt}.
\end{itemize}
The first variant was implemented in the HADES RICH detector.
It has the big advantage of exceptionally clean power supply lines, inducing no additional noise into the sensitive pre-amplifiers.
However, the big disadvantage of this variant is the large supply currents at low voltage levels.
In case of HADES, supply currents of several \SI{100}{\ampere} had to be distributed, potentially causing shifts of ground levels, and significant power loss and heat on the supply cables.
A total of 10 individual LV power supplies are used for HADES in order to power the detector.
The required complexity of the power distribution cabling is enormous.

Variant 2 is much more elegant in this sense, requiring only a single LV power source, with moderate current due to the larger voltage.
However, operating 4 different DC/DC converters on each DIRICH backplane / Power module in very close vicinity to the sensitive analog part of the front-end electronics is a major challenge for proper shielding and filtering. 

A first iteration of the DIRICH Power module design allowed to choose between both powering variants, by comprising DC/DC converters and an additional connector for external powering (with DC/DC converters being switched off).
Initial tests of both schemes revealed major noise issues with the DC/DC variant, which is the reason why for the HADES detector, the external powering scheme was implemented.

However, for the CBM RICH detector, due to the much more confined space constraints and much longer cable lengths, using a set of external LV power supplies for all required voltage levels is excluded. 
Consequently, a second iteration of the DIRICH Power module was developed, using new, recent state-of-the-art “silent-switcher” DC/DC converters (LT8648S from Analog Devices) combined with better filtering- and shielding design, accompanied by an external metal shielding to reduce EMC noise as much as possible.

The following chapter describes a comprehensive test of this new DIRICH Power module, and a direct comparison to the first variant in terms of induced noise into the readout chain.
Aiming for a full qualification of this new variant of Power module for usage in the CBM RICH detector.

\begin{figure}[htb]
\centering
  \includegraphics[width=0.22\textwidth]{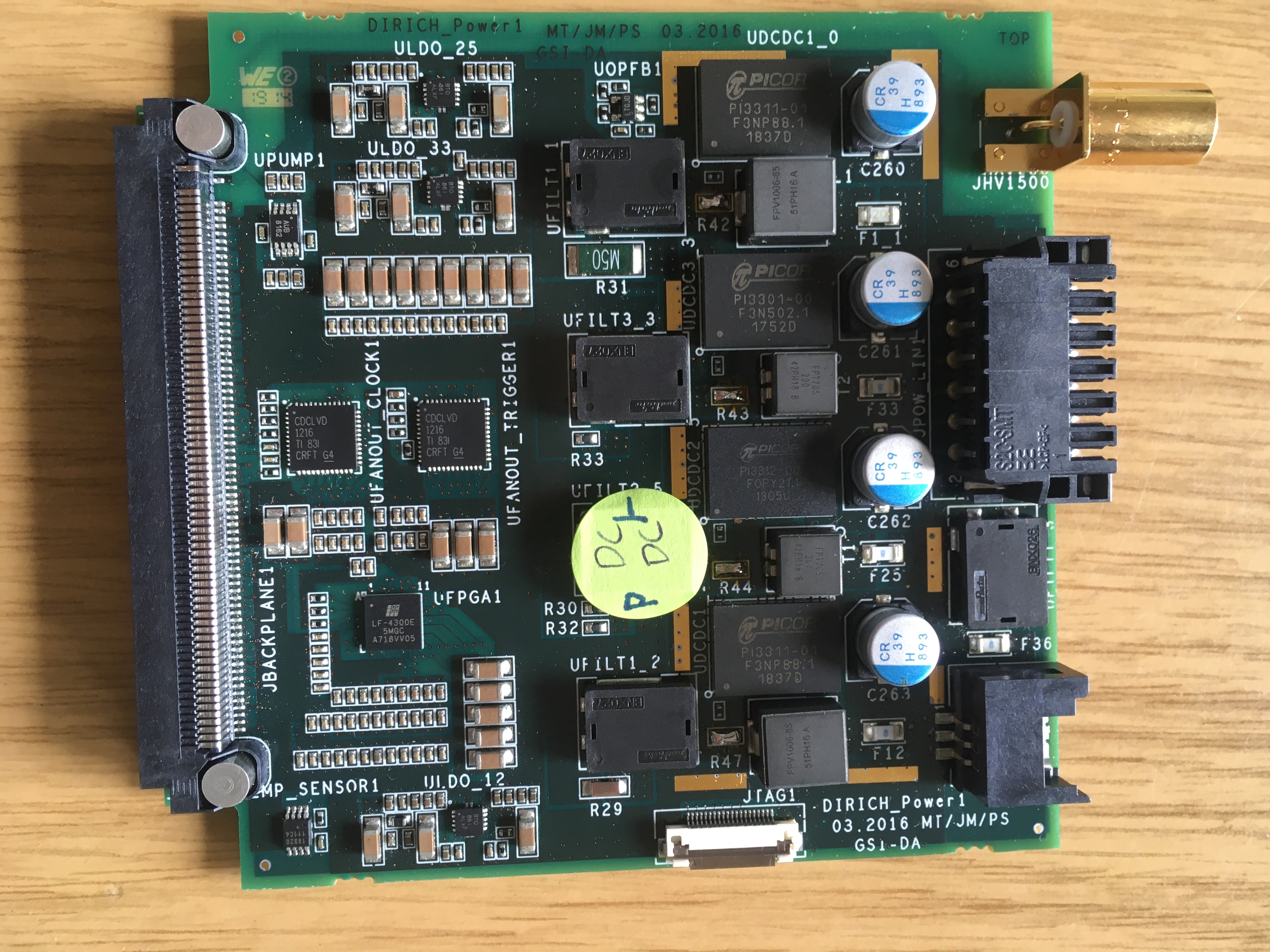}
  \includegraphics[width=0.22\textwidth]{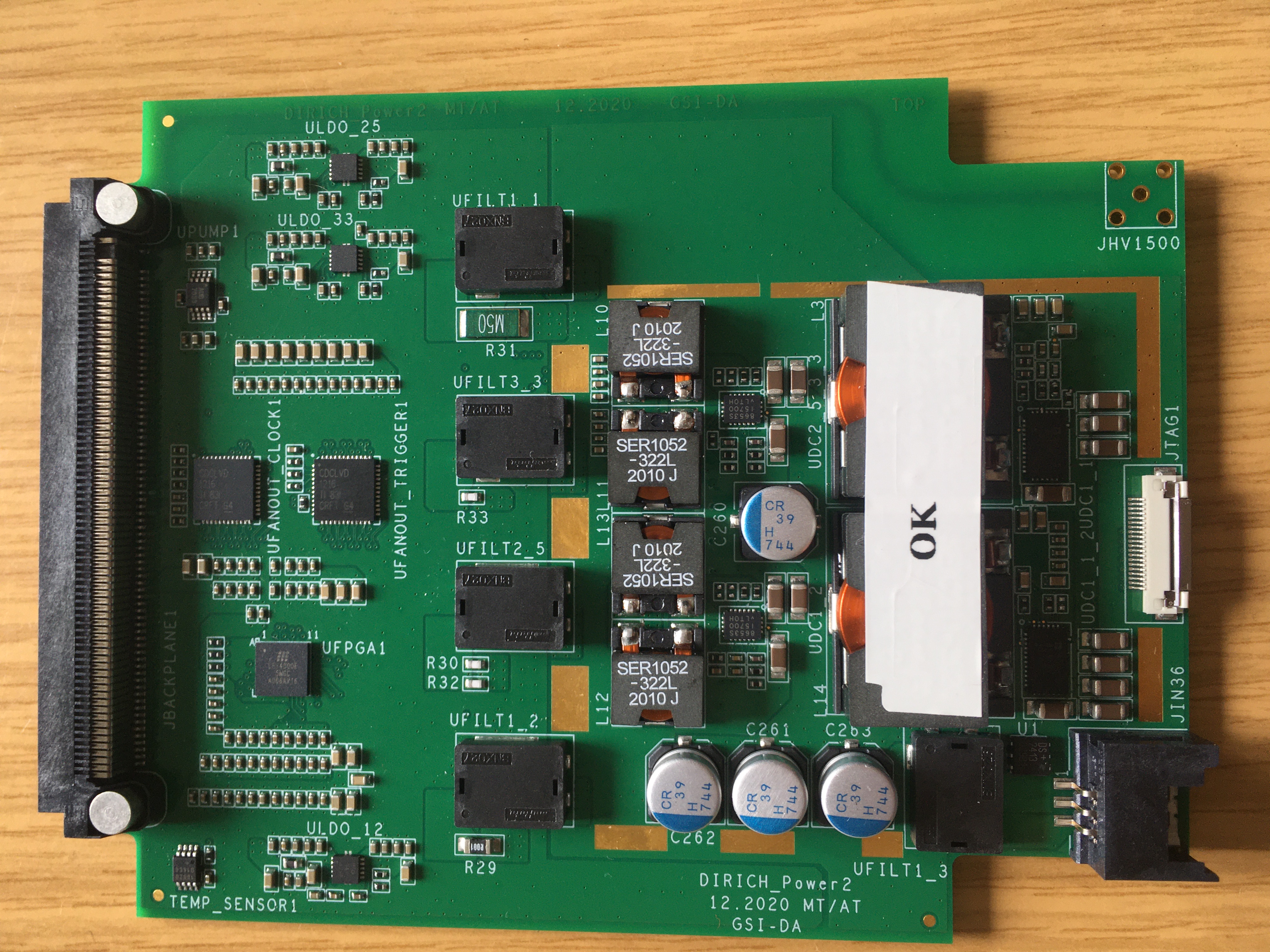}
  \includegraphics[width=0.22\textwidth]{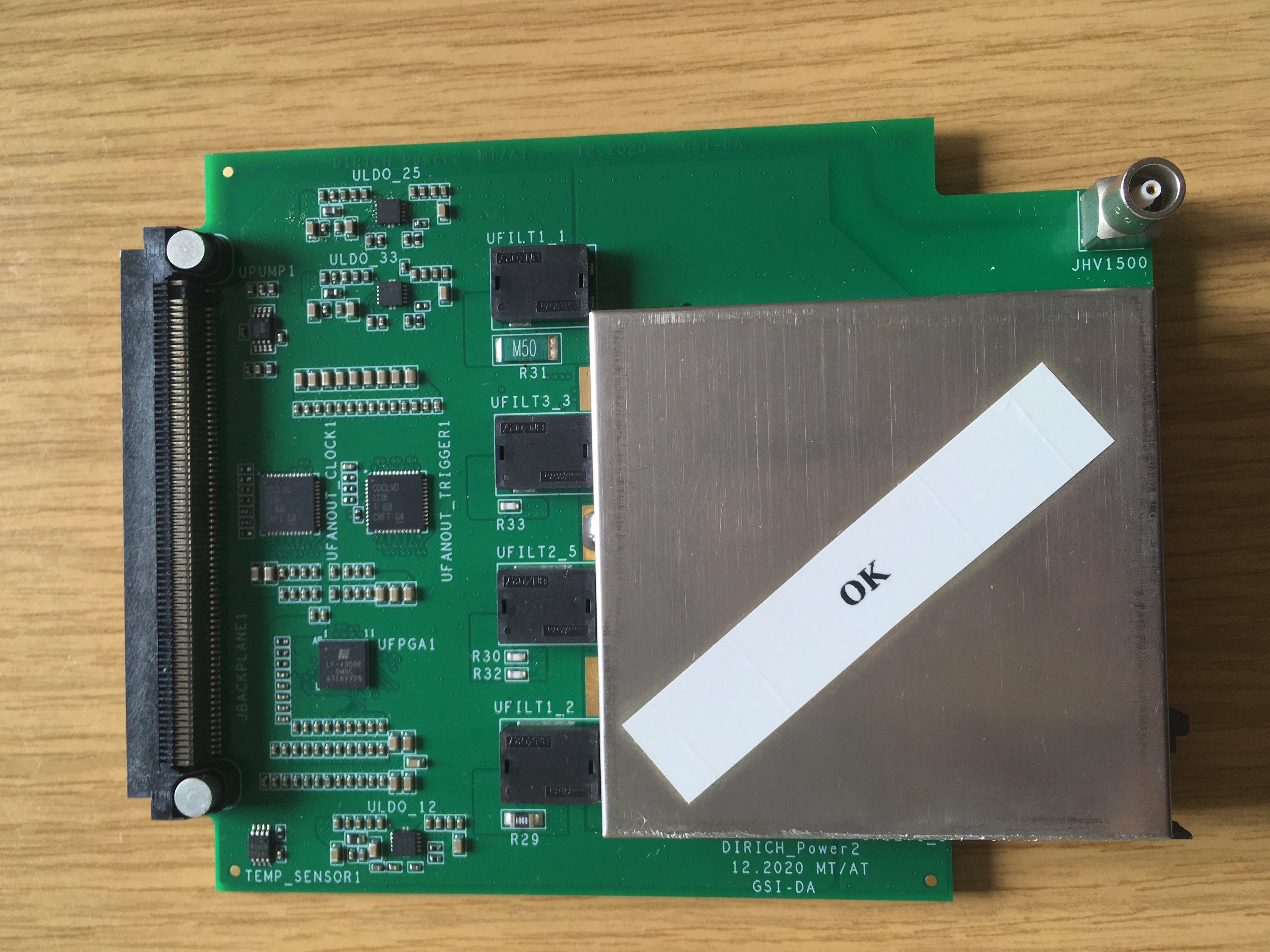}
  \includegraphics[width=0.22\textwidth]{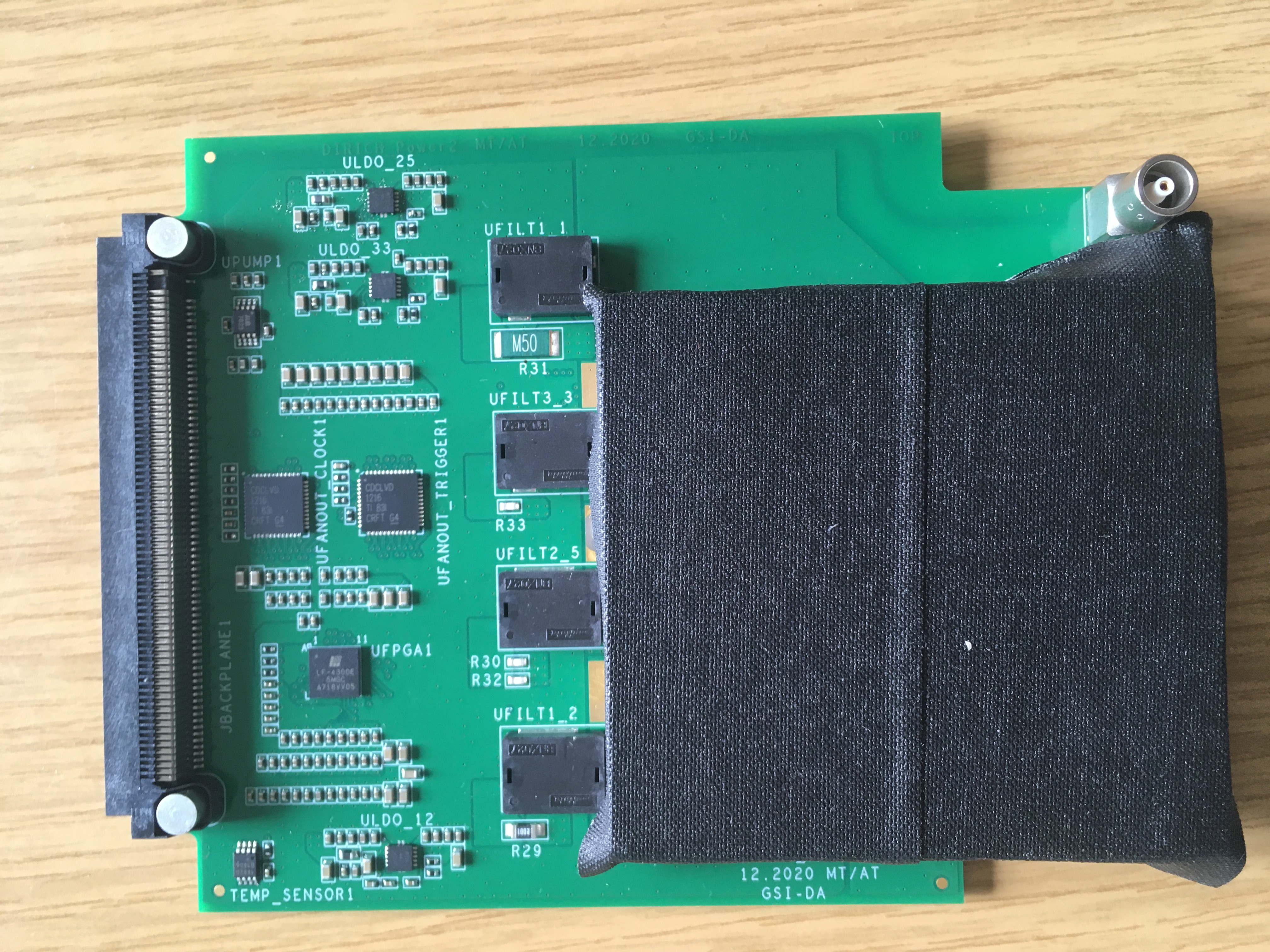}
  \caption{(From Left) Picture 1: First iteration of power module with on-board DC / DC converter, which has both single voltage input (SVI) port and externally regulated LV input port.
  Picture 2: Second iteration of power module with DC/DC converter (having only the SVI port).
  Picture 3: Second iteration of power module with additional shielding box.
  Picture 4: Second iteration of power module with shielding box and EMV foil.}
  \label{fig:2}
\end{figure}

Figure \ref{fig:2} shows the different variants of power modules which were tested:
The version 1 module (with either internal DC/DC or external LV powering), the new, improved version 2 module, the new module including external EMC shield, and the new module with EMC shield plus additional ferrite EMC absorber foil on top.
To reduce contact conduction, the EMV foil is covered with insulating tape.
All these variations were tested both in a dedicated  lab setup as well as inside the mRICH detector setup in mCBM.

\subsection{Measurement of Noise bandwidth}
To compare the effect of different powering variants on the noise performance of the readout, the noise bandwidth of the DIRICH FEB is used as the measurement observable. 
\begin{figure}[]
\centering
  \includegraphics[width=0.45\textwidth]{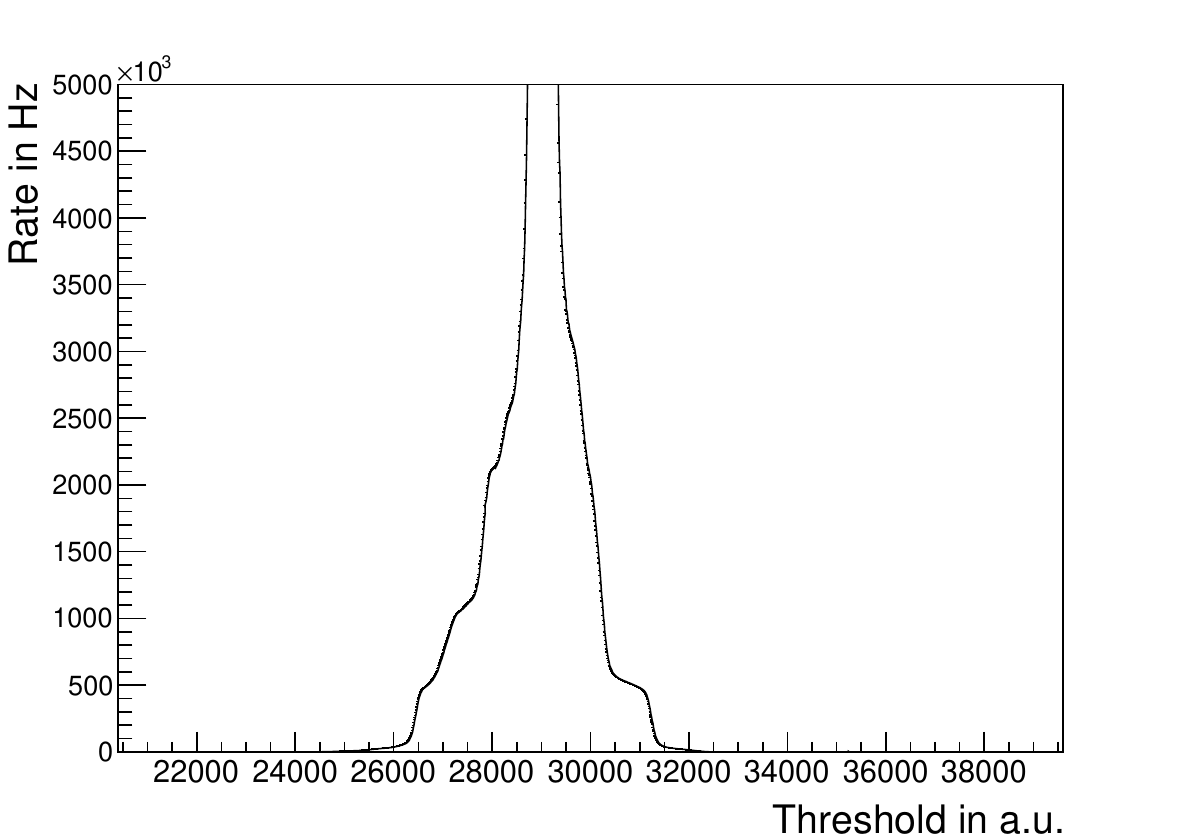}
  \includegraphics[width=0.45\textwidth]{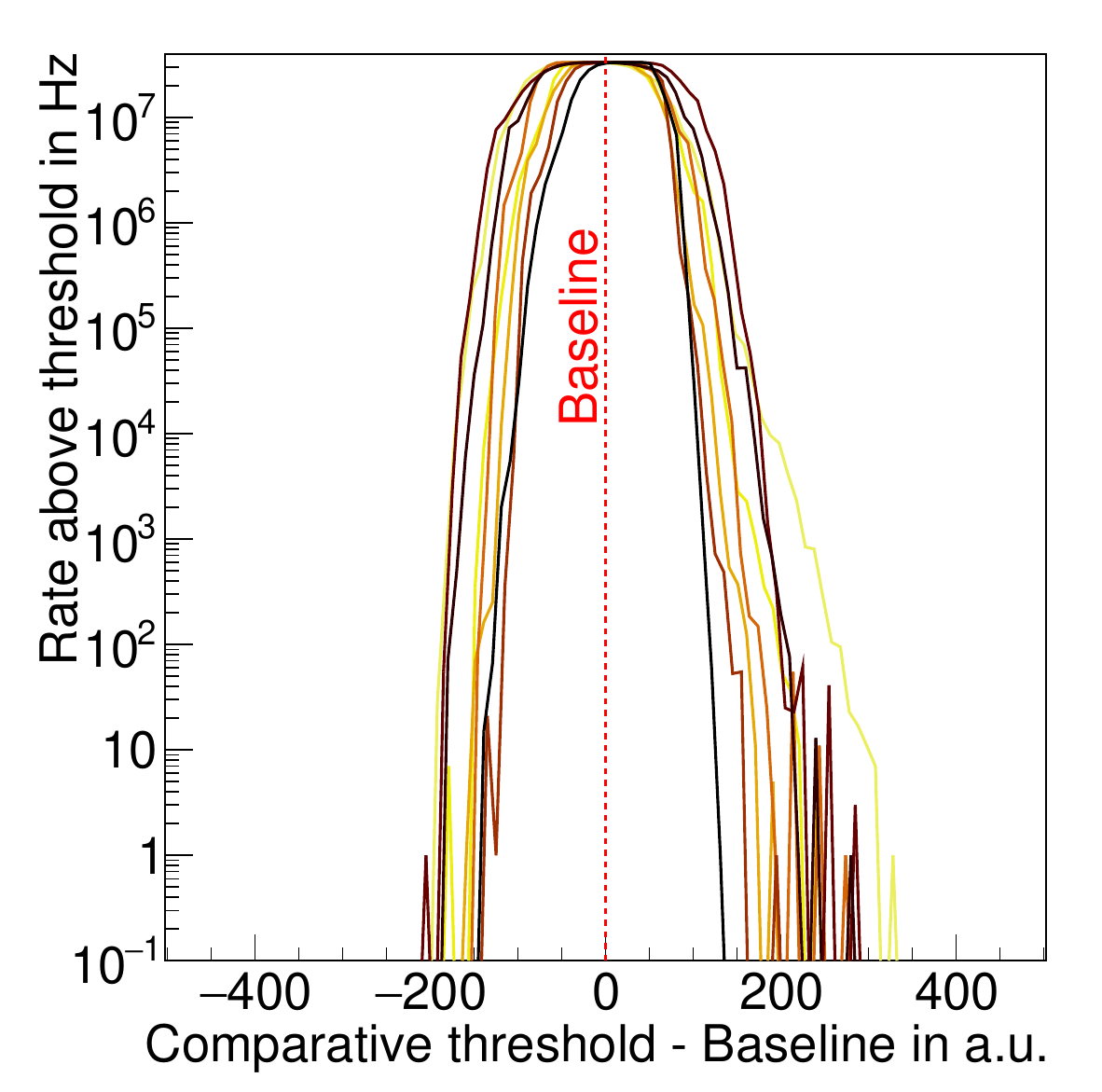}
  
  \caption{Left panel:  Graph depicting the scaler rate for a single channel of DIRICH for different noise reduction threshold voltage. Right panel:  Graph depicting scaler rate measured above the noise reduction threshold (Adopted from ~\cite{jorg-scan}).}
  \label{fig:6}
\end{figure}
Each DIRICH input stage is based on a comparator, comparing the actual input signal to an onboard-generated bias- or threshold voltage (see section~\ref{rich-fee}). 
For measurement of the noise bandwidth, this threshold voltage is systematically varied from negative values, crossing zero, to positive voltages, while measuring the rate of detected threshold crossings (i.e. “hit rate”) as function of threshold for each voltage using the scalers directly connected to the output of the comparators on the DIRICH.
A very steep increase in hit rate by many orders of magnitude (i.e., “noise”) is observed for absolute threshold voltage levels lower than the noise amplitude on the input signal. 
The bias voltage span for which such noise is observed is called “noise bandwidth”. 
The center of this band is defined as “bias level”, corresponding to threshold zero.
For normal DIRICH operation, the threshold must be set well outside this noise band, of which the half-width defines the minimum-usable threshold value.
The typical hit rate dependency on the threshold voltage levels is depicted in figure~\ref{fig:6}.

For comparison of the different power module variants, this noise bandwidth (both average and maximum over all channels of a given FEB) is used as a measure of the noise induced by the power module into this individual module.

\subsection{Laboratory setup}
A fully equipped backplane with 12 DIRICH FEBs, a power module, and a combiner is set up in the lab.
The power drawn for different power modules at different input voltages is shown in table~\ref{table:DC-DC},
\begin{table}[htb]
\centering
\caption{The power drawn by various types of power modules is summarized in this table.}
\begin{tabular}{|p{4.2cm}|p{2.5cm}|p{2.5cm}|p{4cm}|}
\hline
\textbf{Power module variant} & \textbf{Voltage (V)} & \textbf{Current (mA)} & \textbf{Power drawn by backplane (W)} \\ \hline
New (DC / DC)   & 32 & 768      &24.57\\
New (DC / DC) &18 &1242    &22.46\\
Old (DC / DC) &32 &796     &25.47\\
\hline
\end{tabular}
\label{table:DC-DC}
\end{table}

\begin{figure}[]
\centering
  \includegraphics[width=0.37\textwidth ]{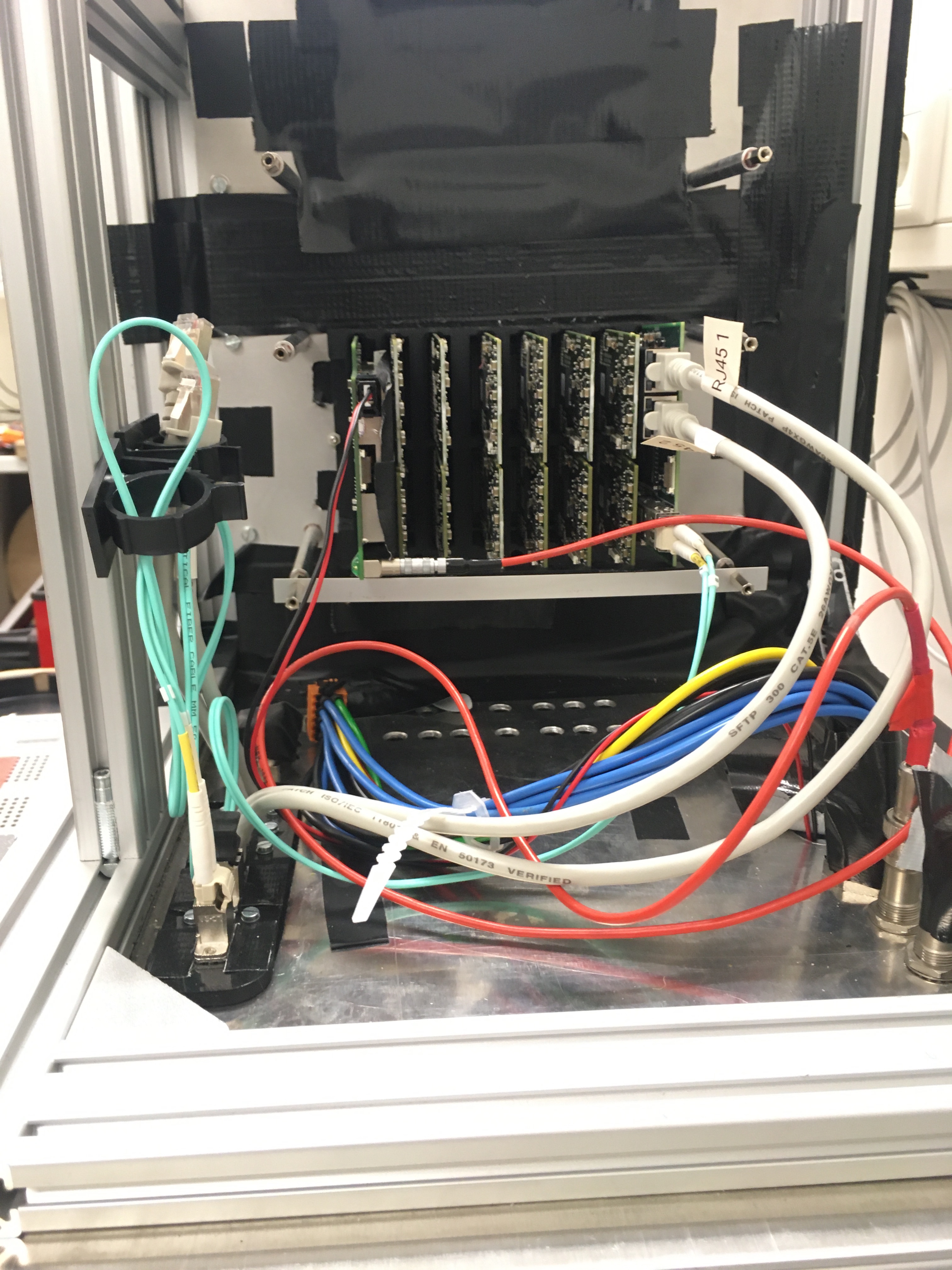}
  \includegraphics[width=0.44\textwidth]{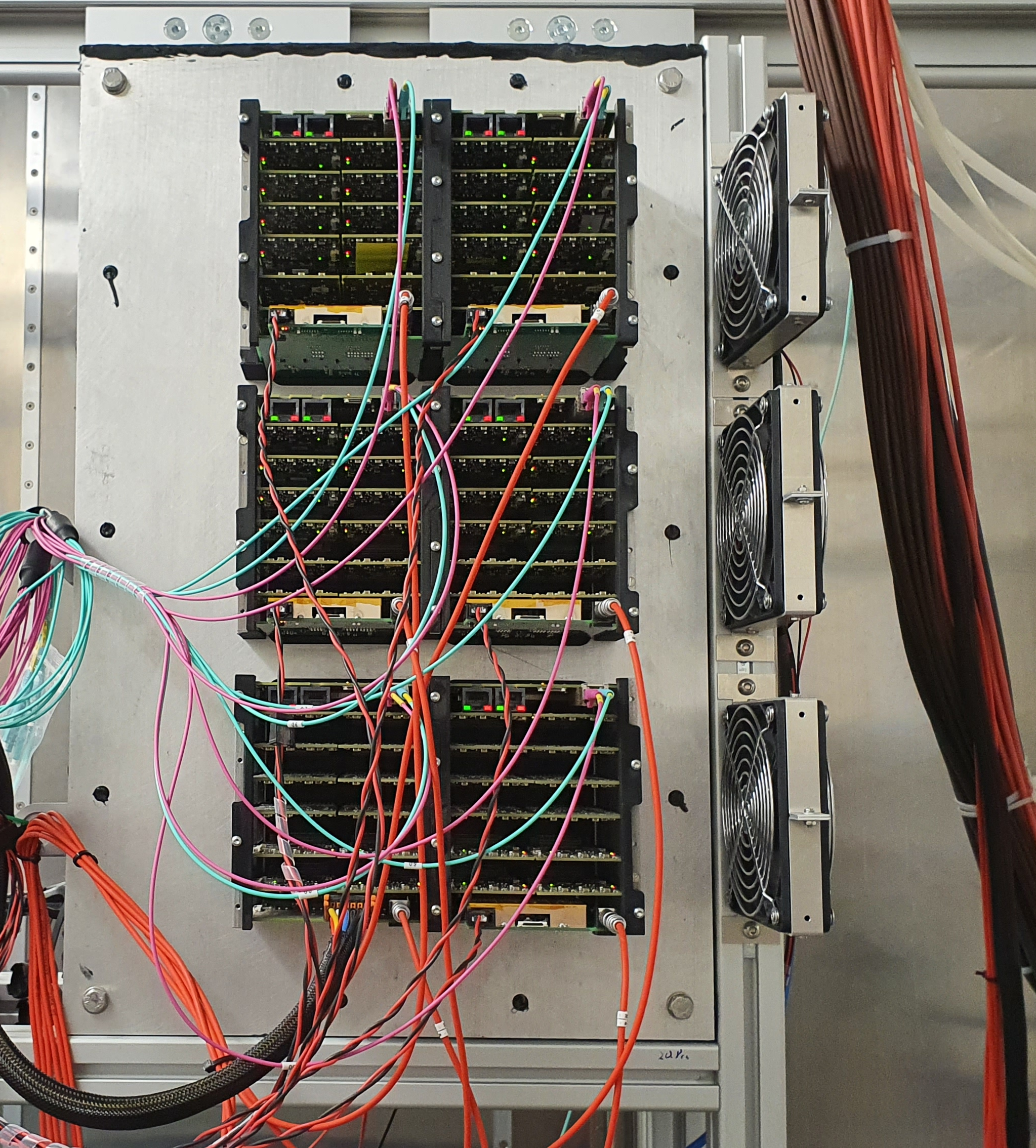}
  
  \caption{Left panel: Photograph showing the laboratory setup, where one backplane with 12 DIRICH FEBs and a power module. Right panel: Picture showing mRICH setup at mCBM, showing six backplanes with 72 DIRICH FEBs and six power modules. Here, FEBs and combiners of five power modules are powered by .
  }
  \label{lab and mrich setup}
\end{figure}

Measurement principle:
\begin{itemize}
    \item To understand the effect of distance from the power module on the emitted EM noise, the DIRICH FEB under tests are kept at different distances (positions relative) to the power module in a backplane.
    \item The backplane is fully loaded by filling vacant places with dummy DIRICH FEBs (having equivalent power consumption as ordinary FEBs).
    \item A noise bandwidth scan is performed.
    \item Average / Maximum half noise bandwidth over/of 32 channels of each DIRICH FEBs is obtained.
    \item To reduce the positional bias, the DIRICH FEB under test is shuffled to three different positions. The average of all three measurements is considered as the final noise bandwidth.
    \item The procedure is performed for the old power module with an external LV supply and using a supply voltage of \SI{32}{\volt}.
    \item The procedure is performed for two new power modules,  without and with using a shielding box and EMV foil.
    \item Since the new DC/DC based  converter can work from 18V to 32V supply voltage, the module is powered with both extreme voltages for comparison. 
\end{itemize}

\subsection{Measurement results and discussion}
\begin{figure}[h!b]
\centering
  \includegraphics[width=1.15\textwidth ]{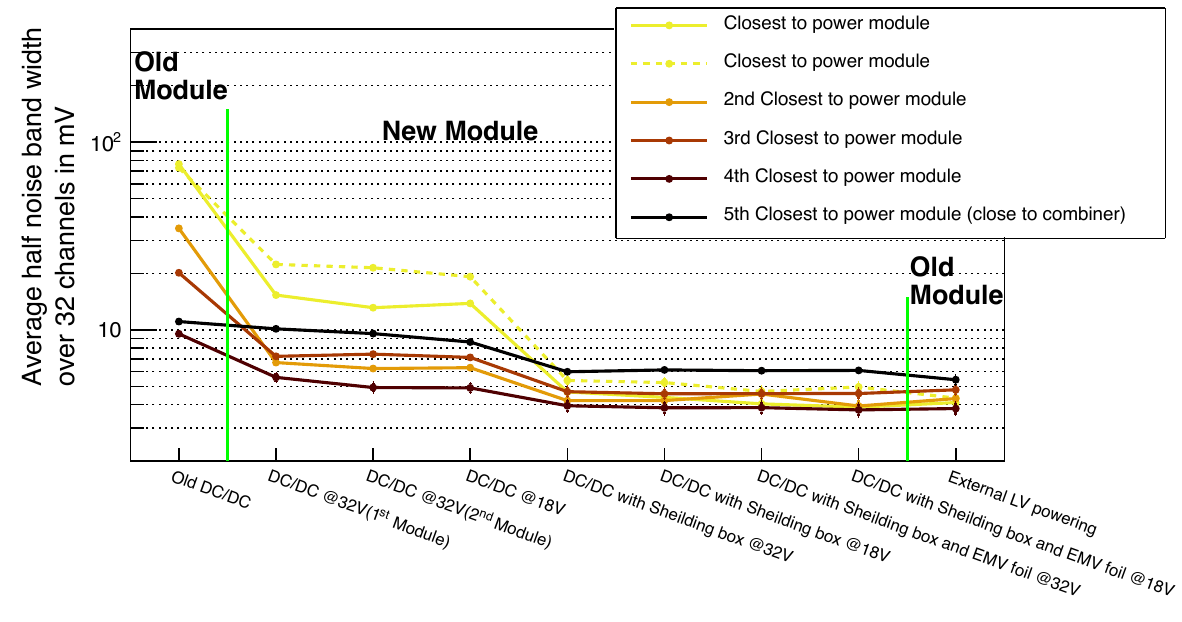}
  \caption{Average half noise bandwidth over 32 channels for different iterations of power modules with different configurations. Here, the green line indicates the separation of the old module from different configurations used in the new module.}
  \label{average half noise bandwidth}
\end{figure}

 For calculation of the noise bandwidth, six DIRICH FEBs at different distances to the power module are considered.
 Figure~\ref{average half noise bandwidth} shows the comparison of average noise bandwidth measured for different power modules with various configurations.
 The noise bandwidth for a DIRICH FEB close to the power module is higher, and reduces as the distance from the power module increases. 
 This dependence upon distance indicates that most of the induced noise is radiated via air and not caused by noise or ripple on the low-voltage side and supply lines (thanks to the extensive filtering on the power module). 
 
The old power module using an external LV supply, with DC/DC converters being switched off, has the lowest-possible noise level (less than \SI{10}{\milli\volt} half-bandwidth, roughly \SI{0.5}{\milli\volt} at the amplifier input), and can be regarded as a reference for evaluation of the new DC/DC-based power module.
Using the DC/DC converters on the old power module results in much larger noise levels (\SIrange{10}{100}{\milli\volt} depending on distance).
It is evident from comparing the first two points in figure~\ref{average half noise bandwidth}, that the new,  DC/DC based power module radiates much less noise (\SIrange{5}{25}{\milli\volt}) compared to the old module using the DC/DC  converter.
The use of the shielding box further reduces the noise (less than \SI{7}{\milli\volt}) for the new module, and adding additional EMV absorber foil on top of the shielding box does not have much effect.
Two different new power modules produce similar results for noise bandwidth.
The noise bandwidth is independent of applied input voltage (\SIrange{18}{32}{\volt}), as is the total power consumption (indicating constant DC/DC efficiency over a wide input voltage range). 
\begin{figure}[h!b]
\centering
  \includegraphics[width=1.15\textwidth ]{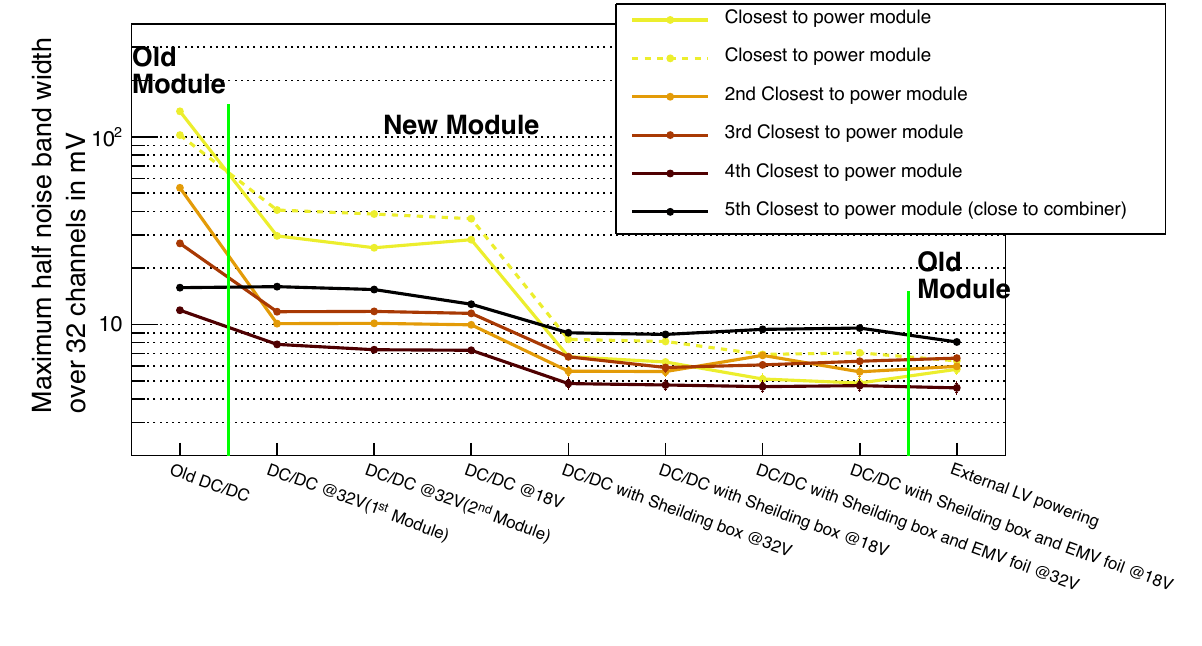}
  \caption{Maximum half noise bandwidth over 32 channels for different iterations of power modules with different configurations.}
  \label{maximum half noise bandwidth}
\end{figure}

With the new iteration of the DC/DC-based power module and the mounting of the additional shielding cover plate, the overall noise half-bandwidth is less than \SI{10}{\milli\volt} which is far less than the intended operating threshold of \SIrange{50}{100}{\milli\volt}, and comparable to the noise observed using external LV powering with the old module.

\subsection{mRICH setup and measurement results}
The lab measurements described above provide a solid, quantitative evaluation of the noise performance of the new DC/DC based power module.
However, these results are derived using a single backplane that is populated with several power dummy DIRICH FEBs due to an insufficient number of available DIRICH modules at the time of measurements.
\begin{figure}[thb]
\centering
  \includegraphics[width=1\textwidth ]{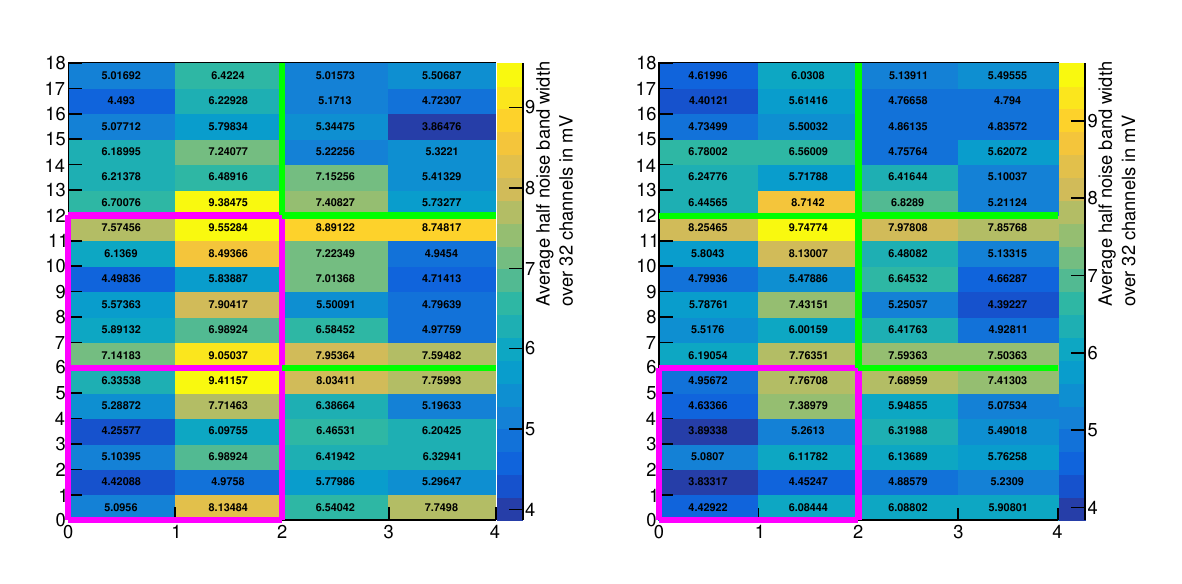}
  \caption{Noise characteristics of 72 DIRICH FEBs on 6 backplanes of the mRICH setup. Left panel: Scan 1 with 4/6 backplanes with new power module. Right panel: Scan 2 with 5/6 backplanes with the new power module. The Green lines indicate the border of each backplane with the new power module, and the purple line indicates the border of the backplane with the old power module. Each box represents a DIRICH FEB, and the values in each box are the average half noise bandwidth over 32 channels in mV. The results reflect the observations from a single backplane (in figure~\ref{average half noise bandwidth}).}
  \label{mrich-result-1}
\end{figure}

In order to confirm these results under realistic conditions, an additional, conceptually similar measurement was carried out using the mRICH prototype detector setup, which is part of the mCBM installation at GSI.
The mRICH setup (see figure~\ref{lab and mrich setup}) consists of  six fully populated backplanes (with six power modules, six combiner modules, and 72 DIRICH FEBs). 
Five (four) of the power modules are of the new, DC/DC-based type, only a single (two) power module (lower left in figure~\ref{lab and mrich setup}) is of the old type and using the external LV supply scheme.
Each of the new power modules was fitted with a shielding cage. 
Using this setup, the noise bandwidth of all 72 DIRICH FEBs on each of the 6 backplanes could be measured in parallel, thus providing a very comprehensive evaluation.
Two scans were performed, where in the first scan four out of six backplanes were fitted with power modules of the new type, and the rest with power modules of the old type powered by an external LV supply.
For the  second scan, one more of the old type modules was replaced with a module of new type.
Thus, one can analyze the noise bandwidth for more DIRICH FEBs at different distances and also one can have a direct comparison of one particular backplane with two different power modules under identical conditions.

The results of the measurement are plotted in figure~\ref{mrich-result-1}. 
The following inferences can be drawn from the two scans:
\begin{itemize}
    \item The average half noise bandwidth of all the DIRICH FEBs connected to the backplane with the new power module is less than \SI{10}{\milli\volt}, and thus comparable to the old power module with external LV supply (the optimum reference standard).
    \item The measured noise bandwidth is far below the  standard operating threshold values for standard operation ($\ge$ \SI{50}{\milli\volt}).
    \item The noise bandwidth in the DIRICH FEB closest to the power module is slightly increased and quickly drops with increasing distance. A similar observation can be made on each of the backplanes.
\end{itemize}

\section{Summary and outlook}
Using a dedicated lab test setup, various measurements on the DIRICH readout chain were carried out in order to qualify the readout chain for the CBM RICH detector, and several quantitative results were obtained:
The neighboring channel charge sharing crosstalk for single photon detection with H12700 MAPMTs connected to DIRICH FEB modules was measured to be less than 2$\%$. 
This crosstalk probability is independent of the applied discrimination threshold in the range of \SIrange{30}{70}{\milli\volt} (well below the average single-photon signal amplitude).
A dedicated test using simultaneously incident photons from a pulsed laser source was performed to check the compatibility of the readout chain to handle high photon occupancy, as it is expected in CBM. 
The large occupancy confirmed to cause significant additional capacitive crosstalk hits, with almost 88\% additional crosstalk hits for 18 signal hits (on the higher side of occupancy).
However, these capacitive crosstalk hits are found to have low ToT characteristic ($\leq$ \SI{3}{\nano\second}), well separated from ToT values for real photon hits.
Hence, these additional hits can be easily removed by imposing a simple ToT-cut on registered hits.
Though the overall data rate does increase due to additional crosstalk hits (at least as long as no additional ToT condition is implemented on the DIRICH FEB itself), the data after applying a ToT-cut will be rather clean.
In the expected maximum MAPMT occupancy range of up to $\approx 28 \%$, an optimal ToT-cut of \SI{3}{\nano\second} is found to be sufficient. 

The lab setup was also used to perform a successful high rate test of the DIRICH front-end electronic readout chain. 
Apart from the well-understood data transfer limitations imposed by the lab setup, it is found that an individual DIRICH FEB channel can handle at least up to \SI{2.2}{\mega\hertz} input rate.
However, the size of the individual channel buffers and the main DIRICH FEB buffer is currently do impose limitations. 
A test of the data quality at large input rate was performed using a pulsed laser as a signal source and a  DC-powered LED providing uncorrelated, adjustable background.
A collimator used in combination with the laser provided a realistic ring-like hit pattern on the MAPMT for this test.
Even at very high uncorrelated background hit rates, the data purity of the signal hits of interest is maintained (if appropriate time cuts are applied during data analysis), and the data is still transmitted reliably. 
Analog signal propagation and digitization show now impairment even under the highest tested rate conditions.

A dedicated measurement was made to characterize the capability of DIRICH FEB to distinguish simultaneous hits in time.
For this measurement, the input fed to the DIRICH FEB(s) is a controlled negative pulse produced using a pulse generator (similar to a single photon response of PMT).
The leading edge time information from two hits from two different channels of same and different DIRICH FEB(s) are used for the evaluation.
It is found that the leading edge accuracy to distinguish both hits in the same DIRICH FEB is about \SI{22}{\pico\second} RMS and in different DIRICH FEBs is about \SI{36}{\pico\second} RMS, hence proving the DIRICH FEB is not the limiting factor in resolving simultaneous hits as compared to the MAPMTs (transit time spread of H12700 MAPMTs is about \SI{149}{\pico\second} RMS).

Using the identical setup, the DIRICH FEB's capability to distinguish double photon hits from single photon hits by utilizing ToT information is evaluated.
It was found that the ToT has a strong correlation with the width of the input pulse, but a weak correlation with the amplitude of the signal.
Hence, presently, the DIRICH FEB is found not suitable to resolve single photon hits from double photon hits.
This is not regarded as a critical issue, as it may offer the added advantage of providing the capability for reconstructing partially overlapping rings, which are in themselves rare.

The new iteration of the power module to be used in the CBM RICH is tested for its noise-emitting characteristics by observing the noise bandwidth measure of DIRICH FEBs on the same backplane, powered by the new module and using DC/DC converters in combination with a single input supply voltage of \SIrange{18}{32}{\volt}.
Using an additional metal shielding covering the DC/DC circuit part, the new power module produces noise comparable to powering via externally regulated low-voltage supplies, as previously used in HADES. 
The observed average noise half-bandwidth is less than \SI{10}{\milli\volt} which is far below the intended operation noise threshold regime (\SIrange{50}{70}{\milli\volt}). 
These results could be confirmed  with the larger (six fully equipped backplanes) mRICH setup.

The measurements described above not only proved valuable for the final qualification of the DIRICH readout chain for CBM RICH but also served to define optimal operating parameters for later routine operation.
The results indicate that the DIRICH electronics proved to be functionally adequate for the CBM RICH detector.

As an outlook, the developed lab setup can now also be used to derive a parametrization of capacitive crosstalk probability as function of distance from the initiating photon hit, and as function of occupancy.
This will help to implement crosstalk in a more realistic way into the CBMROOT simulation framework. 

\bibliographystyle{elsarticle-num}
\bibliography{bib_tn}

@book{cherenkov_effect,
    author = {William R. Leo},
    title = {Techniques for Nuclear and Particle Physics Experiments},
    publisher = {Springer Verlag Berlin Heidelberg},
    year = {1994},
    doi = {10.1007/978-3-642-57920-2},
    url= {https://www.springer.com/de/book/9783540572800}
}

@article{co2_threshold,
Title = {The CBM RICH project},
journal = {Nuclear Instruments and Methods in Physics Research Section A: Accelerators, Spectrometers, Detectors and Associated Equipment},
volume = {766},
pages = {101-106},
year = {2014},
issn = {0168-9002},
doi = {https://doi.org/10.1016/j.nima.2014.05.071},
url = {https://www.sciencedirect.com/science/article/pii/S0168900214006111},
author = {J. Adamczewski-Musch et al.},
}

@article{mapmt_series_testing,
title = {Status of the CBM and HADES RICH projects at FAIR},
journal = {Nuclear Instruments and Methods in Physics Research Section A: Accelerators, Spectrometers, Detectors and Associated Equipment},
volume = {952},
pages = {161970},
year = {2020},
issn = {0168-9002},
doi = {https://doi.org/10.1016/j.nima.2019.03.025},
url = {https://www.sciencedirect.com/science/article/pii/S0168900219303249},
author = {J. Adamczewski-Musch et al.},
}

@article{Michel_2017,
    year = {2017},
    month = {jan},
    publisher = {},
    volume = {12},
    number = {01},
    title = "{Electronics for the RICH detectors of the HADES and CBM experiments}",
    journal = {Journal of Instrumentation},
    author = {J. Michel and M. Faul and J. Friese and C. Höhne and K.-H. Kampert and V. Patel and C. Pauly and D. Pfeifer and P. Skott and M. Traxler and C. Ugur},
    doi = {10.1088/1748-0221/12/01/C01072},
    
}

@unpublished{lebedev_collab_meet_2020,
    author = {S.Lebedev},
    title = {Status RICH simulations},
    note = {36th CBM Collaboration meeting 2020}
}

@TECHREPORT{pauly_2017,
      key          = {209729},
      editor       = {Toia, Alverica and Selyuzhenkov, Ilya},
      title        = {{CBM} {P}rogress {R}eport 2017},
      number       = {CBM Progress Report 2017},
      address      = {Darmstadt},
      publisher    = {GSI Helmholtzzentrum für Schwerionenforschung},
      reportid     = {GSI-2018-00485, CBM Progress Report 2017},
      isbn         = {978-3-9815227-5-4},
      series       = {CBM Progress Report},
      pages        = {146},
      year         = {2018},
      doi          = {10.15120/GSI-2018-00485},
}

@PHDTHESIS{Reinecke:2016grs,
      author       = {Reinecke, Sascha},
      title        = {{C}haracterisation of photon sensors for the {CBM}-{RICH}
                      and its use for the reconstruction of neutral mesons via
                      conversion},
      school       = {Bergischen Universität Wuppertal},
      year         = {2016},
      url          = {https://repository.gsi.de/record/206473},
}

@PHDTHESIS{jorg-scan,
      author       = {Förtsch, Jörg},
      title        = {{U}pgrade of the {HADES} {RICH} photon detector and first performance analyses},
      school       = {Bergischen Universität Wuppertal},
      year         = {2021},
      doi = {https://doi.org/10.25926/69gp-b484}
      
}

@phdthesis{adrian_weber,
    author =  {Weber, Adrian Amatus },
    title = {Development of readout electronics for the RICH detector in the HADES and CBM experiments - HADES RICH upgrade, mRICH detector construction and analysis},
    school = {Justus-Liebig-Universität Gießen},
    year = {2021},
doi={http://dx.doi.org/10.22029/jlupub-288}
}

@misc{thor-labs-nominal,
    Title = {Nominal values of the ND filters, Thor labs},
    url = {https://www.thorlabs.com/newgrouppage9.cfm?objectgroup_id=5011} 
}

@article{BROGNA2006301,
title = {N-XYTER, a CMOS read-out ASIC for high resolution time and amplitude measurements on high rate multi-channel counting mode neutron detectors},
journal = {Nuclear Instruments and Methods in Physics Research Section A: Accelerators, Spectrometers, Detectors and Associated Equipment},
volume = {568},
number = {1},
pages = {301-308},
year = {2006},
note = {New Developments in Radiation Detectors},
issn = {0168-9002},
doi = {https://doi.org/10.1016/j.nima.2006.06.001},
url = {https://www.sciencedirect.com/science/article/pii/S0168900206011247},
author = {A.S. Brogna and S. Buzzetti and W. Dabrowski and T. Fiutowski and B. Gebauer and M. Klein and C.J. Schmidt and H.K. Soltveit and R. Szczygiel and U. Trunk},
}
\clearpage
\section{Appendix}
This appendix contains all additional figures referenced in this paper but not considered important enough to be included in the main text.
\subsection{Calibration of filter with ToT-cut}
\label{app-calibration-with-tot}
\begin{figure}[!htb]
\captionsetup{width=0.5\linewidth}
\centering
  \includegraphics[width=0.7\linewidth]{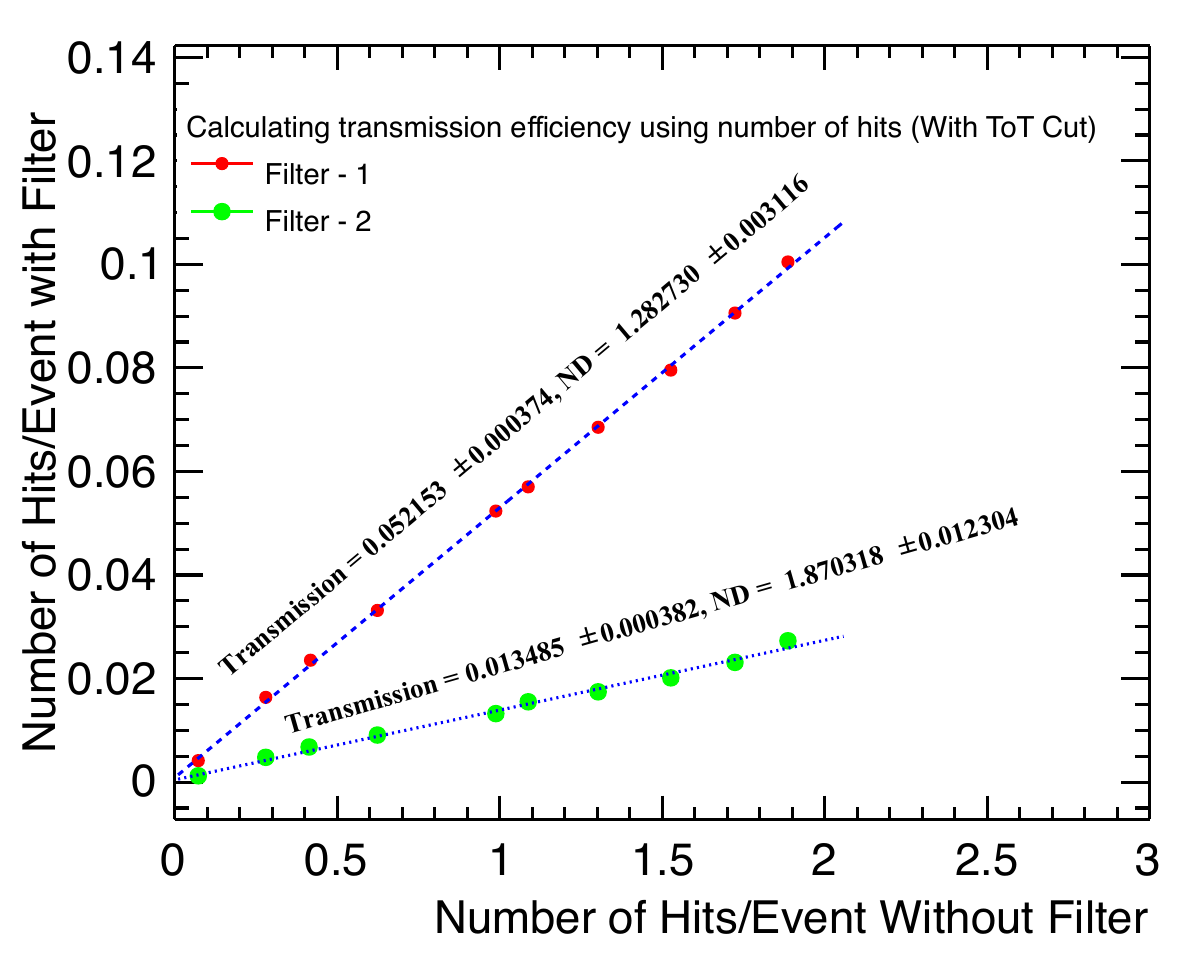}
  \caption{Estimation of the transmission probability of ND filters with cut on ToT ($>$ \SI{3}{\nano\second}) on the registered hits. }
  \label{highoccupancy-calibration-with-tot}
\end{figure}

The procedure described in section~\ref{section-filter-calibration} is repeated for finding the transmission probability of the ND filters by imposing a ToT-cut of $<$ \SI{3}{\nano\second} on the registered hits (shown in figure~\ref{highoccupancy-calibration-with-tot} and tabulated in Table~\ref{table:filter-2}), thus reducing the probability of crosstalk even in low multiplicities ($<$ 3 hits).
The derived transmittance T is used for finding the expected number of hits without filter in section~\ref{section-high-occupancy-result}.

\begin{table}[htb]
\caption{Transmission probability of the ND filters}
\begin{tabular}{lll}
\hline
\textbf{Filter Number} & \textbf{Transmission probability (T)} & \textbf{Neutral density (ND = $\log_{10}\frac{1}{T}$)} \\
\hline
1 &  $0.052153 \pm 0.000374$ & $1.282730 \pm 0.003116$ \\
\hline
2 & $0.013485 \pm 0.00038$ & $1.870318 \pm 0.012304$ \\
\hline
\end{tabular}
\label{table:filter-2}
\end{table}

\newpage
\subsection{Simulation of the double photons}
\label{app-sim-double-photon}
\begin{figure}[h!]
\centering
  \includegraphics[width=0.40\textwidth]{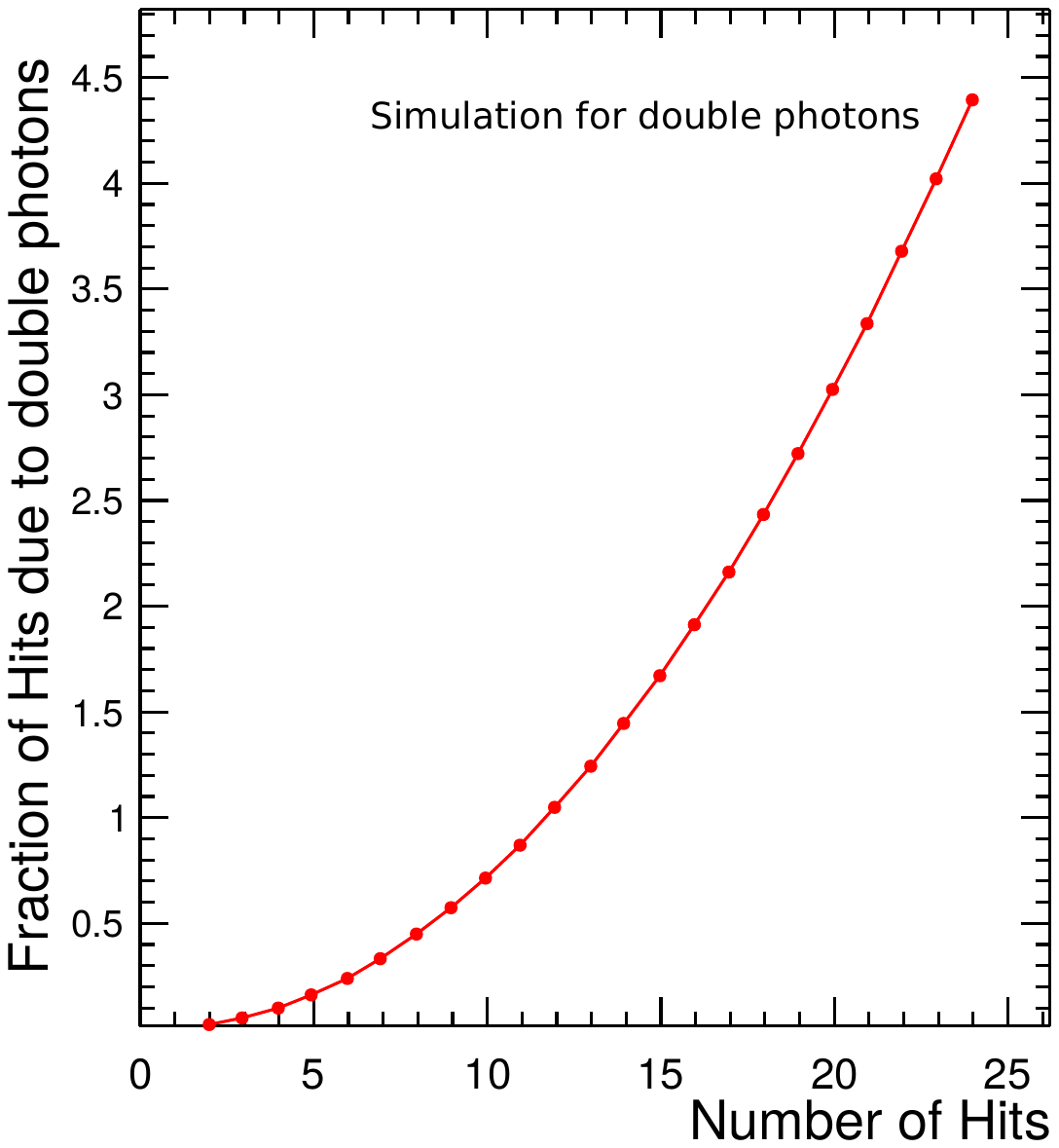}
  \caption{Simulation result for the double photon contribution as a function of hit multiplicity}
  \label{highoccupancy-double-photons}
\end{figure}

 The contribution of double photons is determined through a Monte Carlo simulation.
 This is done by throwing N random numbers corresponding to the (number of hits registered in a MAPMT) into an 8 $\, \times \,$ 8 matrix.
 The probability of finding two values in a single index is found, and considered to be the probability of double photon hits.
 The non-homogeneous illumination of MAPMT, might result in producing double photons often in some regions.
 Therefore, the values in the derived matrix are weighted according to homogeneity of illumination. 
$3 \times 10^6$ events are simulated and plotted in figure~\ref{highoccupancy-double-photons}.

\newpage
\subsection{Scaler rate vs. rate at DAQ for different threshold values}
\label{app-scaler-vs-rate-threshold}
\begin{figure}[!htb]
\centering
  \includegraphics[width=0.5\textwidth ]{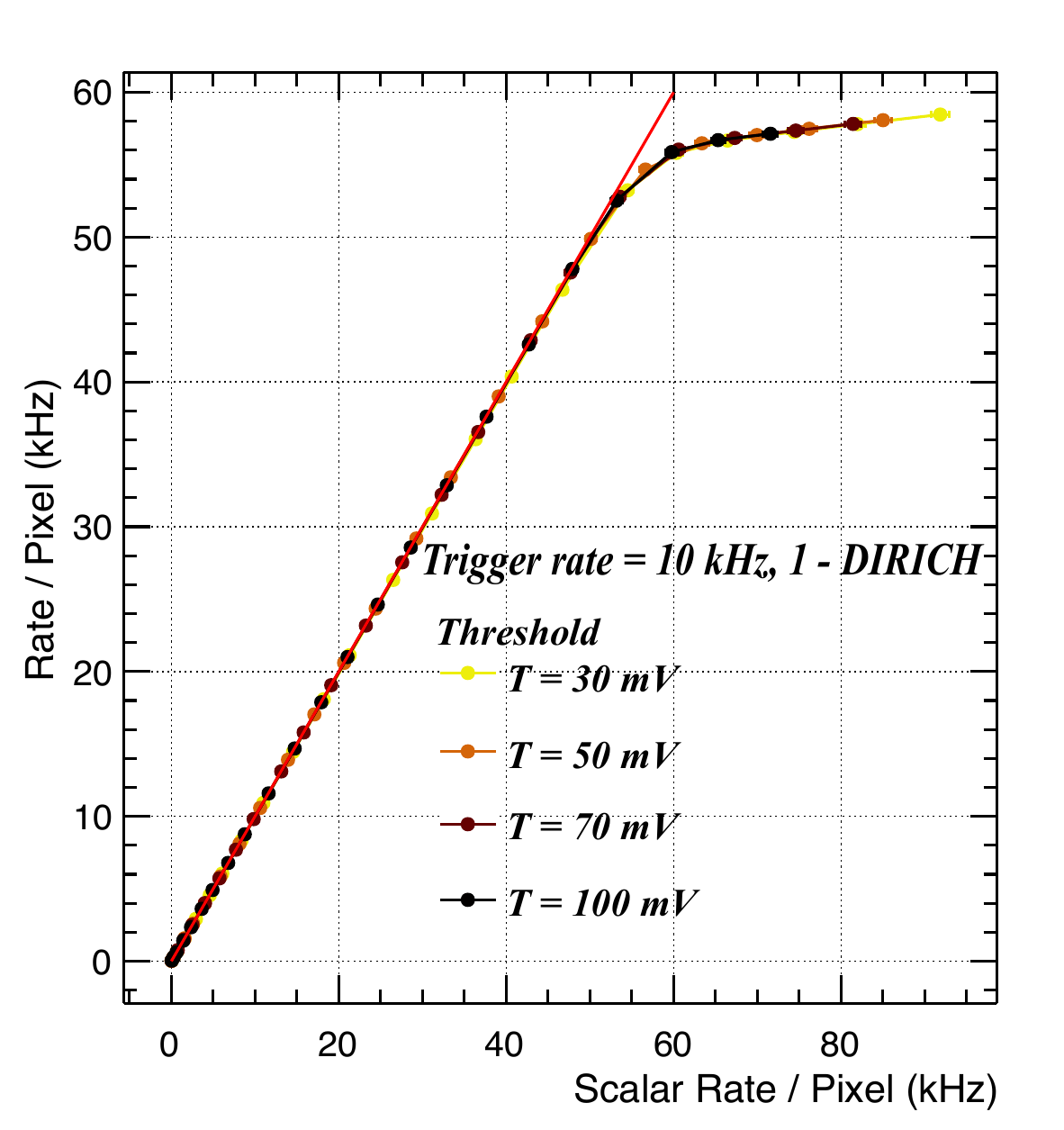}
  \caption{scaler rate vs. rate at DAQ for different threshold values.}
  \label{appendix-rate-diff-threshold}
\end{figure}

A test is performed to check the high rate behavior of the DIRICH FEB for different noise discrimination thresholds.
Since the scaler rate is calculated after the threshold is set, the possibility of any behavior change is not expected, and figure~\ref{appendix-rate-diff-threshold} affirms the assumption.%
\newpage

\subsection{Median noise bandwidth}
\label{app-median-noise}
\begin{figure}[h]
\centering
  \includegraphics[width=1\textwidth ]{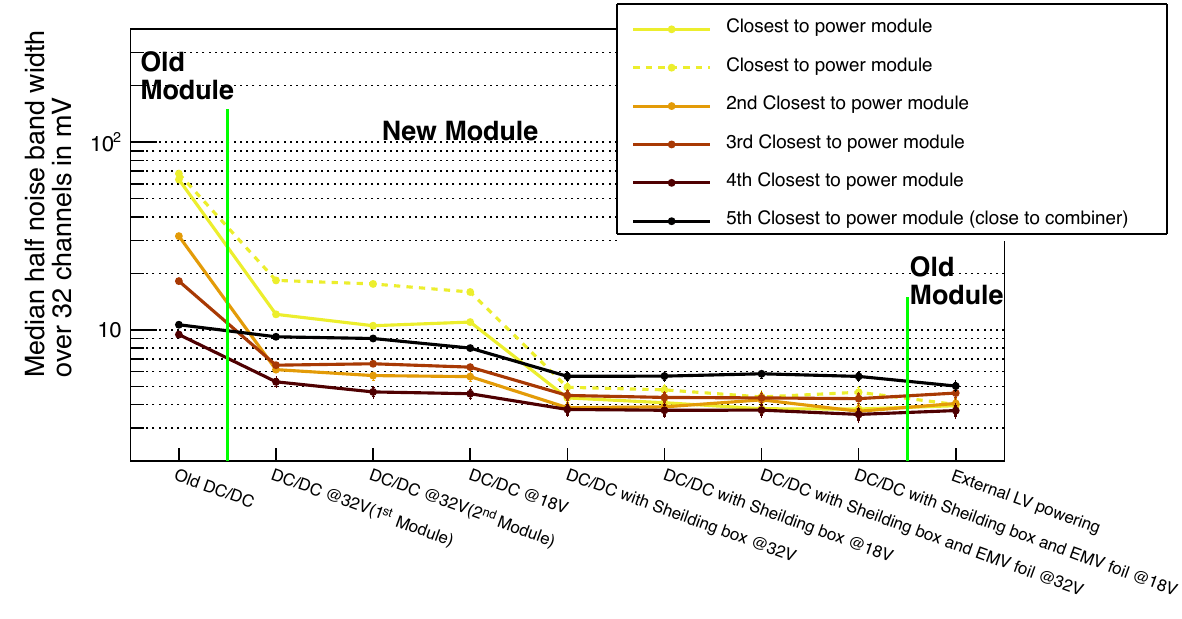}
  \caption{The median half noise bandwidth over 32 channels for different iterations of power modules for different configurations.}
  \label{median half noise}
\end{figure}

To avoid the outlier bias of finding the average and maximum noise bandwidth for each channel in the DIRICH FEB, the median half noise bandwidth for each channel is found (plotted in figure~\ref{median half noise}).
The results indicate, the derived median noise bandwidth is in quite an agreement with the average half noise bandwidth.
\newpage
\subsection{Direct comparison of noise bandwidth due to two power modules at mRICH}
\label{app-direct-comp-mrich}
\begin{figure}[h]
\centering
  \includegraphics[width=0.9\textwidth ]{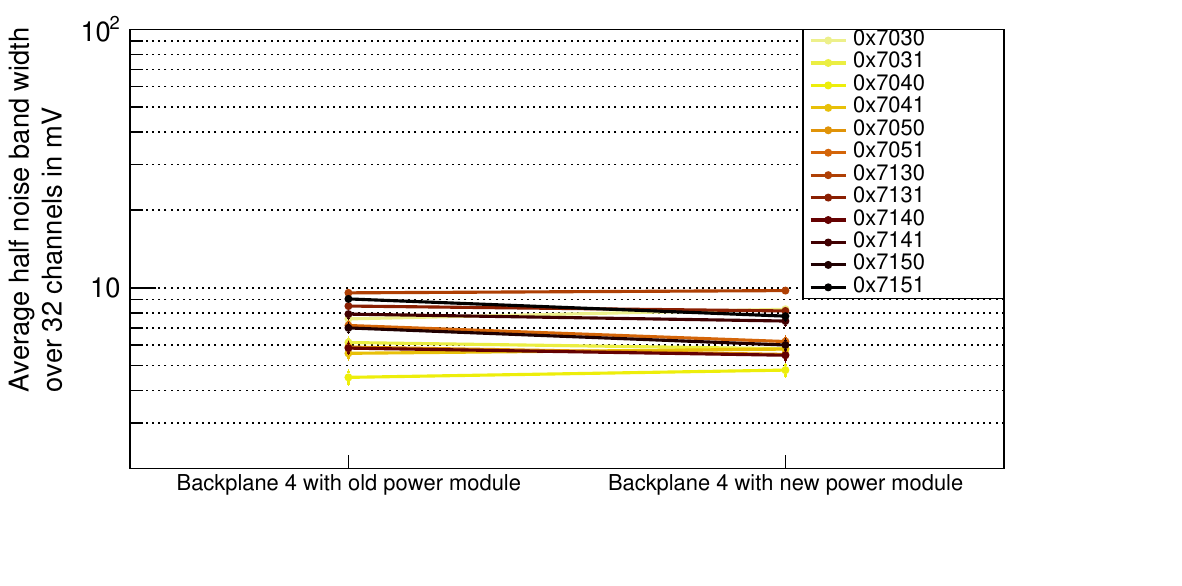}
  \caption{Direct comparison of the noise bandwidth of the DIRICH FEBs due to two different power modules. 
  Backplane 4—Center left at the mRICH setup (see figures ~\ref{lab and mrich setup},~\ref{mrich-result-1}).}
  \label{mcbm half noise}
\end{figure}

The average half noise bandwidth induced on DIRICH FEB by the operation of old and new power modules on the same backplane (derived by performing independent scans) are compared, and the values for each DIRICH FEB are shown in figure~\ref{mcbm half noise}. 
The measured noise levels are consistent with each other, and further the noise bandwidth is lower at the center of the backplane for both the cases. 
\newpage
\subsection{Effect of change in power module to time over threshold measurement}
\label{app-powermodule-tot}
\begin{figure}[h!b]
\centering
  \includegraphics[width=0.8\textwidth ]{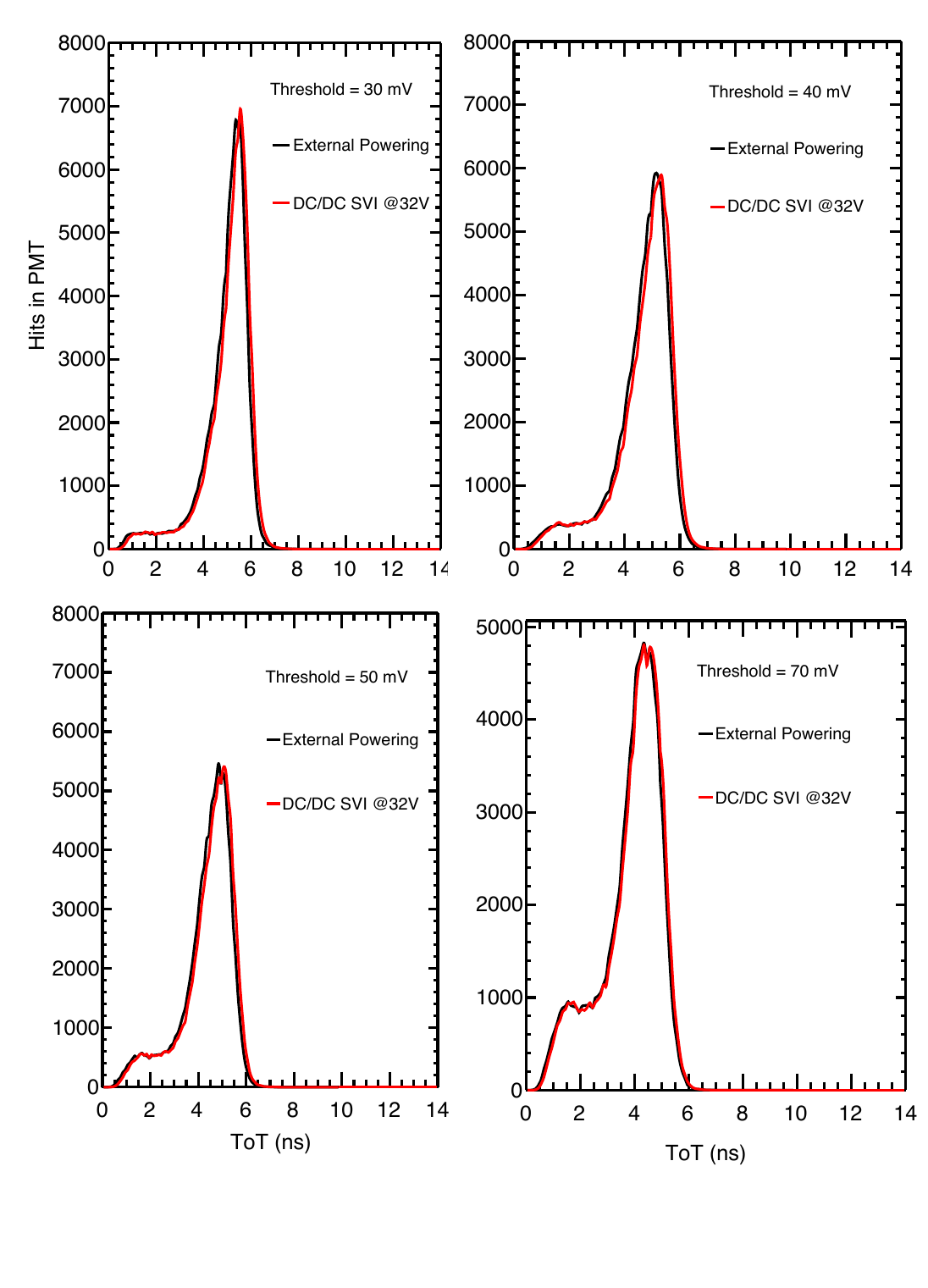}
  \caption{Time over threshold spectra for different thresholds with the external powering and new DC / DC @ \SI{32}{\volt}. }
  \label{tot-power}
\end{figure}

The overall timing performance of DIRICH should be unaffected by any changes to the power module.
In order to verify that a setup as shown in figure~\ref{highoccupancy setup} with the laser is tuned to have less than two hits per event is used.
The ToT of the hits is measured, and the measurement is repeated by changing the power module.
The change in the DC / DC converter has a negligible effect on the overall timing precision of the DIRICH FEB (see figure~\ref{tot-power}).
\end{document}